\documentclass[onecolumn]{elsart3p}
\usepackage{psfrag,amsmath,amssymb,amsfonts,dcolumn,bm,calc,graphicx}
\usepackage[active]{srcltx}
\journal{Journal of Computational Physics}

\begin{document}
\begin{frontmatter}

\title{Interacting particles in two dimensions: numerical solution of the four-dimensional Schr\"odinger equation in a hypercube}

\author{Andr\'as V\'anyolos\corauthref{cor}},
\corauth[cor]{Corresponding author.}
\ead{vanyolos@kapica.phy.bme.hu}
\author{G\'abor Varga}
\address{Department of Physics, Budapest University of Technology and Economics, 1521 Budapest, Hungary}

\begin{abstract}
We study numerically the Coulomb interacting two-particle stationary states of the Schr\"odinger equation, where the particles are
confined in a two-dimensional infinite square well. Inside the domain the particles are subjected to a steeply increasing
isotropic harmonic potential, resembling that in a nucleus. For these circumstances we have developed a fully discretized finite
difference method of the Numerov-type that approximates the four-dimensional Laplace operator, and thus the whole Schr\"odinger
equation, with a local truncation error of $\mathcal{O}(h^6)$, with $h$ being the uniform step size. The method is built on a
89-point central difference scheme in the four-dimensional grid. As expected from the general theorem by Keller
[Num.\ Math.\ \textbf{7}, 412 (1965)], the error of eigenvalues so obtained are found to be the same order of magnitude which we
have proved analytically as well. Based on this difference scheme we have obtained a generalized matrix Schr\"odinger equation by
vectorization. In the course of its numerical solution group theoretical methods were applied extensively to classify energy
eigenvalues and associated two-particle wave functions. This classification scheme inherently accounts for the symmetry property
of the two-particle state under permutation, thereby making it very easy to fully explore the completely symmetric and
antisymmetric subspaces of the full Hilbert space. We have obtained the invariance group of the interacting Hamiltonian and
determined its irreducible representations. With the help of these and Wigner-Eckart theorem, we derived an equivalent block
diagonal form of the eigenvalue equation. In the low-energy subspace of the full Hilbert space we numerically computed the ground
state and many ($\approx200$) excited states for the noninteracting as well as the interacting cases. Comparison with the
noninteracting exact results, which we have found analytically too, reveals indeed that already at a modest resolution of
$h\approx1/15$ our numerical data for the eigenvalues are accurate globally up to three or more digits of precision. Having
obtained the energy eigenvalues and eigenstates we calculated some relevant physical quantities with and without
interaction. These are the static two-particle densities labeled by irreducible representations, one- and two-particle density of
states and some different measures of entanglement, including the reduced density matrix and the von Neumann entropy. All these
quantities signal concordantly that the symmetric states (with respect to permutation of particles) as well as their energies are
affected by the interaction more dramatically then their antisymmetric counterparts.
\end{abstract}

\end{frontmatter}

\section{Introduction}\label{sec:int}
The importance of numerical solution of the Schr\"odinger equation for analytically unsolvable potentials can hardly be
overestimated in any fields of modern physics. In one-dimension (1D) the search for either exact or numerical solutions with ever
increasing accuracy has been in the forefront of research for many decades now \cite{landau-book,kang}. One of the reasons for the
1D case has earned that much attention lies partially in the fact that many higher dimensional problems have spherically symmetric
potentials which then allows separation of variables and the study of the radial equation only. Numerical methods for obtaining
bound state as well as scattering state solutions can be broadly classified into two types: numerical integration or in other
words the shooting methods \cite{johnson,killingbeck93} and matrix methods
\cite{kang,fack85,fack87,hajj82,hajj85,avdelas,kalogiratou}. Also, there are analytical approaches too like perturbative
treatments or the Rayleigh-Ritz variational principle, see Ref.~\cite{fack87} and references therein.

The fact that this fundamental equation of non-relativistic quantum physics is a linear second order differential equation without
a first order term makes it especially suitable to be studied by, beyond the classical finite difference methods
\cite{fack85,fack87}, Numerov's method as well \cite{fack87,chawla,andrew,killingbeck,gonzalez}. As to the classical methods,
these are the most simple finite difference schemes based on the fact that the differential operator representing the kinetic
energy can be expressed as a series of central difference operators \cite{fack85,froberg-book}. Truncating this expansion at the
first, second or third terms yields algebraic eigenvalue equations with tri-, penta- or heptadiagonal matrices,
respectively. Also, it turns out that the local truncation error characterizing the approximation of the continuous differential
equation with the difference equation is, in these cases, $\mathcal{O}(h^n)$, with $n=2$, 4 and 6 and $h$ is the uniform step
size. More interestingly, it can be proven analytically that the errors of eigenvalues so obtained have the same order of
magnitude as the truncation error \cite{keller}. Practically it means that in the heptadiagonal case for example seven to eight
digits of precision can be easily reached. Because of this high accuracy they have been used extensively to calculate the
properties of the ground state and some low energy excited states of many different potentials. These include the anharmonic
oscillator potential, the symmetric double well, the Razavy potential and many others \cite{fack87}.

Despite the success of the classical finite difference schemes mentioned above, the most widely used method for solving the
Schr\"odinger equation is, however, the Numerov method \cite{fack87,chawla,andrew,killingbeck,gonzalez}. This is because the first
approximation already yields very accurate eigenvalues with error at most $\mathcal{O}(h^4)$. This is also true in two and higher
dimensions, though the procedure is slightly complicated by the fact that the recurrence relation involves second-neighbors as
well and so some additional information or physically motivated approximation is needed close to the boundary in order to start
the algorithm \cite{avdelas,gonzalez}. Though its performance, in the sense of accuracy, is obviously not as good as the classical
heptadiagonal approximation, with higher order approximations for the derivatives \cite{killingbeck,allison,jamieson} or with
extrapolation techniques \cite{hajj82,hajj85,avdelas} (Richardson, Aitken) it can be refined easily up to $\mathcal{O}(h^6)$ or
higher.

As to higher dimensions, it is rather interesting that despite the availability of a great deal of numerical methods, developed
primarily for 1D, only a few of them have been adopted to two or higher dimensional Schr\"odinger problems. Perhaps the most
natural and simple approach of all is the classical fully discretized five-point method with an error of
$\mathcal{O}(h^2)$ \cite{hajj82,hajj85,rozsa-book}. This was successfully applied in two-dimensions (2D) for potentials that lack
rotational symmetry. Later both the standard Numerov method and its extended version have been transported to 2D in order to
obtain, respectively, $\mathcal{O}(h^4)$ and $\mathcal{O}(h^6)$ accurate results for the lowest eigenvalues of some Coulombic
potentials \cite{avdelas}. Also, partial discretization methods have been proposed recently for the study of the highly anisotropic
Henon-Heiles potential \cite{kalogiratou}.

Our aim in this paper is to study numerically an interacting two-particle problem in two-dimensions. The geometry of the problem
and the number of particles imply naturally that we have to face a four-dimensional Schr\"odinger equation. The method we apply is
the full discretization of the differential equation which we solve with a generalized Numerov-type approach. Generalized in the
sense that (i) the algorithm is $\mathcal{O}(h^6)$ accurate, and (ii) to our knowledge Numerov's method has not been ported to
dimensions greater than two \cite{gabor}. From the physics point of view we are of course primarily interested in the study of
interaction and so the mathematical background we have developed along the way could be considered as a necessity. Nevertheless,
we feel that it is alone very interesting because it reveals the connection of the high-dimensional finite difference method with
the theory of matrices composed of commuting blocks. It could be utilized elsewhere and in other dimensions as well. The paper is
organized as follows. In Section~\ref{sec:sch} we start with the precise formulation of the quantum mechanical problem, then
develop the necessary algebra. In Section~\ref{sec:group} we study the group theoretical aspects of the
problem. Sections~\ref{sec:nointres} and~\ref{sec:intres} are devoted to our numerical results without and with interaction,
respectively. Finally, Section~\ref{sec:conclusions} summarizes our conclusions.

\section{Schr\"odinger equation and stationary states}\label{sec:sch}
\subsection{Problem formulation}\label{sub:problem}
Let us start by the formulation of the quantum mechanical problem of two identical interacting particles. The quantum theory we
are about to explore is non-relativistic and therefore the spin degrees of freedom of particles will not show up explicitly in the
Schr\"odinger equation. For the same reason neither spin-orbit coupling will be considered here. The total two-particle wave
function is thus separable as
$\Psi(\mathbf{r}_1,\sigma_1;\mathbf{r}_2,\sigma_2)=\psi(\mathbf{r}_1,\mathbf{r}_2)\chi(\sigma_1,\sigma_2)$, where $\chi$ is the
spin dependent component for the construction of which the knowledge of statistics (Fermi or Bose) and the appropriate spin
quantum number is essential, while $\psi$ is the orbital part governed by the usual time independent Schr\"odinger equation
\begin{subequations}\label{qm-problem}
\begin{equation}
\left(-\frac{\hbar^2}{2m}(\Delta_1+\Delta_2)+U(\mathbf{r}_1)+U(\mathbf{r}_2)
+V(|\mathbf{r}_1-\mathbf{r}_2|)\right)\psi(\mathbf{r}_1,\mathbf{r}_2)=E\psi(\mathbf{r}_1,\mathbf{r}_2).\label{schrodinger}
\end{equation}
Here $U$ is the external potential that acts on both particles separately and $V$ is the pair interaction depending only on the
distance between them. In order to determine possible energy eigenvalues $E$ and the corresponding states $\psi$ unambiguously,
that is the spectrum of the Hamiltonian, one needs to supplement Eq.~\eqref{schrodinger} with a boundary condition. In the paper
we consider a two-dimensional square shaped potential well centered on the origin
\begin{equation}
D_2=[-b,b]\times[-b,b],\label{domain}
\end{equation}
inside of which $U(\mathbf{r})$ is a well-behaved smooth function, but outside $U(\mathbf{r})=U_0\to\infty$. This infinite
potential barrier gives a constraint on the wave function $\psi(\mathbf{r}_1,\mathbf{r}_2)$: for both arguments on the boundary
\begin{equation}
\psi|_{\partial D_2}=0.\label{constraint}
\end{equation}
\end{subequations}
Now that the dimensionality of the problem is specified, the Laplacian in Eq.~\eqref{schrodinger} is expressed as
$\Delta_i=\partial^2/\partial x_i^2+\partial^2/\partial y_i^2$, $i=1,2$. Clearly, it is the Descartes coordinate system that best
suits the rectangular domain. From purely mathematical point of view Eqs.~\eqref{qm-problem} define a (self-consistent) boundary
value problem, where the parameter $E$ has to be determined as well. It is its precise numerical solution that is the primary
object of the present investigation.

\subsection{Method of errors $\mathcal{O}(h^6)$}\label{sub:errors}
From now on, to facilitate numerical calculations, we use atomic units where the unit of length is the Bohr-radius
$a_0=\hbar^2/(me^2)$ and energy is measured in the Rydberg unit $E_\text{Ryd}=e^2/(2a_0)=\hbar^2/(2ma_0^2)$. It is also convenient
to introduce unified notation for space variables as
\begin{equation}
(x_1,y_1,x_2,y_2)\to(x_1,x_2,x_3,x_4)=\mathbf{x},\label{newvar}
\end{equation}
and an effective potential with
\begin{equation}
\tilde U(\mathbf{x})=U(\mathbf{r}_1)+U(\mathbf{r}_2)+V(|\mathbf{r}_1-\mathbf{r}_2|).\label{effectivepot}
\end{equation}
Equation~\eqref{newvar} shows that in the unified four-dimensional configuration space the first two dimensions composed of
$(x_1,x_2)$ belong to one particle, say the ``first'', and the complementary subspace to the other, the ``second''.  With all
these the Schr\"odinger equation can be reformulated as a Dirichlet problem for Poisson's equation
\begin{subequations}\label{poisson}
\begin{align}
\Delta\psi(\mathbf{x})&=-f(\mathbf{x}),\label{sch}\\
f(\mathbf{x})&=(E-\tilde U(\mathbf{x}))\psi(\mathbf{x}),\label{f}
\end{align}
where $\Delta=\sum_{i=1}^4\partial^2/\partial x_i^2$, the region of confinement is a four-dimensional (4D) hypercube
\begin{equation}
D_4=[-b,b]\times[-b,b]\times[-b,b]\times[-b,b],\label{hypercube}
\end{equation}
and the boundary condition in Eq.~\eqref{constraint} now reads as
\begin{equation}
\psi|_{\partial D_4}=0.\label{4Dconstraint}
\end{equation}
\end{subequations}
It is clear that Eqs.~\eqref{poisson} can also be thought of as a one-particle Schr\"odinger equation for an abstract
four-dimensional particle that is confined in a box where the ``external'' potential is given by Eq.~\eqref{effectivepot}. This
picture will turn out to be very useful when we turn our attention to the representation theory of the invariance group of the
Hamiltonian in Section~\ref{sec:group}. Before that, however, we develop the necessary algebra: based on a generalized
Numerov-type approach we derive an equivalent matrix Schr\"odinger equation that provides the wave function on a finite grid
\cite{fack87,avdelas,kalogiratou}. To this end we divide the interval $[-b,b]$ into $(n+1)$ equal segments with increment $h$
given by $h=2b/(n+1)$. By this procedure a linearly spaced 4D cubic grid is obtained, where the $p$th lattice point in the $i$th
dimension is $x_{i,p}=-b+ph$, $p=1,\dots,n$ and $i=1,\dots,4$. Further, we use the simplified notation
\begin{equation}
v(x_{1,p},x_{2,i},x_{3,k},x_{4,l})=v_{pikl},\qquad p,i,k,l=1,2,\dots,n,\label{vgrid}
\end{equation}
for any multivariable function $v$. Now define first-, second-, third- and fourth-neighbor difference operators,
respectively. These are the natural generalizations of the central difference operators (three-point, five-point, \dots) applied
commonly in one-dimension \cite{fack85,fack87} and two-dimensions \cite{hajj82,hajj85,avdelas,rozsa-book}. In 4D the first-neighbor
difference operator reads
\begin{subequations}\label{diffop}
\begin{equation}
\square_1\psi_{pikl}=\sum_{\alpha=\pm1}(\psi_{p+\alpha,ikl}+\psi_{p,i+\alpha,kl}+\psi_{pi,k+\alpha,l}
+\psi_{pik,l+\alpha})-8\psi_{pikl}.\label{first}
\end{equation}
On the right hand side $g_1=8$ is the total number of first-neighbor sites. For clarity we shall note here that $\square_1$ acts
on the function and not on its values, thus on the left hand side $\{\square_1\psi\}_{pikl}$ would be more precise. Nevertheless,
we stick with the shorter notation as there should be no confusion by dropping those extra braces. Similarly, the second-neighbor
difference operator is
\begin{equation}
\square_2\psi_{pikl}=\sum_{\alpha,\beta=\pm1}(\psi_{p+\alpha,i+\beta,kl}+\psi_{p+\alpha,i,k+\beta,l}
+\psi_{p+\alpha,ik,l+\beta}+\psi_{p,i+\alpha,k+\beta,l}+\psi_{p,i+\alpha,k,l+\beta}+\psi_{pi,k+\alpha,l+\beta})
-24\psi_{pikl},\label{second}
\end{equation}
where $g_2=24$ is the total number of second-neighbor sites in 4D. In the third-neighbor case
\begin{equation}
\square_3\psi_{pikl}=\sum_{\alpha,\beta,\gamma=\pm1}(\psi_{p+\alpha,i+\beta,k+\gamma,l}+\psi_{p+\alpha,i+\beta,k,l+\gamma}
+\psi_{p+\alpha,i,k+\beta,l+\gamma}+\psi_{p,i+\alpha,k+\beta,l+\gamma})-32\psi_{pikl},\label{third}
\end{equation}
with $g_3=32$, and finally
\begin{equation}
\square_4\psi_{pikl}=\sum_{\alpha,\beta,\gamma,\delta=\pm1}\psi_{p+\alpha,i+\beta,k+\gamma,l+\delta}
-16\psi_{pikl}+[\square_1\psi_{pikl}]_{h\to2h}.\label{fourth}
\end{equation}
\end{subequations}
Note that the fourth-neighbor sites of a given lattice point are at a distance
$\sqrt{4}h=\sqrt{\alpha^2+\beta^2+\gamma^2+\delta^2}h$, so they can be situated along the 4D space diagonal as well as along the
coordinate axes. Hence, there are altogether $g_4=24$ of them. This fact is represented by the last term in
Eq.~\eqref{fourth}. With these expressions at hand we can make a linear combination with arbitrary constants $\alpha,\beta,\gamma$
and $\delta$, and a subsequent Taylor expansion in $h$ up to order $h^6$ yields
\begin{align}
(\alpha\square_1+\beta\square_2+\gamma\square_3+&\delta\square_4)\psi_{pikl}=h^2(\alpha+6\beta+12\gamma+12\delta)\Delta\psi\notag\\
&\phantom{=}+\frac{h^4}{12}([\alpha+6\beta+12\gamma+24\delta](\psi_{x_1^4}+\dots+\psi_{x_4^4})
+[12\beta+48\gamma+48\delta](\psi_{x_1^2x_2^2}+\dots))\notag\\
&\phantom{=}+\frac{h^6}{360}([\alpha+6\beta+12\gamma+72\delta]
(\psi_{x_1^6}+\dots+\psi_{x_4^6})+[360\gamma+720\delta](\psi_{x_1^2x_2^2x_3^2}+\dots)\notag\\
&\phantom{=}\phantom{+\frac{h^6}{360}(}+[30\beta+120\gamma+120\delta]
(\psi_{x_1^2x_2^4}+\psi_{x_1^4x_2^2}+\psi_{x_1^2x_3^4}+\psi_{x_1^4x_3^2}+\dots)).\label{basic}
\end{align}
Here, all derivatives on the right hand side must be evaluated at the lattice point $p,i,k,l$. As to the partial derivatives, we
used shorthand notation, for example
\begin{equation}
\psi_{x_i^px_j^q\dots}=\left(\frac{\partial^p}{\partial x_i^p}\frac{\partial^q}{\partial x_j^q}\dots\right)\psi.\label{deriv}
\end{equation}
The first term on the right hand side of Eq.~\eqref{basic} is of order $h^2$ and consists of only second derivatives. Similarly,
the second (third) term is of order $h^4$ ($h^6$), consisting of only fourth (sixth) order partial derivatives. Moreover, the
orders of the derivatives with respect to each variable are even. The fact that this must be so and that there are no terms
proportional to odd powers of $h$ are justified by the definitions of $\square_i$. Namely, all the neighborhoods of a given
lattice point appearing in Eqs.~\eqref{diffop} are in a sense complete: they are all symmetric under 4D spatial inversion and
reflections, thus odd terms must drop out indeed. With simple combinatorial arguments it is easy to see that there are 6, 4 and 12
terms exhibiting structure like $\psi_{x_1^2x_2^2}$, $\psi_{x_1^2x_2^2x_3^2}$ and $\psi_{x_1^2x_2^4}$,
respectively. Equation~\eqref{basic} holds for every analytic function and so are
\begin{subequations}\label{identities}
\begin{gather}
\Delta^2\psi=\psi_{x_1^4}+\dots+\psi_{x_4^4}+2(\psi_{x_1^2x_2^2}+\psi_{x_1^2x_3^2}+\dots),\label{seged1}\\
\Delta^3\psi=\psi_{x_1^6}+\dots+\psi_{x_4^6}
+3(\psi_{x_1^2x_2^4}+\psi_{x_1^4x_2^2}+\psi_{x_1^2x_3^4}+\psi_{x_1^4x_3^2}+\dots)+6(\psi_{x_1^2x_2^2x_3^2}+\dots),\label{seged2}
\end{gather}
and
\begin{equation}
(\partial_{x_1^2x_2^2}+\dots+\partial_{x_3^2x_4^2})\Delta\psi=3(\psi_{x_1^2x_2^2x_3^2}+\dots)+
\psi_{x_1^2x_2^4}+\psi_{x_1^4x_2^2}+\psi_{x_1^2x_3^4}+\psi_{x_1^4x_3^2}+\dots.\label{seged3}
\end{equation}
\end{subequations}
These identities can be verified easily by direct calculation.

So far the calculations might seem rather artificial and the formulas apply to all differentiable function including the true (yet
unknown) physical solution of Eqs.~\eqref{poisson}. Now, the crucial recognition that validates prior work is the fact that one
can choose the parameters of Eq.~\eqref{basic} in such a way that its right hand side is a differential form of $\Delta\psi$
only. It is easy to see that the term of order $h^4$ is much like Eq.~\eqref{seged1}, one only has to choose the prefactors
appropriately. In a similar manner one can also realize that the term of order $h^6$ consists of derivatives that appear in
Eqs.~\eqref{seged2} and \eqref{seged3}. Therefore, with a suitable choice for $\alpha,\beta,\gamma$ and $\delta$ we (possibly) can
achieve that to be proportional to
\begin{equation}
\Delta^3\psi+\eta(\partial_{x_1^2x_2^2}+\dots+\partial_{x_3^2x_4^2})\Delta\psi,\label{eta}
\end{equation}
where $\eta$ is introduced as a fifth unknown. With all these findings a system of linear equations is obtained
\begin{subequations}\label{firstset}
\begin{align}
2(\alpha+6\beta+12\gamma+24\delta)&=12\beta+48\gamma+48\delta,\\
3(\alpha+6\beta+12\gamma+72\delta)&=30\beta+120\gamma+120\delta-\eta,\\
6(\alpha+6\beta+12\gamma+72\delta)&=360\gamma+720\delta-3\eta,
\end{align}
\end{subequations}
with solution
\begin{subequations}
\begin{align}
\alpha&=12\gamma,\\
\beta&=\gamma+8\delta,\\
\eta&=60\gamma. 
\end{align}
\end{subequations}
As there are only three equations it is not surprising that among the five unknowns two, say $\gamma$ and $\delta$, remained
undetermined. The question how to fix these degrees of freedom is very interesting from both purely mathematical and physical
points of view. We will consider this issue in detail in Subsection~\ref{sub:free} and we will see that it is rather nontrivial
and requires very careful analysis. Now, putting all these expressions together and remembering the fact that the physical
solution satisfies Eq.~\eqref{sch} we end up with
\begin{subequations}\label{majdnem}
\begin{equation}
-\frac{1}{30h^2}\left(12\gamma\square_1+(\gamma+8\delta)\square_2+\gamma\square_3+\delta\square_4\right)\psi_{pikl}=w_{pikl},
\qquad p,i,k,l=1,\dots,n,\label{majdnemegyenlet}
\end{equation}
where
\begin{equation}
w_{pikl}=(\gamma+2\delta)f_{pikl}+h^2\left(\frac{\gamma}{12}+\frac{\delta}{5}\right)(\Delta
f)_{pikl}+ \frac{h^4}{12}\left(\left(\frac{\gamma}{30}+\frac{2\delta}{15}\right)(\Delta^2f)_{pikl}+
\frac{\gamma}{15}(f_{x_1^2x_2^2}+\dots)_{pikl}\right).\label{w}
\end{equation}
\end{subequations}
This is the equivalent of Poisson's equation, now discretized on a 4D grid \cite{rozsa-book}. If we were to solve a true Poisson's
equation numerically with a given differentiable function $f$, this would be the starting point. Then of course we could keep the
formalism as easy as possible by taking $\delta=0$ and $\gamma=1$, without affecting precision. However, in our case $f$ is
unknown as it is itself determined by $\psi$, see Eq.~\eqref{f}. It follows at once that this equation in the present form is not
(completely) a linear difference equation, because $w_{pikl}$ still involves continuous derivatives. This difficulty, however, can
be easily overcome with the technique developed so far: one has to express the right hand side of Eq.~\eqref{w} as a difference
operator acting on $f$, with the constraint that the error so introduced must not exceed $\mathcal{O}(h^6)$. This is because the
local truncation error (we have already made by cutting the series expansion in Eq.~\eqref{basic}) on the right hand side of
Eq.~\eqref{majdnemegyenlet} is at most $\mathcal{O}(h^6)$. Making use of Eq.~\eqref{basic} again, now written for a function $f$
and the requirements regarding the error of discretization, a second system of linear equations is found
\begin{subequations}\label{secondset}
\begin{align}
\alpha'+6\beta'+12\gamma'+12\delta'&=\frac{\gamma}{12}+\frac{\delta}{5},\label{elso}\\
\alpha'+6\beta'+12\gamma'+24\delta'&=\frac{\gamma}{30}+\frac{2\delta}{15},\\
12\beta'+48\gamma'+48\delta'&=2\left(\frac{\gamma}{30}+\frac{2\delta}{15}\right)+\frac{\gamma}{15},
\end{align}
\end{subequations}
whose solution is
\begin{subequations}
\begin{align}
\alpha'&=12\gamma'-\frac{\gamma}{30},\\
\beta'&=-4\gamma'+\frac{\gamma}{36}+\frac{2\delta}{45},\\
\delta'&=-\frac{\gamma}{240}-\frac{\delta}{180}.\label{nontrivial}
\end{align}
\end{subequations}
As in Eqs.~\eqref{firstset} before, the number of equations is again less than that of unknowns (with primes), and as such one
variable, say $\gamma'$, varies freely. Putting these results together leads finally to
\begin{multline}
-\frac{1}{30h^2}\left(12\gamma\square_1+(\gamma+8\delta)\square_2+\gamma\square_3+\delta\square_4\right)\psi_{pikl}\\
=\left(\gamma+2\delta+\left(12\gamma'-\frac{\gamma}{30}\right)\square_1+
\left(\frac{\gamma}{36}+\frac{2\delta}{45}-4\gamma'\right)\square_2+\gamma'\square_3-
\left(\frac{\gamma}{240}+\frac{\delta}{180}\right)\square_4\right)f_{pikl},
\label{vegso}
\end{multline}
where, as required, the error is at most $\mathcal{O}(h^6)$. This is again a discretized version of Poisson's equation that
provides $\psi_{pikl}$ given the set of $f_{pikl}$ is known. Though its solution is the same as that of Eqs.~\eqref{majdnem}, in
certain cases this form might be better suited for the particular problem. Suppose for example that the values of $f$ are only
available at lattice points, and as such performing the differentiation prescribed in Eq.~\eqref{w} cannot be carried out
explicitly. For that, first a smooth multivariable interpolation would be necessary, but this intermediate (auxiliary) step is
totally superfluous, as Eq.~\eqref{vegso} yields the same result, thereby revealing the true usefulness of this formula. As to the
solution of the Schr\"odinger equation, which is our primary object, one must remember that the values of $f$ will indeed be
available at the lattice points only, as according to Eq.~\eqref{f}
\begin{equation}
f_{pikl}=(E-\tilde U_{pikl})\psi_{pikl},\label{f2}
\end{equation}
and this results in the usual self-consistent equation, the quantum mechanical eigenvalue problem.

Equation~\eqref{vegso} is the main result of this subsection. In order to find the eigenstates and the energy eigenvalues we need
to say something about the remaining parameters, because they might affect the outcome. We will see indeed that they do. Therefore
now we proceed with the analysis of this question.

\subsection{Free parameters and matrix representation}\label{sub:free}
In this subsection our main concern is the issue of free parameters $\gamma$, $\delta$ and $\gamma'$, that are left undetermined
in the Schr\"odinger equation, Eq.~\eqref{vegso}. This equation, in conjunction with Eq.~\eqref{f2}, is in fact a huge coupled
system of linear difference equations, where the indices of lattice points run in the range $1,\dots,n$. These points make up a
dense cubic grid inside the 4D hypercube. Lattice sites, that are on its surface, have coordinates where at least one of $p,i,k$
or $l$ is equal to $0$ or $n+1$. According to the boundary condition of Eq.~\eqref{4Dconstraint}, here $\psi_{pikl}=0$. Further,
the physical wave function must also vanish outside the boundary because of the infinite potential barrier we have imposed. This
we know from the analytical solution. We have already pointed out these before, but at this point it is worth repeating and having
them in mind because of the following intriguing property of Eq.~\eqref{vegso}. The fourth-neighbor operator, when acting on a
function, takes into account neighboring sites that are two steps away along the coordinate axes from a given point. If we
evaluate $\square_4\psi_{pikl}$ right next to the boundary, that is say $p=1$, the result involves values of $\psi$ taken outside
the boundary, in this case $\psi_{-1,ikl}$. Function values taken at grid points with indices equal to $-1$ or $n+2$ are
fictitious. If we were to write down the equations that apply to them, new values were introduced that are located even farther
from the boundary, leading to an infinite hierarchy of equations. We would like to emphasize that though it is very tempting to
set these values to zero (because of the potential barrier), and thereby cutting the hierarchy, from numerical point of view this
procedure is in principle not adequate. The true physical wave function does indeed vanish identically outside the domain, the
non-zero fictitious values of $\psi$, however, serve to determine the field inside consistently. The only condition we can make
use of when solving the equations for the field inside is the boundary condition.

In order to cut the infinite hierarchy and resolve this problem one can invoke Lagrange interpolation and express fictitious
values with those that lie inside the domain. To retain full consistency in the sense that errors introduced in different ways are
at most $\mathcal{O}(h^6)$ a six-point interpolation is necessary, for example \cite{rozsa-book}
\begin{equation}
\psi_{-1,ikl}=\sum_{j=1}^6(-1)^{j-1}\binom{6}{j}\psi_{j-1,ikl}.\label{lagrange}
\end{equation}

The outlined interpolation technique should only be applied when Eq.~\eqref{vegso} is considered right next to the boundary and
even in this case the extra precision it provides compared to that when all fictitious values are set to zero by hand is
presumably negligible, given the fact that $\psi$ vanishes on the boundary anyway. Therefore, hereafter we neglect this and take
$\psi_{pikl}=0$ everywhere outside the surface of the hypercube. It will turn out soon that luckily this step does not affect the
required $\mathcal{O}(h^6)$ accuracy of our calculations, at least as far as the eigenvalues are concerned. We note that these
kind of difficulties arising at the boundary are well known in lower dimensions as well, not just in Numerov's method
\cite{avdelas,gonzalez} but also in the usual classical fourth-order and sixth-order methods \cite{fack85}.

According to what has been said so far, it is clear that among the difference operators $\square_4$ is somewhat special. Moreover,
as Eq.~\eqref{nontrivial} suggests, the prefactors $\delta$ and $\delta'$ cannot be made vanish at the same time, otherwise
$\gamma=0$ would hold too and we would be left with only the trivial solution of Eq.~\eqref{firstset}. So, at this point we have a
degree of freedom how we distribute $\square_4$ on the two sides of Eq.~\eqref{vegso}. We find it convenient to set $\delta=0$,
thereby eliminating it from the left side. Then, knowing $\gamma\ne0$ must be, we can take without loss of generality $\gamma=1$
to obtain
\begin{multline}
-\frac{1}{30h^2}\left(12\square_1+\square_2+\square_3\right)\psi_{pikl}\\
=\left(1+\left(12\gamma'-\frac{1}{30}\right)\square_1+
\left(\frac{1}{36}-4\gamma'\right)\square_2+\gamma'\square_3-\frac{1}{240}\square_4\right)f_{pikl},\qquad p,i,k,l=1,\dots,n.
\label{vegso2}
\end{multline}

Now it is easy to see that all four difference operators appear indeed in the equation, neither of them can be completely
eliminated, provided of course that we insist on the $\mathcal{O}(h^6)$ accuracy. From this and from Eqs.~\eqref{diffop} it
follows that the method we have just developed is actually a fully discretized 89-point central difference scheme on the 4D
grid. As mentioned in the introduction in Section~\ref{sec:int}, in the 2D Schr\"odinger problem the full and partial
discretization methods have already been applied for the study of anisotropic Coulombic potentials
\cite{hajj82,avdelas,kalogiratou}. However, in dimensions greater than two the literature is not very helpful on this, to our
knowledge no detailed calculations have been performed so far \cite{gabor}. Our approach might be one of the first to fill up this
gap.

Setting values of two parameters turned out to be quite easy, but it is not the case with the last one and the rest of this
subsection is devoted to this issue. In order to show how nontrivial the result is compared to $\gamma$ and $\delta$, we shall
anticipate it here
\begin{equation}
\gamma'=\frac{23}{3840}\approx0.0059.\label{gammav}
\end{equation}
Before we go into details of its proof, we believe it is worthwhile and instructive to examine how Eq.~\eqref{vegso2} would be
modified if we followed the standard approach and derived it with an error of $\mathcal{O}(h^2)$. This would then be the analogue
of the most simple five-point scheme in two-dimensions \cite{hajj82,avdelas,rozsa-book}. It is not hard to see that in this case
only the first term remains in the large parentheses. This corresponds to the usual case of an eigenvalue problem where there is
only the identity operator on the right hand side. As opposed to this, had we wanted to reach an error of at most
$\mathcal{O}(h^4)$, it would have been sufficient to stop at the second term in Eq.~\eqref{w} and to express that with difference
operators. This result leads us to the following important conclusion: whenever the required precision reaches $\mathcal{O}(h^4)$,
or higher as in Eq.~\eqref{vegso2}, the usual eigenvalue problem turns into a generalized one because even the right hand side,
where the eigenvalue $E$ shows up, contains difference operator(s). This observation characteristic to Numerov's method is well
known in lower dimensions as well \cite{fack87,avdelas,kalogiratou}.

\begin{figure}
\begin{center}
\includegraphics[width=0.22\textwidth]{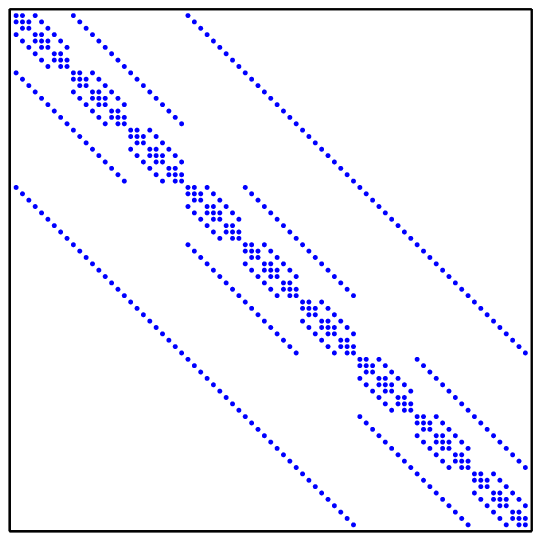}\hfill
\includegraphics[width=0.22\textwidth]{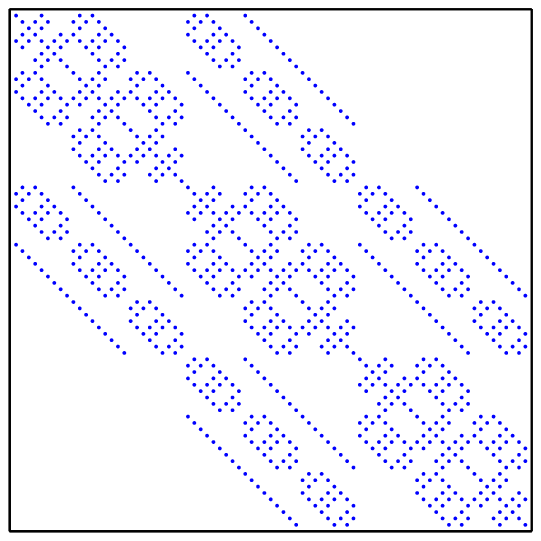}\hfill
\includegraphics[width=0.22\textwidth]{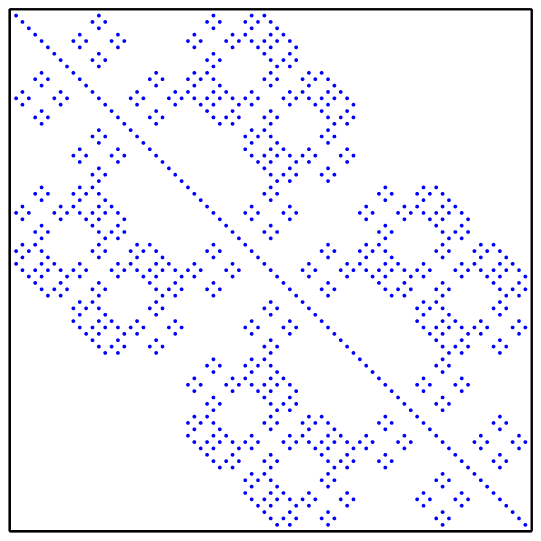}\hfill
\includegraphics[width=0.22\textwidth]{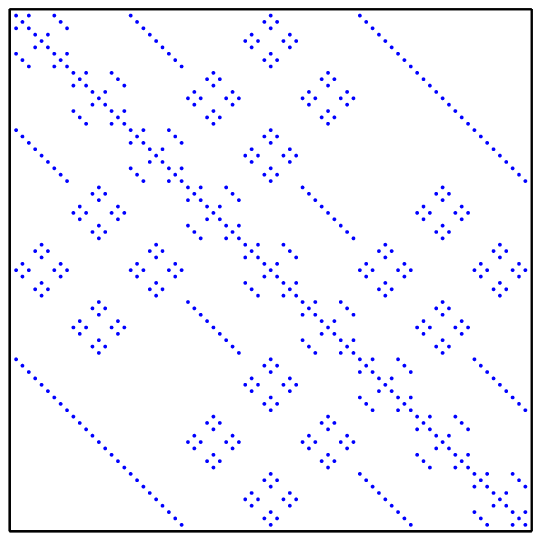}
\end{center}
\caption{\label{fig:sparsity}(Color online) Sparsity patterns of $\mathbf{M}_1$, $\mathbf{M}_2$, $\mathbf{M}_3$ and $\mathbf{M}_4$
  (from left to right) at a very low resolution of $n=3$. Their total size is $n^4=81$.}
\end{figure}

The difference equation~\eqref{vegso2} means in fact $n^4$ coupled linear equations. In order to handle them together we form
large column vectors, so-called stacks of dimension $n^4$ as
\begin{equation}
v_\mu=v_{pikl},\label{vectorindex}
\end{equation}
where the one-to-one correspondence between lattice point indices and the vector index $(\mu=1,\dots,n^4)$ is given by
\begin{equation}
\mu=p+n(i-1)+n^2(k-1)+n^3(l-1).\label{indextransform}
\end{equation}
By this procedure any function $v$ defined on the 4D grid can be mapped into a vector $\mathbf{v}\in\mathbb{R}^{n^4}$ and vice
versa. Equation~\eqref{indextransform} might remind one of the relation between indices of a Kronecker-product (direct-product) of
matrices and those of the constituents. This is not surprising as $\mathbf{v}$ can be written as
\begin{equation}
\mathbf{v}=\sum_{p,i,k,l=1}^nv_{pikl}\,\mathbf{l}\otimes\mathbf{k}\otimes\mathbf{i}\otimes\mathbf{p},\label{vector}
\end{equation}
where $\mathbf{p,i,k}$ and $\mathbf{l}$ are column vectors of size $n$ whose elements differ from zero only at the $p$th, $i$th,
$k$th and $l$th position, respectively, where they all equal one.\footnote{Note that two definitions are used commonly in the
  literature for the Kronecker-product of matrices, $\mathbf{C}=\mathbf{A}\otimes\mathbf{B}$. We use the one that places the
  second matrix in the first, that is $\mathbf{C}$ is a large block matrix (sometimes called hypermatrix), where the $i$th block
  in the $j$th column is $a_{ij}\mathbf{B}$.} Now that we know how to map multivariable functions to column vectors, it is obvious
that linear operators acting on grid functions are isomorph to linear transformations of $\mathbb{R}^{n^4}$. As a result, the
difference operators of Eq.~\eqref{diffop} can also be mapped isomorphically to $\mathbf{M}_i\in M_{n^4}[\mathbb{R}]$ matrices,
where $i=1,\dots,4$ and $M_{n^4}[\mathbb{R}]$ denotes the set of all $n^4$-by-$n^4$ real matrices. Straightforward but lengthy
calculations yield
\begin{align}
\square_1\longleftrightarrow\mathbf{M}_1&=\mathbf{A}\otimes\mathbf{E}_{n^3}+\mathbf{E}_n\otimes\mathbf{A}\otimes\mathbf{E}_{n^2}+
\mathbf{E}_{n^2}\otimes\mathbf{A}\otimes\mathbf{E}_n+\mathbf{E}_{n^3}\otimes\mathbf{A}-8\mathbf{E}_{n^4}\label{m1}
\end{align}
for the first-neighbor,
\begin{align}
\square_2\longleftrightarrow\mathbf{M}_2&=
\mathbf{A}\otimes\mathbf{A}\otimes\mathbf{E}_{n^2}+\mathbf{A}\otimes\mathbf{E}_n\otimes\mathbf{A}\otimes\mathbf{E}_n
+\mathbf{A}\otimes\mathbf{E}_{n^2}\otimes\mathbf{A}+\mathbf{E}_n\otimes\mathbf{A}\otimes\mathbf{A}\otimes\mathbf{E}_n\notag\\
&\phantom{=}+\mathbf{E}_n\otimes\mathbf{A}\otimes\mathbf{E}_n\otimes\mathbf{A}
+\mathbf{E}_{n^2}\otimes\mathbf{A}\otimes\mathbf{A}-24\mathbf{E}_{n^4}\label{m2}
\end{align}
for the second-neighbor and
\begin{align}
\square_3\longleftrightarrow\mathbf{M}_3&=
\mathbf{A}\otimes\mathbf{A}\otimes\mathbf{A}\otimes\mathbf{E}_n+\mathbf{A}\otimes\mathbf{A}\otimes\mathbf{E}_n\otimes\mathbf{A}
+\mathbf{A}\otimes\mathbf{E}_n\otimes\mathbf{A}\otimes\mathbf{A}+\mathbf{E}_n\otimes\mathbf{A}\otimes\mathbf{A}\otimes\mathbf{A}
-32\mathbf{E}_{n^4}\label{m3}
\end{align}
for the third-neighbor difference operator, respectively. Here $\mathbf{E}_k$ is the unit matrix of size $k$ and $\mathbf{A}$ is
the $n$-by-$n$ tridiagonal matrix
\begin{equation}
\mathbf{A}=
\begin{pmatrix}
0 & 1\\
1 & 0 & 1\\
\hdotsfor{5}\\
& & 1 & 0 & 1\\
& & & 1 & 0
\end{pmatrix}.\label{a}
\end{equation}
Before we proceed with $\mathbf{M}_4$, we would like to call the attention to an interesting observation. Namely, the matrices
above, if written as direct-products, exhibit similar ``patterns'' as the corresponding expressions in Eqs.~\eqref{diffop}. It is
not so hard to see that the very simple tridiagonal form of $\mathbf{A}$ corresponds to that when values of $\psi_{pikl}$ are
considered and added at first-neighbor sites in one given dimension. Then, by means of direct-product, linear combinations of any
neighbors of a given site in any lower dimension subspace can be constructed. Coming back to the fourth-neighbor difference
operator, this reasoning leads us to
\begin{align}
\square_4\longleftrightarrow\mathbf{M}_4&=\mathbf{A}\otimes\mathbf{A}\otimes\mathbf{A}\otimes\mathbf{A}
-16\mathbf{E}_{n^4}\notag\\
&\phantom{=}+\mathbf{A'}\otimes\mathbf{E}_{n^3}+\mathbf{E}_n\otimes\mathbf{A'}\otimes\mathbf{E}_{n^2}+
\mathbf{E}_{n^2}\otimes\mathbf{A'}\otimes\mathbf{E}_n+\mathbf{E}_{n^3}\otimes\mathbf{A'}-8\mathbf{E}_{n^4},\label{m4}
\end{align}
where $\mathbf{A'}$ is $n$-by-$n$ and corresponds to combination of second-neighbors along a given coordinate axis. The second
line on the right hand side is the matrix representation of the last term in Eq.~\eqref{fourth}. Comparison with Eqs.~\eqref{m1}
and~\eqref{a} suggests $\mathbf{A'}$ should be pentadiagonal
\begin{equation}
\begin{pmatrix}
0 & 0 & 1\\
0 & 0 & 0 & 1\\
1 & 0 & 0 & 0 & 1\\
\hdotsfor{6}\\
& & 1 & 0 & 0 & 0\\
& & & 1 & 0 & 0
\end{pmatrix}.\label{av}
\end{equation}
Here, in the first and last two rows there is only one nonzero element. This is because along any given coordinate axis the first
two and last two grid points each have only one such second-neighbor that is also inside the domain. Right on the boundary
$\psi_{pikl}=0$ because of the boundary condition, whereas, as pointed out in the beginning of this subsection, fictitious values
of $\psi_{pikl}$ were taken to be zero by hand. It was also emphasized there that this procedure does not affect required
precision, which we will prove exactly at the end of this subsection. Now we make another presumably very accurate approximation
that does not affect precision either: instead of Eq.~\eqref{av} we take
\begin{equation}
\mathbf{A'}=\mathbf{A}^2-2\mathbf{E}_n=
\begin{pmatrix}
-1 & 0 & 1\\
0 & 0 & 0 & 1\\
1 & 0 & 0 & 0 & 1\\
\hdotsfor{6}\\
& & 1 & 0 & 0 & 0\\
& & & 1 & 0 & -1
\end{pmatrix},\label{avgood}
\end{equation}
that is $A'_{11}=A'_{nn}=-1$, otherwise $\mathbf{A'}$ is the same as Eq.~\eqref{av}. The sparsity of $\mathbf{M}_i$ can be seen in
Fig.~\ref{fig:sparsity}.

Having obtained these important matrices we are now in a position to vectorize Eq.~\eqref{vegso2}. Taking into account
Eq.~\eqref{f2} as well we finally arrive to the $\mathcal{O}(h^6)$ matrix form of the discretized four-dimensional Sch\"odinger
equation
\begin{equation}
(h^{-2}\mathbf{M}+\mathbf{N}\text{diag}(\mathbf{\tilde U})){\bm\psi}=E\mathbf{N}\mathbf{\bm\psi}.\label{vegsosch}
\end{equation}
Here, $\text{diag}(\mathbf{\tilde U})$ is a diagonal matrix composed of the vector $\mathbf{\tilde U}$ and the matrices
$\mathbf{M}$ and $\mathbf{N}$, which include all information supplied by the boundary condition, are
\begin{equation}
\mathbf{M}=-\frac{1}{30}(12\mathbf{M}_1+\mathbf{M}_2+\mathbf{M}_3),\label{m}\\
\end{equation}
and
\begin{equation}
\mathbf{N}=\mathbf{E}_{n^4}+\left(12\gamma'-\frac{1}{30}\right)\mathbf{M}_1
+\left(\frac{1}{36}-4\gamma'\right)\mathbf{M}_2+\gamma'\mathbf{M}_3-\frac{1}{240}\mathbf{M}_4.\label{n}
\end{equation}
Apropos of Eq.~\eqref{vegso2} we have already mentioned that from linear algebraic point of view Eq.~\eqref{vegsosch} constitutes
a generalized eigenvalue problem because $\mathbf{N}\ne\mathbf{E}_{n^4}$. In what follows, we collect its basic mathematical
properties as well as the necessary requirements it must satisfy in order to best represent the continuous Schr\"odinger equation,
Eq.~\eqref{poisson}. We shall see that these will lead us to the very specific choice of $\gamma'$ anticipated in
Eq.~\eqref{gammav}.

\begin{enumerate}
\item Properties of the Kronecker-product and the symmetry of $\mathbf{A}$ imply that $\mathbf{M}_i$ $(i=1,\dots,4)$ as well as
  $\mathbf{M}$ and $\mathbf{N}$ are all symmetric matrices.
\item As shown in Appendix~\ref{appsec:matrix}, the matrices $\mathbf{M}_i$ are all negative definite. From this, using the
  properties of definite\footnote{We say that a matrix is definite if it is either positive or negative definite.} matrices, it
  follows that $\mathbf{M}$ is positive definite.
\item Let $\gamma'$ be such that $\mathbf{N}$ is non-singular. Then multiplying Eq.~\eqref{vegsosch} by $\mathbf{N}^{-1}$ one ends
  up with the usual form of a quantum mechanical eigenvalue problem where the kinetic energy operator $-\Delta$ is represented by
  $h^{-2}\mathbf{N}^{-1}\mathbf{M}$. It is also shown in Appendix~\ref{appsec:matrix} that $\mathbf{M}$ and $\mathbf{N}$ commute:
  $[\mathbf{M},\mathbf{N}]=\mathbf{0}$. This is a very important and central result for it assures that the matrix Hamiltonian
  ($h^{-2}\mathbf{N}^{-1}\mathbf{M}+\text{diag}(\mathbf{\tilde U})$) is symmetric leading to real energy eigenvalues.
\item According to (iii),\ $h^{-2}\mathbf{N}^{-1}\mathbf{M}$ represents the negative Laplace operator. It is known from functional
  analysis that $-\Delta$ is a nonnegative and non-bounded linear operator, thus $\gamma'$ must be chosen so that
  $\mathbf{N}^{-1}\mathbf{M}$ is positive definite \cite{chawla}. Again, using the properties of definite matrices and taking
  advantage of the fact that the terms of this product commute, beyond the requirement for $\mathbf{N}$ to be non-singular it must
  also be positive definite. In addition to that, asymptotically, as $n\to\infty$, the spectrum of
  $h^{-2}\mathbf{N}^{-1}\mathbf{M}$ must approach that of $-\Delta$ supplied by the boundary condition
  \begin{equation}
  \lim_{n\to\infty}\rho\left((n+1)^2\mathbf{N}^{-1}\mathbf{M}\right)=
  \left\{\,\pi^2\sum_{i=1}^4k_i^2\biggm|k_i=1,2,\dots\,\right\}.\label{spektrumlimit}
  \end{equation}
  Note that the right hand side is nothing else but the set of allowed discrete energy levels of the four-dimensional ``particle
  in a cubic box'' problem.
\end{enumerate}
In Appendix~\ref{appsec:matrix} we present detailed derivations of the statements in (ii) and (iii). Also, we give analytical
proofs of the followings. The necessary requirement in (iv) for $\mathbf{N}$ to be positive definite can be satisfied if $\gamma'$
is taken from the interval
\begin{equation}
\gamma'\le\frac{23}{3840}.\label{range}
\end{equation}
Besides, the same direct polynomial structure of $\mathbf{M}$ and $\mathbf{N}$ imply that all $n^4$ eigenvalues of
$\mathbf{N}^{-1}\mathbf{M}$ are of the form $\theta(\mathbf{k})/\lambda(\mathbf{k})$, where $\theta(\mathbf{k})$ and
$\lambda(\mathbf{k})$ are eigenvalues of $\mathbf{M}$ and $\mathbf{N}$, respectively. Explicit expressions for these quantities
are given in Eqs.~\eqref{ms} and~\eqref{ns}. Here, $\mathbf{k}=(k_1,k_2,k_3,k_4)$ is a composite vector index with components
taking integer values in the range $1,\dots,n$. With these results at hand we can perform asymptotic expansion for large $n$, and
by Eq.~\eqref{asymptlaplace} we find
\begin{equation}
(n+1)^2\frac{\theta(\mathbf{k})}{\lambda(\mathbf{k})}=\epsilon(\mathbf{k})
\left(1-\frac{a(\mathbf{k})-b(\mathbf{k})\gamma'}{n^6}+\dots\right),
\end{equation}
where
\begin{equation}
\epsilon(\mathbf{k})=\pi^2\sum_{i=1}^4k_i^2,\qquad k_i=1,2,\dots\label{boxe}
\end{equation}
is the energy level in Eq.~\eqref{spektrumlimit}. This result forms the cornerstone of our $\mathcal{O}(h^6)$ theory, for it
assures that (i) as $n$ goes to infinity, the spectrum of the discrete Laplace operator evolves into that of $-\Delta$, (ii) the
leading error is at most $\mathcal{O}(n^{-6})$. We see now finally that the approximations we applied by setting fictitious
function values to zero by hand and using Eq.~\eqref{avgood} instead of Eq.~\eqref{av} are indeed that accurate as were
expected. They did not affect the required precision. This is a very satisfactory situation because it suggests that higher order
calculations could be performed in exactly the same manner.

Tuning $\gamma'$ within the range of Eq.~\eqref{range} the order of error cannot be reduced any further. However, we can still try
to minimize its prefactor. According to Eqs.~\eqref{minab} and~\eqref{gfinal} this can be achieved successfully by taking the
largest value allowed
\begin{equation}
\gamma'=\frac{23}{3840},\label{finalg}
\end{equation}
and this is exactly what we have anticipated in Eq.~\eqref{gammav}. It might be of interest that with this choice $\mathbf{N}$
becomes asymptotically singular, suggesting numerical implementation should avoid manipulation with the inverse. For more details
please, see Subsection~\ref{sub:n} of Appendix~\ref{appsec:matrix}.

Now that we have successfully derived the matrix Schr\"odinger equation and built up the relevant matrices, in the last subsection
we give a very brief discussion on the memory demand of the storage of these matrices.

\subsection{Sparsity properties, memory consumption}\label{sub:memory}
Figure~\ref{fig:sparsity} shows the sparsity patterns of the matrices $\mathbf{M}_i$, that are the basic building rocks of the
Schr\"odinger equation, Eq.~\eqref{vegsosch}. As for the resolution, for illustration purposes we used a very low value of $n=3$,
which means that the total size of the matrices is $n^4=81$. Remarkably, we found that due to the direct-product structure of
these matrices the number of nonzero elements and therefore the density or the sparsity can be calculated exactly for arbitrary
$n$. Since the number of nonzero elements in $\mathbf{A}$ equals $N(\mathbf{A})=2(n-1)$ it follows form Eq.~\eqref{m1} that for
$\mathbf{M}_1$ it is $8(n-1)n^3+n^4$. Similarly, for $\mathbf{M}_2$, $\mathbf{M}_3$ and $\mathbf{M}_4$ the number of nonzero
elements are exactly $24(n-1)^2+n^4$, $32(n-1)^3n+n^4$ and $16(n-1)^4+8(n-2)n^3+n^4$, respectively. Knowing these values it is now
straightforward to obtain respective results for $\mathbf{M}$ and $\mathbf{N}$ and they read
\begin{subequations}
\begin{align}
N(\mathbf{M})&=8(n-1)n^3+24(n-1)^2n^2+32(n-1)^3n+n^4\approx65n^4,\label{melem}\\
N(\mathbf{N})&=8(n-1)n^3+24(n-1)^2n^2+32(n-1)^3n+16(n-1)^4+8(n-2)n^3+n^4\approx89n^4,\label{nelem}
\end{align}
\end{subequations}
where the approximate values are the leading terms for large $n$. It also turns out that the matrix in the parentheses on the left
hand side of Eq.~\eqref{vegsosch} has exactly the same structure as $\mathbf{N}$ (from the sparsity point of view), which means
that for large $n$ one has to allocate memory for roughly $178n^4$ nonzero real matrix elements in order to simply store the
equation as a whole. For example at a modest resolution of $n=20$ and with double precision arithmetic on a 64 bit computer the
memory consumption is roughly 0.37 gigabytes.

In this Section we have established the algebra of the two-particle problem, it is now time to proceed with the group theoretical
aspects. These, as expected, result in further substantial simplifications that will facilitate numerical implementation.

\section{Group theoretical aspects: classification of energy levels}\label{sec:group}
In the previous section we have developed a new finite difference method for the numerical solution of the two-particle stationary
Schr\"odinger equation. The most noticeable feature of the theory is that it provides the energy levels very accurately. The
physical problem we are considering in this paper is the interaction of two identical particles in two-dimensions and with closed
boundary condition. A closed boundary in our case means that both particles are confined in an infinite square well, which leads
to the Dirichlet condition in Eq.~\eqref{4Dconstraint}. For these circumstances we have derived a matrix Schr\"odinger equation in
Eq.~\eqref{vegsosch}. We showed analytically that as the step size $h$ goes to zero it converges to the continuous Schr\"odinger
equation with a local truncation error of $\mathcal{O}(h^6)$. The problem can be thought of as an eigenvalue problem of an
abstract four-dimensional particle in a 4D cubic box. From this and from the details of derivation one can see that the algorithm
can be applied in any dimension, or, if necessary, the precision can be refined further as well.

Despite of the favorable high precision there are, however, difficulties too. When it comes to numerical solution, we face a
generalized eigenvalue problem of dimension $n^4$. As we saw in Subsection~\ref{sub:memory}, due to the multiple direct-product
structure even a modest resolution in $n$ results in huge matrices that, though being sparse, still require lots of memory. Also,
what is more important, the classification of energy levels and wave functions is unknown at this stage. It might happen that we
unnecessarily calculate degenerate levels many times and cannot reach higher energy excited states that might be of interest. The
good thing is that these problems can be entirely overcome as there exists a powerful theoretical tool, group theory, that was
developed exactly for these needs.

Before jumping into details we would like to emphasize already at this point that the group theoretical analysis we are about to
present here is not by all means necessary in order to solve Eq.~\eqref{vegsosch}. With a sufficiently powerful workstation the
diagonalization at this stage is in principle straightforward \cite{gabor}. However, exploiting all the hidden information that
lies in the symmetry of Eq.~\eqref{vegsosch} is very instructive an far reaching as we shall see right away.

\subsection{Group of the Schr\"odinger equation: general remarks}\label{sub:intro}
The very first step in the development of representation theory is the exploration of the maximal invariance group $\mathcal{G}$
of the Hamiltonian $H$. In quantum mechanical description of a system $\mathcal{G}$ consists of all unitary operators of the
Hilbert space that commute with $H$. In many cases it turns out that these groups are, in group theoretical sense, linear Lie
groups and it is the generators of the associate real Lie algebra that represent the physical quantities that are constants of
motion. As the generators are Hermitian operators that also commute with $H$, they are indeed the usual conserved quantities of
the system. It is a very interesting observation that the exploration of the maximal invariance group can be quite complicated
sometimes \cite{mcintosh}. Perhaps the two most famous problems of quantum theory are the hydrogen atom (also known as the quantum
mechanical Kepler problem) and the isotropic harmonic oscillator \cite{fradkin,fradkin-prog}. As a matter of fact, both problems
can be considered in any dimension $n$. These problems have an obvious geometrical symmetry: due to rotational symmetry of the
central force, $\mathcal{G}$, whatever it may be, must contain the $n$-dimensional orthogonal group $\text{O}(n,\mathbb{R})$ as a
subgroup.\footnote{Hereafter we drop extra notation $\mathbb{R}$, because we only consider real matrices throughout the paper and
  there will be no confusion by simply writing $\text{O}(n)$.} This immediately explains degeneracy of the energy levels of both
systems in the angular hypermomenta $m_i$, $i=1,\dots,n-2$, but cannot account for the degeneracy in the total angular momentum
$l$. Detailed investigations show that behind the substantial ``accidental'' degeneracies there are larger symmetry groups: in
case of hydrogen atom it is $\text{O}(n+1)$, whereas for the isotropic oscillator $\text{SU}(n)$ is found
\cite{fradkin,jauch,baker,montgomery}. Even the respective classical Hamiltonians possess these enlarged invariance groups. The
extra symmetries spanning the larger groups, that were initially overlooked, are the so-called hidden or dynamical symmetries that
cannot be associated with apparent symmetries of the configuration space \cite{mcintosh,pogosyan}. They are related to the fact
that the Schr\"odinger equation can be separated in other coordinate systems as well as in spherical polars. The additive
generators of the respective Lie algebras are in turn the Runge-Lenz vector operator and a symmetric tensor operator in case of
the hydrogen atom and the oscillator, respectively \cite{fradkin,hughes}.

Our main concern in this section is the exploration of the symmetry group and development of representation theory of the
interacting two-particle problem under focus. This, as we have already noted before, can be looked at as a one-particle problem in
a four-dimensional space where the particle is trapped in an impenetrable cubic box. In addition to that there is a well defined
potential inside the domain too. The problem has an obvious geometrical symmetry due to the cubic structure of the
cavity. Therefore $\mathcal{G}$ will certainly contain discrete symmetry operations related to the symmetry of the configuration
space and the external potential. The more interesting question, however, addresses the possible existence of dynamical
symmetries. In order to give a full classification scheme of the levels and wave functions it is necessary to incorporate these
symmetries as well, if there are any. It turns out that the Coulomb interaction we are about to apply excludes the possibility of
such hidden symmetries and we are left with a discrete symmetry group consisting of geometrical symmetries only. In this specific
case the unitary operators of the Hilbert space that build up $\mathcal{G}$ are the so-called scalar transformation operators
\cite{cornwell-book}. We give detailed analysis of them in Subsection~\ref{sub:scalar}. These operators are known to be isomorphic
to the geometrical symmetry operations of the system, thus $\mathcal{G}$ can equally be considered as the group of abstract
coordinate transformations of the four-dimensional space. In what follows we find it more convenient to use this latter
designation.

Without interaction the problem, though elementary from quantum mechanical point of view, from the viewpoint of group theory
becomes much more elaborate. The lack of interaction leads to substantially larger number of discrete symmetries and more
interestingly to the appearance of dynamical symmetries as well. This issue will be partly explored in Section~\ref{sec:nointres}.

\subsection{Group of the Schr\"odinger equation of the confined interacting particles}\label{sub:inv}
According to what has been said in the second paragraph of the previous subsection, this group can be considered as consisting of
those four-dimensional coordinate transformations $T$ that leave $H$ invariant. To put this requirement in a mathematical formula
recall that $H$ is now
\begin{equation}
H(\mathbf{x})=-\sum_{i=1}^4\frac{\partial^2}{\partial x_i^2}+\tilde U(\mathbf{x}),\label{ham}
\end{equation}
where the potential is given by Eq.~\eqref{effectivepot}. As is known from quantum mechanics, the $n$-dimensional Laplace operator
is invariant under pure rotations of the orthogonal group $\text{O}(n)$. In our case $n=4$ of course. Let $T$ be a coordinate
transformation that introduces a new set of mutually orthogonal axes $(x_1',x_2',x_3',x_4')$, but which leaves the origin
unmoved. Now there exists a unique matrix $\mathbf{R}\in\text{O}(4)$, with which the coordinates of a fixed point with respect to
this new system can be expressed as $\mathbf{x'}=\mathbf{Rx}$, with $\mathbf{x}$ being the coordinates with respect to the
original frame $(x_1,x_2,x_3,x_4)$. The transformation $T$ is then said to leave the Hamiltonian invariant if
\begin{equation}
H(\mathbf{R}\mathbf{x})=H(\mathbf{x}).\label{inv}
\end{equation}
The group of abstract transformations $T$ is isomorphic to the group of matrices $\mathbf{R}$, hence we will refer to coordinate
transformations as matrices and $\mathcal{G}$ will hereafter be considered as the group of the corresponding matrices.

Due to the presence of the potential and the strict boundary condition our system is not homogeneous, pure translations do not
leave $H$ invariant. Further, if $\mathbf{R}\in\mathcal{G}$ is a symmetry transformation, then (i) it must be a symmetry of the
four-dimensional hypercube and (ii) satisfy
\begin{equation}
\tilde U(\mathbf{Rx})=\tilde U(\mathbf{x}).\label{potinv}
\end{equation}
As to the symmetries of the domain, the $n$-dimensional hypercube has altogether $2^nn!$ symmetry transformations, all of which
can be faithfully represented by $n$-by-$n$ orthogonal matrices having only one nonzero element, $+1$ or $-1$ in each row and
column. Such matrices are called signed permutation matrices. It follows there are 384 such matrices in 4D, these constitute the
so-called four-dimensional cubic group $\text{O}_4$ \cite{dai}. To single out those that make up $\mathcal{G}$ we have to specify
the potential too. The external field the particles are subjected to is chosen as
\begin{equation}
U(\mathbf{r})=\frac{1}{4}\omega^2(x^2+y^2),\label{harmonic}
\end{equation}
where $\omega>0$ is a dimensionless parameter responsible for its scale. We use quadratic potential because for this the
noninteracting problem has exact analytical solution
\begin{subequations}\label{eosc}
\begin{align}
E&=\omega\left(\nu_{k_1}+\nu_{k_2}+\nu_{k_3}+\nu_{k_4}+2\right),\label{exactosc}\\
\psi_\mathbf{k}(\mathbf{x})&=\phi_{k_1}(x_1)\phi_{k_2}(x_2)\phi_{k_3}(x_3)\phi_{k_4}(x_4).\label{exactoscwave}
\end{align}
\end{subequations}
Here $\mathbf{k}=(k_1,k_2,k_3,k_4)$ and each quantum number takes integer values: $k_i=0,1,2,\dots$. The wave functions $\phi_k$
and the first sixteen $\nu_k$'s are shown in Eqs.~\eqref{ptlanosc}, \eqref{psosc} and in Table~\ref{tab:osc} of
Appendix~\ref{appsec:osc}, respectively.  Besides it is not singular as opposed to atomic potentials (the 2D hydrogen atom of
Ref.~\cite{yang} for example), which is also convenient from numerical point of view \cite{kang,jamieson,eichten}. With this
choice we find that the first two terms of $\tilde U$ together are invariant under all rotations of $\text{O}(4)$. In contrast to
this, the pair interaction $V$ constitutes a much stronger constraint: it is invariant under such orthogonal transformations for
which
\begin{equation}
|\mathbf{r}_1-\mathbf{r}_2|^2=\mathbf{x}^\text{T}\mathbf{Qx}=\text{inv.},\label{intinv}
\end{equation}
where
\begin{equation}
\mathbf{Q}=
\begin{pmatrix}
 1 & 0 &-1 & 0\\
 0 & 1 & 0 &-1\\
-1 & 0 & 1 & 0\\
 0 &-1 & 0 & 1
\end{pmatrix}.\label{q}
\end{equation}
Now, a transformation $\mathbf{R}$ can only be a member of $\mathcal{G}$ if $[\mathbf{R},\mathbf{Q}]=\mathbf{0}$, which in turn
restricts the 384 potential candidates to a subgroup of order 32. This can be verified either by direct calculation or by the
following alternative method. Consider the group $\mathcal{A}$ of order 4 with members
\begin{alignat}{4}
\mathbf{R}_1&=
\begin{pmatrix}
1 & & &\\
& 1 & &\\
& & 1 &\\
& & & 1
\end{pmatrix},
&\quad
\mathbf{R}_2&=
\begin{pmatrix}
& & 1 & \\
& & & 1\\
1 & & &\\
& 1 & &
\end{pmatrix},
&\quad
\mathbf{R}_3&=
\begin{pmatrix}
& & 1 &\\
& 1 & &\\
1 & & &\\
& & & 1
\end{pmatrix},
&\quad
\mathbf{R}_4&=
\begin{pmatrix}
1 & & &\\
& & & 1\\
& & 1 &\\
& 1 & &
\end{pmatrix}.\label{amembers}
\end{alignat}
Only the nonzero elements are shown for better readability. These matrices correspond to permutations of coordinates of the
particles. For example $\mathbf{R}_2$ permutes the two particles $(x_1,x_2)=(x_1,y_1)\longleftrightarrow(x_2,y_2)=(x_3,x_4)$,
whereas $\mathbf{R}_3$ and $\mathbf{R}_4$ permutes only the $x$ and $y$ coordinates, respectively. On the other hand, the
symmetries of the confining two-dimensional square well form another group $\mathcal{B}$ of order 8 with elements
\begin{alignat}{4}
\mathbf{R}_1&=
\begin{pmatrix}
1 & & &\\
& 1 & &\\
& & 1 &\\
& & & 1
\end{pmatrix},
&\quad
\mathbf{R}_2&=
\begin{pmatrix}
& 1 & &\\
1 & & &\\
& & & 1\\
& & 1 &
\end{pmatrix},
&\quad
\mathbf{R}_3&=
\begin{pmatrix}
& -1 &  &\\
1  & &  &\\
&  & & -1\\
&  & 1  &
\end{pmatrix},
&\quad
\mathbf{R}_4&=
\begin{pmatrix}
1  & &  &\\
& -1 &  &\\
&  & 1  &\\
&  & & -1
\end{pmatrix},\notag\\
\mathbf{R}_5&=-\mathbf{R}_1,
&\quad
\mathbf{R}_6&=-\mathbf{R}_2,
&\quad
\mathbf{R}_7&=-\mathbf{R}_3,
&\quad
\mathbf{R}_8&=-\mathbf{R}_4.\label{bmembers}
\end{alignat}
We can observe the block diagonal structure of these matrices with the same $2$-by-$2$ blocks repeated twice in the diagonal. This
reflects the fact that the same two-dimensional transformation must be performed on both particles. We note that $\mathcal{B}$ is
essentially a four-dimensional faithful representation of the group of point symmetries of a square, denoted commonly by $C_{4v}$
in solid state literature. Every member of $\mathcal{A}$ and $\mathcal{B}$ obeys Eq.~\eqref{inv} and so do their products. This
results in the rather plausible observation that the group of the Schr\"odinger equation is expressible as
\begin{equation}
\mathcal{G}=\left\{\mathbf{R}_a\mathbf{R}_b\mid\mathbf{R}_a\in\mathcal{A},
\mathbf{R}_b\in\mathcal{B}\right\}.\label{invgroup}
\end{equation}
It is indeed a group of order 32, where $\mathcal{A}$ and $\mathcal{B}$ are trivial subgroups. Closer inspection shows that
$\mathcal{A}$ is an invariant Abelian subgroup of $\mathcal{G}$ and isomorphic to $C_2^2$, with $C_2$ being the cyclic group of
order 2. This feature, in conjunction with the defining equation~\eqref{invgroup} and the fact that $\mathcal{A}$ and
$\mathcal{B}$ have only the identity element in common lead to an important result: the full invariance group of the Hamiltonian
has a semi-direct product structure \cite{cornwell-book}
\begin{equation}
\mathcal{G}=\mathcal{A}\circledS\mathcal{B}\sim C_2^2\circledS C_{4v}.\label{semi}
\end{equation}

This will come in handy when we proceed with the determination of the irreducible representations of $\mathcal{G}$. Technical
details concerning this can be found in Appendix~\ref{appsec:irr}.

\subsection{Scalar transformation operators}\label{sub:scalar}
Now that we have successfully explored the group of the Schr\"odinger equation we go on and construct the scalar transformation
operators. From group representation theory it is known that these linear operators are absolutely necessary for the explicit
determination of symmetry adapted basis functions \cite{cornwell-book}. Let $\mathbf{R}\in\mathcal{G}$ a symmetry transformation
of the Hamiltonian, then the scalar transformation operator $P(\mathbf{R})$ is defined by
\begin{equation}
P(\mathbf{R})v(\mathbf{x})=v(\mathbf{R}^{-1}\mathbf{x}),\label{scalartrafo}
\end{equation}
where $v(\mathbf{x})$ is any multivariable function. Recall now that in the numerical analysis the values of a function are only
available at 4D grid points, see Eq.~\eqref{vgrid}. Thus a function is essentially nothing else than a huge set of real numbers
conveniently mapped to column vectors by means of Eq.~\eqref{vector}. Accordingly, every linear operator acting on this space
stands in one-to-one correspondence with a matrix of $M_{n^4}[\mathbb{R}]$, for which we have already seen examples in
Subsection~\ref{sub:free} by the construction of the matrices $\mathbf{M}_i$. As to the scalar transformation operators we find
\begin{equation}
\left\{\mathbf{P}(\mathbf{R})\right\}_{\mu\nu}=1,\label{defp}
\end{equation}
where
\begin{align}
\mu&=p'+n(i'-1)+n^2(k'-1)+n^3(l'-1),\label{sorindex}\\
\nu&=p+n(i-1)+n^2(k-1)+n^3(l-1),\label{oszlopindex}
\end{align}
and the relation between indices is given by
\begin{equation}
\begin{bmatrix}
p'\\i'\\k'\\l'
\end{bmatrix}
=\frac{n+1}{2}(\mathbf{E}_4-\mathbf{R})
\begin{bmatrix}
1\\1\\1\\1
\end{bmatrix}
+\mathbf{R}
\begin{bmatrix}
p\\i\\k\\l
\end{bmatrix}.\label{indexrelation}
\end{equation}
Running the indices $p,i,k$ and $l$ in the range $1,\dots,n$ the whole matrix can be constructed. Further, by Eq.~\eqref{defp}
these have exactly one nonzero element in each row and column, they are necessarily orthogonal and describe permutations. This
must be so as $\mathbf{P}(\mathbf{R})\mathbf{v}$ is by definition the same scalar field as $\mathbf{v}$, but now it is looked at
from within the transformed coordinate system introduced by $\mathbf{R}$. In other words we can say that every such matrix is a
unitary linear transformation of $\mathbb{R}^{n^4}$. Similarly, the $P(\mathbf{R})$'s are unitary operators of the Hilbert space
\cite{cornwell-book}. In complete analogy with Eq.~\eqref{vector} they can be expressed with multiple direct-products too
\begin{equation}
\mathbf{P}(\mathbf{R})=\sum_{p,i,k,l=1}^n
\left(\mathbf{l'}\circ\mathbf{l}^\text{T}\right)\otimes
\left(\mathbf{k'}\circ\mathbf{k}^\text{T}\right)\otimes
\left(\mathbf{i'}\circ\mathbf{i}^\text{T}\right)\otimes
\left(\mathbf{p'}\circ\mathbf{p}^\text{T}\right),\label{opkron}
\end{equation}
where $\circ$ stands for the dyadic product (outer product).

One of the most important theorems of group representation theory, at least from quantum mechanical point of view, is that the
scalar transformation operators form a group isomorphic to the group $\mathcal{G}$ of transformations $\mathbf{R}$
\cite{cornwell-book}. Further, as Eqs.~\eqref{inv} and~\eqref{scalartrafo} suggest, $H$ and $P(\mathbf{R})$ commute for all
$\mathbf{R}\in\mathcal{G}$. As we pointed out in Subsection~\ref{sub:intro}, in quantum mechanics the group of such unitary
operators is usually identified as the invariance group of the system. Now it is not surprising that the matrices representing
these operators in the finite difference method obey a very same equation
\begin{equation}
\left[h^{-2}\mathbf{N}^{-1}\mathbf{M}+\text{diag}(\mathbf{\tilde U}),\mathbf{P}(\mathbf{R})\right]
=\mathbf{0}.\label{hpcommute}
\end{equation}
Beyond this we can also prove that $\mathbf{P}(\mathbf{R})$ not only commutes with the full Hamiltonian, but does it also with
each term separately. The fact it is true for the potential is almost trivial from Eq.~\eqref{potinv}, while the proof for
$\mathbf{M}$ and $\mathbf{N}$ is as follows. First remember that the basic building blocks of the finite difference method are the
difference operators given by Eqs.~\eqref{diffop}. Every such operator can be expressed concisely as
\begin{equation}
\square_j v(\mathbf{x})=\sum_\mathbf{d}v(\mathbf{x}+\mathbf{d})-g_jv(\mathbf{x}),\label{diffcommon}
\end{equation}
where the sum is over all $j$th neighbors $\mathbf{d}$ that are at a well defined distance from the given lattice point
$\mathbf{x}$. Let their total number be denoted by $g_j$. In the four-dimensional grid we use for numerics this is equivalent to
the constraint $\mathbf{d}^2=h^2(\alpha^2+\beta^2+\gamma^2+\delta^2)=\text{const.}$, with $\alpha,\dots,\delta$ being integer
coordinates of $\mathbf{d}$. This expression is clearly invariant under any permutation of coordinates with sign changes
included. In other words it is invariant under all 384 orthogonal transformations of $\text{O}_4$ and as such for all
$\mathbf{R}\in\mathcal{G}$ as well. This leads immediately to the fact that the neighborhood is completely symmetric:
$\left\{\mathbf{d}_1,\dots,\mathbf{d}_{g_j}\right\}=\left\{\mathbf{Rd}_1,\dots,\mathbf{Rd}_{g_j}\right\}$. This is to be
interpreted as an equation of sets where the order of elements is irrelevant. With these findings and the definition in
Eq.~\eqref{scalartrafo} it is now easy to verify that
\begin{equation}
\left[\square_j,P(\mathbf{R})\right]=0.\label{squarep}
\end{equation}
Taking into account Eqs.~\eqref{m} and~\eqref{n} and the fact that the commutator is a real bilinear function, commutativity of
the respective matrices is found
\begin{align}
\mathbf{P}(\mathbf{R})\mathbf{M}\mathbf{P}(\mathbf{R})^{-1}&=\mathbf{M},\\
\mathbf{P}(\mathbf{R})\mathbf{N}\mathbf{P}(\mathbf{R})^{-1}&=\mathbf{N}.
\end{align}
This fundamental result, which we shall make use of shortly in Wigner-Eckart theorem \cite{cornwell-book} in the following
subsection, shows that beyond the potential and the full Hamiltonian these are also irreducible tensor operators (matrices) of the
completely symmetric irreducible representation ${\bm\Gamma}^{11}$ of $\mathcal{G}$. Representations and their labeling
convention can be found in Appendix~\ref{appsec:irr}. In particular, ${\bm\Gamma}^{11}$ is defined by Eq.~\eqref{compsymm}.

The final step in the development of group theoretical background is the determination of projection operators and the
construction of a set of symmetry adapted basis functions.

\subsection{Projection operators, basis functions of irreducible representations and the low energy subspace}\label{sub:pro}
In the previous subsection we have obtained matrix representations of the scalar transformation operators. Also, in
Appendix~\ref{appsec:irr} we have elaborated all irreducible representations ${\bm\Gamma}^{qp}$ of the group of the
Schr\"odinger equation. With this knowledge we can now construct a special orthogonal basis of the Hilbert space, every member of
which is a basis function (in group theoretical sense) transforming as some row of some irreducible representation of
$\mathcal{G}$. This is achieved by means of the projection operator method. Let's see how!

The projection operators are defined by \cite{cornwell-book}
\begin{equation}
\mathbf{P}^{qp}_{ij}=(d_{qp}/g)\sum_{\mathbf{R}\in\mathcal{G}}
\Gamma^{qp}(\mathbf{R})_{ij}\mathbf{P}(\mathbf{R}),\label{project}
\end{equation}
where $g=32$ is the order of the group and $d_{qp}$ is the dimension of ${\bm\Gamma}^{qp}$. In general, these objects as well
as the scalar transformation operators are linear operators acting on functions of the infinite dimensional Hilbert space
$\mathcal{H}$. In the finite difference method, however, since $\mathcal{H}=\mathbb{R}^{n^4}$ they become large but finite
dimensional sparse matrices. According to developments of representation theory if $\mathbf{v}$ is a vector of this space such
that $\mathbf{w}^{qp}_j=\mathbf{P}^{qp}_{jj}\mathbf{v}$, $j=1,\dots,d_{qp}$ are all nonzero, then they form a basis for
${\bm\Gamma}^{qp}$
\begin{equation}
\mathbf{P}(\mathbf{R})\mathbf{w}_{j}^{qp}=\sum_{i=1}^{d_{qp}}\Gamma^{qp}(\mathbf{R})_{ij}
\mathbf{w}_{i}^{qp}.\label{repbasis}
\end{equation}
Furthermore, the powerful Wigner-Eckart theorem says that if $\mathbf{S}$ is any irreducible tensor operator transforming as the
completely symmetric irreducible representation ${\bm\Gamma}^{11}$, that is commutes with all $\mathbf{P}(\mathbf{R})$, then a
remarkable simplification of the matrix elements occur
\begin{equation}
\left(\mathbf{w}_j^{qp},\mathbf{S}\mathbf{w}_{j'}^{q'p'}\right)=\delta_{jj'}\delta_{qp,q'p'}
\left(\mathbf{w}_j^{qp},\mathbf{S}\mathbf{w}_j^{qp}\right),\label{we}
\end{equation}
and the scalar product on right hand side does not depend on the row index $j$. As to the real scalar product it is defined as
usual: $(\mathbf{a},\mathbf{b})=\sum_ia_ib_i$. In the previous subsection we showed that each term appearing in the matrix
Schr\"odinger equation exhibits the same transformation property as $\mathbf{S}$ above. Now it follows that the eigenvalue
equation of dimension $n^4$ can be transformed into an equivalent block diagonal form where each block belongs to some row of some
irreducible representation of $\mathcal{G}$. Also, it is shown in Appendix~\ref{appsec:irr} that $\mathcal{G}$ possesses 8
one-dimensional and 6 two-dimensional representations. In the block diagonal decomposition there will be therefore $8+2\times6=20$
blocks among which only 14 are different. The solution of the eigenvalue problem is then equivalent to the solution of each block
separately.

Let us now define a basis (in linear algebraic sense) of the Hilbert space. This will serve as input from which the symmetry
adapted basis will be projected out. It reads
\begin{equation}
v_{pikl}(\mathbf{k})=\left(\prod_{i=1}^4\sum_{p=1}^n\phi^2_{k_i}(x_{i,p})\right)^{-\frac12}
\phi_{k_1}(x_{1,p})\phi_{k_2}(x_{2,i})\phi_{k_3}(x_{3,k})\phi_{k_4}(x_{4,l}),\label{linalgbasis}
\end{equation}
where the quantum numbers $k_i$ take integer values in the range $0,\dots,m-1\le n$. Again, as in Eq.~\eqref{exactoscwave}, the
functions $\phi_k(x)$ are the exact eigenfunctions of the confined one-dimensional harmonic oscillator. Their analytical form is
given in Appendix~\ref{appsec:osc} in Eqs.~\eqref{ptlanosc} and~\eqref{psosc}. From these grid values $\mathbf{v}(\mathbf{k})$ is
obtained by Eq.~\eqref{vector}. If $m-1=n$ they constitute an orthonormal basis of the Hilbert space. On the other hand, if
$m-1<n$, they generate a subspace $\mathcal{H}_m\subset\mathcal{H}$ of dimension $m^4$, that is physically of most interest. In
what follows we solve the discrete Schr\"odinger equation in this low energy subspace. This we can do because we are primarily
interested in the ground state and some higher energy excited states only and for this purpose it is completely sufficient to
restrict our analysis to the low energy subspace. Of course, with sufficiently powerful computing facilities one can increase $m$
and thereby expand the computational subspace.

In the basis of $\mathbf{v}(\mathbf{k})$ the equation looks like
\begin{equation}
\sum_\mathbf{k'}\left(\mathbf{v}(\mathbf{k}),\left(h^{-2}\mathbf{M}+\mathbf{N}\text{diag}(\mathbf{\tilde U})\right)
\mathbf{v}(\mathbf{k'})\right)\psi(\mathbf{k}')=
E\sum_\mathbf{k'}\left(\mathbf{v}(\mathbf{k}),\mathbf{N}\mathbf{v}(\mathbf{k'})\right)\psi(\mathbf{k'}).\label{schk}
\end{equation}
This is actually the $\mathbf{k}$th row of the Schr\"odinger equation for the unknown coefficients $\psi(\mathbf{k})$ and energy
$E$. The corresponding state is then obtained as
\begin{equation}
{\bm\psi}=\sum_\mathbf{k}\psi(\mathbf{k})\mathbf{v}(\mathbf{k}).\label{psik}
\end{equation}
Due to the scalar product the matrices of Eq.~\eqref{schk} are not sparse anymore and still quite large. Also, the level
classification is still unsolved at this stage, hence this is the appropriate point to appeal to the symmetry adapted basis
$\{\mathbf{w}_j^{qp}(\mathbf{k})\}$ instead of $\{\mathbf{v}(\mathbf{k})\}$. This is obtained by applying the orthogonal
projections
\begin{equation}
\mathbf{w}_j^{qp}(\mathbf{k})=\mathbf{P}_{jj}^{qp}\mathbf{v}(\mathbf{k}).\label{newbasis}
\end{equation}
From what has been said so far in this subsection it is clear that this newly generated vector is necessarily contained in
$\mathcal{H}$ and, if not zero, transforms as the $j$th row of ${\bm\Gamma}^{qp}$. However, it is not obvious that (i) it is
also a member of $\mathcal{H}_m$ and (ii) the maximal linearly independent set of these is orthogonal and span
$\mathcal{H}_m$. These important properties are proved in detail in Appendix~\ref{appsec:basis}.

Consider now a representation ${\bm\Gamma}^{qp}$. In Appendix~\ref{appsec:basis} we give an algorithm to generate a maximal
set of independent vectors that all belong to the $j$th row of this particular representation. These will be denoted by
$\mathbf{w}_j^{qp}(\mathbf{k}_s)$, $s=1,\dots,r^{qp}$. With all these the Schr\"odinger equation in this symmetry channel is found
to be
\begin{equation}
\sum_{t=1}^{r_{qp}}\left(\mathbf{w}_j^{qp}(\mathbf{k}_s),\left(h^{-2}\mathbf{M}+\mathbf{N}\text{diag}(\mathbf{\tilde U})\right)
\mathbf{w}_j^{qp}(\mathbf{k}_t)\right)\psi_t=
E\sum_{t=1}^{r_{qp}}\left(\mathbf{w}_j^{qp}(\mathbf{k}_s),\mathbf{N}\mathbf{w}_j^{qp}(\mathbf{k}_t)\right)\psi_t,
\qquad s=1,\dots,r^{qp},\label{schqpj}
\end{equation}
and the actual wave function of this symmetry is
\begin{equation}
{\bm\psi}_j^{qp}=\sum_{t=1}^{r^{qp}}\psi_t\mathbf{w}_j^{qp}(\mathbf{k}_t).\label{psiqpj}
\end{equation}
Equation~\eqref{schqpj} is one block out of the twenty of the block diagonal form of Eq.~\eqref{schk}. As the scalar products do
not depend on $j$, energy levels obtained from this will be exactly $d_{qp}$-fold degenerate. In the numerical implementation we
used $n=30$ and $m=8$, see Sections~\ref{sec:nointres} and~\ref{sec:intres}. This means $\mathcal{H}$ and $\mathcal{H}_m$ are of
dimensions $n^4=810000$ and $m^4=4096$, respectively. This latter dimension decomposes into smaller blocks of sizes $r^{qp}$ in
our case, each labelled by a certain representation. These sizes can be be determined either theoretically by means of of
Appendix~\ref{appsec:basis}, or numerically. Our computations showed that these are at most a few hundred ($4096/20\approx200$ in
average), which are very easy to handle numerically. Precise values are shown in the second row of Table~\ref{tab:noint}.

\subsection{Permutation symmetry of the two-particle wave function: Fermi or Bose statistics}\label{sub:fermibose}
So far we were not concerned about the nature of particles we are dealing with. That is we were not interested in whether they
obey Fermi or Bose statistics. The only constraint we ordered during the problem formulation in Subsection~\ref{sub:problem} was
that they are completely indistinguishable. From this it follows that the Schr\"odinger equation has both symmetric and
antisymmetric solutions. Also, as is well known from quantum mechanics, fermions have a total wave function (including spin), that
changes sign whenever two particles are interchanged. In contrast to this the total wave function of bosons is completely
symmetric. However, since in this paper spin related effects are not considered, we omitted the factorizable spin state in the
very beginning and concentrated on the real space component only. Having developed the group representations its permutation
symmetry is now easy to check as follows.


The group element in $\mathcal{G}$ describing permutation of particles is $\mathbf{R}_2\in\mathcal{A}$, see
Eq.~\eqref{amembers}. The scalar transformation matrix for this operation is in turn
\begin{equation}
\mathbf{P}(\mathbf{R}_2)=\sum_{p,i,k,l=1}^n
\left(\mathbf{i}\circ\mathbf{l}^\text{T}\right)\otimes
\left(\mathbf{p}\circ\mathbf{k}^\text{T}\right)\otimes
\left(\mathbf{l}\circ\mathbf{i}^\text{T}\right)\otimes
\left(\mathbf{k}\circ\mathbf{p}^\text{T}\right).\label{perm}
\end{equation}
Knowing the irreducible representations of $\mathcal{G}$ (from Appendix~\ref{appsec:irr}), one can easily verify that
\begin{equation}
{\bm\Gamma}^{qp}(\mathbf{R}_2)=
\begin{cases}
-\mathbf{E}_2,& \text{if $q=2$ and $p=1,\dots,4$},\\
+\mathbf{E}_{d_{qp}},& \text{otherwise},
\end{cases}\label{anti}
\end{equation}
where $\mathbf{E}_k$, as before, denotes the unit matrix of dimension $k$. These results in conjunction with Eqs.~\eqref{repbasis}
and~\eqref{psiqpj} show that all solutions of the discretized Schr\"odinger equation satisfy the transformation property
\begin{equation}
\mathbf{P}(\mathbf{R}_2){\bm\psi}^{qp}_j=\mp{\bm\psi}^{qp}_j.\label{trafo}
\end{equation}
From this we see that it is the representations that unambiguously distinguish between solutions with different permutation
symmetry: eigenvectors transforming as some row of ${\bm\Gamma}^{2p}$ are all antisymmetric and span the antisymmetric
subspace of the total two-particle Hilbert space. On the other hand, all other solutions that belong to the remaining ten
irreducible representations are symmetric and span the symmetric complementary subspace. Note that each of these subspaces can be
physically important: for example if the two particles under consideration are fermions of spin one-half (electrons, protons,
neutrons, \dots), then the spin singlet and triplet states are accompanied by symmetric and antisymmetric wave functions,
respectively.

\section{Numerical results for noninteracting particles}\label{sec:nointres}
This and the next sections are devoted to the demonstration of numerical results that are based on the theory presented so
far. Here the noninteracting problem is concerned. In this case the quantum mechanical problem of the two identical particles can
be traced back to a single-particle problem and analytic solution is easy to find. Though the lack of interaction turns the
problem into exactly solvable, in group theoretical sense it has some peculiarities. Namely, the symmetry group of the system is
enlarged considerably and contains dynamical symmetries as well. These and other geometrical symmetries will be broken by the
interaction.

\subsection{Enlarged symmetry group: dynamical symmetries}\label{sub:dynamical}
Numerical solution of Eq.~\eqref{schqpj} in each symmetry channel provides the level structure and eigenstates of the system. For
noninteracting particles these are already known analytically in Eqs.~\eqref{eosc}. Exact energy eigenvalues $E/\omega$ are shown
in the first column of Table~\ref{tab:noint} to six digits of precision for parameter values of $\omega^2/2=500$ and $b=1$. They
were calculated from Eq.~\eqref{exactosc} and Table~\ref{tab:osc} of Appendix~\ref{appsec:osc}. Numbers in round brackets indicate
the degrees of degeneracy of each level. At this point we would like to call the attention to the fact that because of the
four-dimensional cubic structure of the boundary condition the eigenvalues themselves as well as their degeneracies differ from
those of the usual 4D isotropic oscillator. There the energy would be $E=\omega(n+2)$, with $n$ a nonnegative integer, and the
degeneracy of the $n$th level is \cite{jauch,baker}
\begin{equation}
g(n)=
\begin{pmatrix}
n+3\\
n
\end{pmatrix}.\label{degosc}
\end{equation}
The highly degenerate nature of the levels is in complete accordance with $\text{SU}(4)$ symmetry of the problem
\cite{fradkin}. In our case, however, the confining potential well breaks down full $\text{O}(4)$ rotational symmetry and leads to
the finite four-dimensional cubic group $\text{O}_4$. This partly explains for example why the degeneracy of the third level is
not 10, but splits into $6+4$. Similar splittings can be observed further down the first column of Table~\ref{tab:noint}. In solid
state terminology this is nothing else then a crystal field splitting, where the perturbation in our case playing the role of the
crystal field is the infinite potential barrier.

\begin{table}
\caption{\label{tab:noint}Classification of the first few energy eigenvalues $E/\omega$ of the two noninteracting particles in the
  confined harmonic potential. Scaling factor $\omega$ and the half box size $b$ were chosen as $\omega^2/2=500$ and $b=1$. First
  column shows the exact theoretical results to six digits of precision calculated from Eq.~\eqref{exactosc} and
  Table~\ref{tab:osc} of Appendix~\ref{appsec:osc}. Numbers in parentheses indicate the degrees of degeneracy of each level. In
  the second column numerical data are given that were obtained from Eq.~\eqref{schqpj} for $m=8$ (the dimension of the low energy
  subspace is therefore $m^4=4096$) and for a resolution of $n=30$. The step size is $h=2b/(n+1)=0.0645$. Further columns show the
  level distribution across representations of $\mathcal{G}$, whereas the last one accounts for the total degeneracy of the given
  level. Note that representations written in boldface are two-dimensional. The numbers in the second row mean the total number of
  independent vectors in the given representation. These we denoted by $r^{qp}$ in the text. Thus $\sum_{qp}d_{qp}r^{qp}=4096$.}
\renewcommand*{\arraystretch}{1.1}
\begin{tabular*}{\textwidth}{@{\extracolsep{\fill}}ll*{14}{c}r}
\hline
\multicolumn{2}{c}{$E/\omega$} & $\Gamma^{11}$ & $\Gamma^{12}$ & $\Gamma^{13}$ & $\Gamma^{14}$ & ${\bm\Gamma}^{15}$ &
${\bm\Gamma}^{21}$ & ${\bm\Gamma}^{22}$ & ${\bm\Gamma}^{23}$ & ${\bm\Gamma}^{24}$ & $\Gamma^{41}$ &
$\Gamma^{42}$ & $\Gamma^{43}$ & $\Gamma^{44}$ & ${\bm\Gamma}^{45}$ & deg.\\
theoretical & numerical & 210 & 190 & 120 & 136 & 320 & 240 & 256 & 192 & 320 & 78 & 66 & 120 & 136 & 192\\
\hline
2.000004\,(1) & 2.000000 &1&&&&&&&&&&&&&&1\\
\hline
3.000020\,(4) & 3.000011 &&&&&1&&&&1&&&&&&4\\
\hline
4.000036\,(6) & 4.000019 &1&1&&1&&&1&&&&&&1&&6\\
4.000238\,(4) & 4.000207 &1&&&&&&&&&&&&&&1\\
& 4.000212 &&1&&&&1&&&&&&&&&3\\
\hline
5.000052\,(4) & 5.000020 &&&&&1&&&&1&&&&&&4\\
5.000253\,(12) & 5.000208 &&&&&1&&&&1&&&&&&4\\
& 5.000220 &&&&&1&&&1&1&&&&&1&8\\
5.001948\,(4) & 5.001874 &&&&&1&&&&1&&&&&&4\\
\hline
6.000068\,(1) & 6.000008 &1&&&&&&&&&&&&&&1\\
6.000270\,(12) & 6.000200 &1&1&&1&&&1&&&&&&1&&6\\
& 6.000214 &&&1&&&1&1&&&&&1&&&6\\
6.000472\,(6) & 6.000382 &1&&&&&&&&&&&&&&1\\
& 6.000410 &&1&&&&1&&&&&&&&&3\\
& 6.000423 &1&&&&&&&&&1&&&&&2\\
6.001963\,(12) & 6.001870 &1&1&&1&&&1&&&&&&1&&6\\
& 6.001879 &&&1&&&1&1&&&&&1&&&6\\
6.010901\,(4) & 6.010716 &1&1&&&&1&&&&&&&&&4\\
\hline
\end{tabular*}
\end{table}

A more interesting observation is, however, that the representation theory of $\text{O}_4$ cannot explain completely all
degeneracies found in the level structure, because the dimensions of its irreducible representations are 1, 2, 3, 4, 6 and 8
\cite{dai}. This means, again looking at Table~\ref{tab:noint}, that the twelve times degenerate energy levels either exhibit
accidental degeneracies or beyond geometrical symmetries there are dynamical symmetries as well and the invariance group is
actually larger than $\text{O}_4$. Given the simplicity of Eq.~\eqref{exactosc} and the fact that since the $\nu_n$'s are related
to the zeros of the confluent hypergeometric function they are definitely not integers, one might have the impression that
accidental degeneracies are very rare or more likely do not occur at all. In the latter case simple combinatorial reasoning shows
that the degree of degeneracy of a particular level can only be 1, 4, 6, 12 or 24. Following Ref.~\cite{shaw} we found indeed that
extra degeneracies are related to the existence of a larger symmetry group and it is the irreducible representations of this group
that are associated with the energy eigenvalues. This group will be explored next.

The Hamiltonian without interaction can be written as $H=\sum_iH_i$, where
\begin{equation}
H_i=-\frac{\partial^2}{\partial x_i^2}+\frac14\omega^2x_i^2,\qquad i=1,\dots,4,\label{1dham}
\end{equation}
and as is known
\begin{equation}
[H_i,H_j]=0.\label{hcomm}
\end{equation}
This signals it is not just the total energy that is conserved but also each component separately. In other words, due to the lack
of interaction the one-dimensional sub-oscillators do not exchange energy quanta during the motion and this is essentially the
reason for that the Schr\"odinger equation can be separated in Descartes coordinates. Remember that the Schr\"odinger equation is
supplemented with a boundary condition of Eq.~\eqref{4Dconstraint} that prevents separation in spherical or cylindrical polars,
thus 4D angular momentum is not a constant of motion. Neither are the offdiagonal components of the symmetric tensor operator
found to be responsible for the extra degeneracies of the unconfined oscillator \cite{fradkin,fradkin-prog}. In this subsection we
refer to the invariance group $\mathcal{G}'$ of the noninteracting Hamiltonian as the group consisting of all unitary operators of
the Hilbert space that commute with $H$. We know so far there is a discrete subgroup, call it $\mathcal{G}_2$, that is related to
geometrical symmetries: it involves the scalar transformation operators $P(\mathbf{R})$, where
$\mathbf{R}\in\text{O}_4$. Commutativity of oscillator quanta with the total Hamiltonian suggests that beyond this there is a four
parameter continuous subgroup as well consisting of
\begin{align}
U(\lambda_1,\lambda_2,\lambda_3,\lambda_4)=e^{i\sum_j\lambda_j H_j}.\label{fourlie}
\end{align}
Due to commutativity of the generators in Eq.~\eqref{hcomm} the associated real Lie algebra is not semi-simple and the
representation theory of the problem becomes quite cumbersome. Nonetheless, there exists a discrete group sufficient for our
purposes. Let us define certain linear operators of the Hilbert space \cite{shaw}
\begin{subequations}\label{li}
\begin{align}
\Lambda_0\psi_\mathbf{k}(\mathbf{x})&=\text{sign}(+\nu_{k_1}+\nu_{k_2}+\nu_{k_3}+\nu_{k_4})\psi_\mathbf{k}(\mathbf{x}),\label{l0}\\
\Lambda_1\psi_\mathbf{k}(\mathbf{x})&=\text{sign}(+\nu_{k_1}+\nu_{k_2}-\nu_{k_3}-\nu_{k_4})\psi_\mathbf{k}(\mathbf{x}),\label{l1}\\
\Lambda_2\psi_\mathbf{k}(\mathbf{x})&=\text{sign}(+\nu_{k_1}-\nu_{k_2}+\nu_{k_3}-\nu_{k_4})\psi_\mathbf{k}(\mathbf{x}),\label{l2}\\
\Lambda_3\psi_\mathbf{k}(\mathbf{x})&=\text{sign}(-\nu_{k_1}+\nu_{k_2}+\nu_{k_3}-\nu_{k_4})\psi_\mathbf{k}(\mathbf{x}),\label{l3}
\end{align}
\end{subequations}
where $\psi_\mathbf{k}(\mathbf{x})$ is the exact eigenstate of Eq.~\eqref{exactoscwave}. These functions form an orthonormal basis
of the Hilbert space, therefore Eqs.~\eqref{li} in conjunction with linearity define the action of $\Lambda_i$ on each function of
the space. It is easy to see that since $\nu_i$ are positive $\Lambda_0$ is the identity operator $I$. Its definition reminds of
the action of the Hamiltonian, albeit here it is not the energy that multiplies the eigenstate but its sign only. In a similar
manner the other operators can be put in correspondence with linear combinations of $H_i$. Equations~\eqref{li} lead to the
following properties: (i) $[\Lambda_i,H]=0$, (ii) $[\Lambda_i,\Lambda_j]=0$, (iii) $\Lambda_i^2=I$ and (iiii) $\Lambda_i$ are
unitary and self-adjoint operators at the same time. Now the full invariance group possesses the following Abelian subgroup of
order 16
\begin{equation}
\mathcal{G}_1=\pm\left\{I,\Lambda_1,\Lambda_2,\Lambda_3,\Lambda_1\Lambda_2,\Lambda_1\Lambda_3,
\Lambda_2\Lambda_3,\Lambda_1\Lambda_2\Lambda_3\right\},\label{dynamical}
\end{equation}
which is isomorphic to the direct product of cyclic groups of order 2
\begin{equation}
\mathcal{G}_1\sim\{I,-I\}\otimes\{I,\Lambda_1\}\otimes\{I,\Lambda_2\}\otimes\{I,\Lambda_3\}\sim C_2^4.\label{dync2}
\end{equation}
The elements of $\mathcal{G}_1$ will be called dynamical symmetries. With all these $\mathcal{G}'$ can then be expressed as the
group of products of unitary operators
\begin{equation}
\mathcal{G}'=\left\{\Lambda P\mid\Lambda\in\mathcal{G}_1,P\in\mathcal{G}_2\right\}.\label{fullnoint}
\end{equation}
This is a finite group of order $16\times384=6144$. Development of its representation theory can be based on the very important
observation that this is in fact a semi-direct product group, because it can be shown that $\mathcal{G}_1$ is actually an
invariant subgroup. Consequently we can write
\begin{equation}
\mathcal{G}'=\mathcal{G}_1\circledS\mathcal{G}_2\sim C_2^4\circledS\text{O}_4.
\end{equation}
We have already pointed out in the previous section and in particularly Appendix~\ref{appsec:irr}, that this sort of group
structure is rather convenient when one wants to explore the representations. Convenient, because the method of induction
\cite{cornwell-book} can be applied in order to induce irreducible representations of $\mathcal{G}'$ from those of the
constituents $\mathcal{G}_1$ and $\mathcal{G}_2$, respectively. The method does indeed verify that the irreducible representations
of $\mathcal{G}'$ have dimensions 1, 4, 6, 12 and 24, so these are to be associated with the energy levels of the confined 4D
harmonic oscillator (first column of Table~\ref{tab:noint}). The corresponding noninteracting wave functions of
Eq.~\eqref{exactoscwave} form in turn bases for some (equivalent) forms of these representations.

\subsection{Noninteracting level structure: single- and two-particle density of states}\label{sub:nointlevel}
We now turn to the results of the numerical solution. In Subsection~\ref{sub:pro} we introduced the low energy subspace
$\mathcal{H}_m$ and constructed a symmetry adapted basis for it. In this every member transforms as some row of some irreducible
representation. The first few energy levels computed numerically from Eq.~\eqref{schqpj} are tabulated in the second column of
Table~\ref{tab:noint}. For numerics we used $\omega^2/2=500$, $b=1$, $m=8$ and $n=30$. Consequently the dimension of the whole
Hilbert space is $n^4=810000$, whereas that of the subspace is $m^4=4096$. The results are represented to six digits of precision
and as we can see, compared to the exact theoretical values in the first column, remarkable accuracy is found. As a matter of fact
this is not that surprising as the basis elements of Eq.~\eqref{linalgbasis} are the exact solutions themselves. What is notable
though is that the computed data are accurate up to three or more digits already at the modest resolution of
$h=2b/(n+1)=0.0645$. This is apparently due to the very small $\mathcal{O}(h^6)$ error of our difference scheme. This accuracy
might give us hope for when we finally incorporate the Coulomb interaction, for which we do not have exact analytical results to
compare with, the numerical data will be adequate. From the fourth column on we can see the distribution of eigenvalues across
representations. The last column summarizes the degeneracy of a given level. Note that representations written in boldface are
two-dimensional.

\begin{figure}
\psfrag{x}[t][b]{$n$}
\psfrag{y}[b][t]{$(E_\infty-E_n)/\omega$}
\psfrag{a}{$b=5.27\pm0.07$}
\psfrag{b}{$b=6.29\pm0.08$}
\psfrag{c}{$b=5.95\pm0.03$}
\psfrag{d}{$b=5.84\pm0.04$}
\begin{center}
\includegraphics[width=0.5\textwidth]{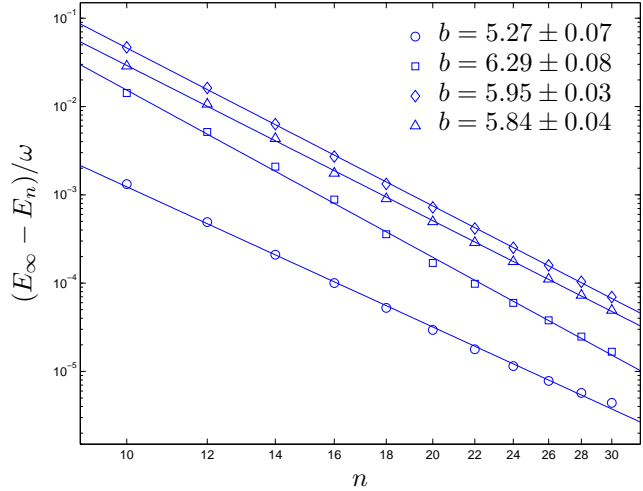}
\end{center}
\caption{\label{fig:konv}(Color online) Some energy eigenvalues belonging to $\Gamma^{11}$ as a function of the resolution
  $n$. The circles, squares, diamonds and triangles are the raw computed data representing the lowest, the second, the fifth and
  the seventh largest eigenvalue of this symmetry, respectively. The solid lines are linear fittings and the legend shows the
  obtained exponents $b$, see Eq.~\eqref{tendency} in the text. Note that both axes are in logarithmic scale.}
\end{figure}

So far we have performed the numerical computations at only one step size corresponding to the resolution $n=30$. In order to
verify the conjecture that the eigenvalues do indeed converge fast to the theoretical predictions, with a leading error of
$\mathcal{O}(h^6)$, we have made test runs with increasing $n$ from $n=10$ to 30 with a step of 2. Meanwhile, though the dimension
of the full Hilbert space grows as $n^4$, the dimension of the low energy subspace was held fixed as specified
above. Figure~\ref{fig:konv} shows the dependence of the raw data on the resolution that were obtained from
Eq.~\eqref{schqpj} for $\Gamma^{11}$. The circles denote the development of the lowest energy eigenvalue in this symmetry
channel. Similarly, the squares, diamonds and the triangles denote the second, fifth and the seventh largest eigenvalues,
respectively. As a function of $n$ each of these follows an algebraic tendency
\begin{equation}
E_n=E_\infty+\frac{a}{n^b}+\dots,\label{tendency}
\end{equation}
where $E_\infty$ stands for the asymptotic limit, i.e.\ the exact result tabulated in the first column of
Table~\ref{tab:noint}. The coefficient $a$ is unknown and also is the exponent $b$. However, according to what has been said so
far we expect $b$ to be close to 6.  Linear fittings on the raw data were performed and are shown in the figure by solid
lines. From these we found that the exponent for the lowest energy (circles) is $b=5.27\pm0.07$, for the second largest energy
(squares) it is $b=6.29\pm0.08$, for the fifth largest energy (diamonds) it is $b=5.95\pm0.03$ and finally for the seventh largest
eigenvalue (triangles) $b=5.84\pm0.04$ was obtained. We consider these representative results satisfying. The deviations from the
theoretically expected exponent could be caused for example by the fact that at this relatively low scale of $n$ higher order
terms in Eq.~\eqref{tendency} (not shown explicitly) can contribute a little. A more important cause, however, might be the fact
that we have ``truncated'' the full Hilbert space considerably, which we did on physical grounds. Increasing the dimension of the
low energy subspace $\mathcal{H}_m$ would result supposedly in more accurate results and exponents closer to 6.

The level structure of a noninteracting system is usually characterized by the single-particle density of states (DOS). In our
case, recalling Eq.~\eqref{exactosc}, it reads
\begin{equation}
g_1(\epsilon)=\sum_{i=1}^2\sum_{n_i=0}^\infty\delta\left(\epsilon-\omega(\nu_{n_1}+\nu_{n_2}+1)\right).\label{dos1}
\end{equation}
From this the total noninteracting two-particle density of states can be expressed as a convolution
\begin{equation}
g_2^\text{tot}(\epsilon)\equiv\sum_{i=1}^4\sum_{n_i=0}^\infty
\delta\left(\epsilon-\omega(\nu_{n_1}+\nu_{n_2}+\nu_{n_3}+\nu_{n_4}+2)\right)
=\int\text{d}\epsilon'g_1(\epsilon-\epsilon')g_1(\epsilon').\label{dos2tot}
\end{equation}
This quantity is, however, rather artificial because the spectrum it measures is actually that of an abstract four-dimensional
particle, the symmetry of its quantum state we do not have restrictions for. Mathematically speaking, the Hilbert space of this
entity is the direct sum of the symmetric and antisymmetric two-particle subspaces. In reality what we have is the pair of two
identical particles and the symmetric and antisymmetric solutions could describe completely different states. For example consider
a spin singlet two-electron state. We know that the spatial wave function can only be symmetric in this case, whatever its energy
is, and so must reside somewhere in the symmetric subspace. As opposed to this the triplet state can only be associated with
antisymmetric wave functions which then obviously restricts the possible energies of the pair. Accordingly, in order to obtain
true two-particle DOS, $g_2^\text{tot}$ has to be split into two terms, loosely speaking, with respect to their symmetries under
permutation. However, the form in Eq.~\eqref{dos2tot} is not appropriate for this as the quantum numbers $n_i$ are not related
directly to symmetries. Irreducible representations provide good quantum numbers for this purpose and we can write
\begin{equation}
g_2(\epsilon)=\sideset{}{'}\sum_{qp}\sum_{r=1}^{r^{qp}}d_{qp}\delta(\epsilon-E^{qp}_r).\label{dos2sym}
\end{equation}
Here $E^{qp}_r$ is the $r$th largest eigenvalue of Eq.~\eqref{schqpj}, associated with the state ${\bm\psi}^{qp}_{j,r}$. The first
few noninteracting data are shown in the second column of Table~\ref{tab:noint}. As we have already pointed out, group theory says
that all energy levels belonging to a certain representation, say ${\bm\Gamma}^{qp}$, are exactly $d_{qp}$-fold degenerate and
this explains the weight of the dirac-delta. Most importantly, the sum running through representations now involves only those for
which the basis functions ${\bm\psi}^{qp}_{j,r}$ are symmetric (or antisymmetric) under permutation, see
Subsection~\ref{sub:fermibose}. This constraint is denoted by a prime.

The insets in the left and right panels of Fig.~\ref{fig:dos2} show the noninteracting two-particle DOS associated with the
antisymmetric and symmetric subspaces, respectively. Both $g_1$ and $g_2$ will be modified by the interaction. This we will
discuss in Subsection~\ref{sub:intdos}.

\subsection{Noninteracting two-particle densities}\label{sub:densnoint}
\begin{figure}[p]
\psfrag{x}{$x$}
\psfrag{y}{$y$}
\psfrag{a}{$\Gamma^{11}$}
\psfrag{b}{$\Gamma^{12}$}
\psfrag{c}{$\Gamma^{13}$}
\psfrag{d}{$\Gamma^{14}$}
\psfrag{e}{${\bm\Gamma}^{15}$ $j=1$}
\psfrag{f}{${\bm\Gamma}^{15}$ $j=2$}
\psfrag{g}{${\bm\Gamma}^{21}$ $j=1$}
\psfrag{h}{${\bm\Gamma}^{21}$ $j=2$}
\psfrag{i}{${\bm\Gamma}^{22}$ $j=1$}
\psfrag{j}{${\bm\Gamma}^{22}$ $j=2$}
\psfrag{k}{${\bm\Gamma}^{23}$ $j=1$}
\psfrag{l}{${\bm\Gamma}^{23}$ $j=2$}
\psfrag{m}{${\bm\Gamma}^{24}$ $j=1$}
\psfrag{n}{${\bm\Gamma}^{24}$ $j=2$}
\psfrag{o}{$\Gamma^{41}$}
\psfrag{p}{$\Gamma^{42}$}
\psfrag{q}{$\Gamma^{43}$}
\psfrag{r}{$\Gamma^{44}$}
\psfrag{s}{${\bm\Gamma}^{45}$ $j=1$}
\psfrag{t}{${\bm\Gamma}^{45}$ $j=2$}
\begin{center}
\includegraphics[width=0.4\textwidth]{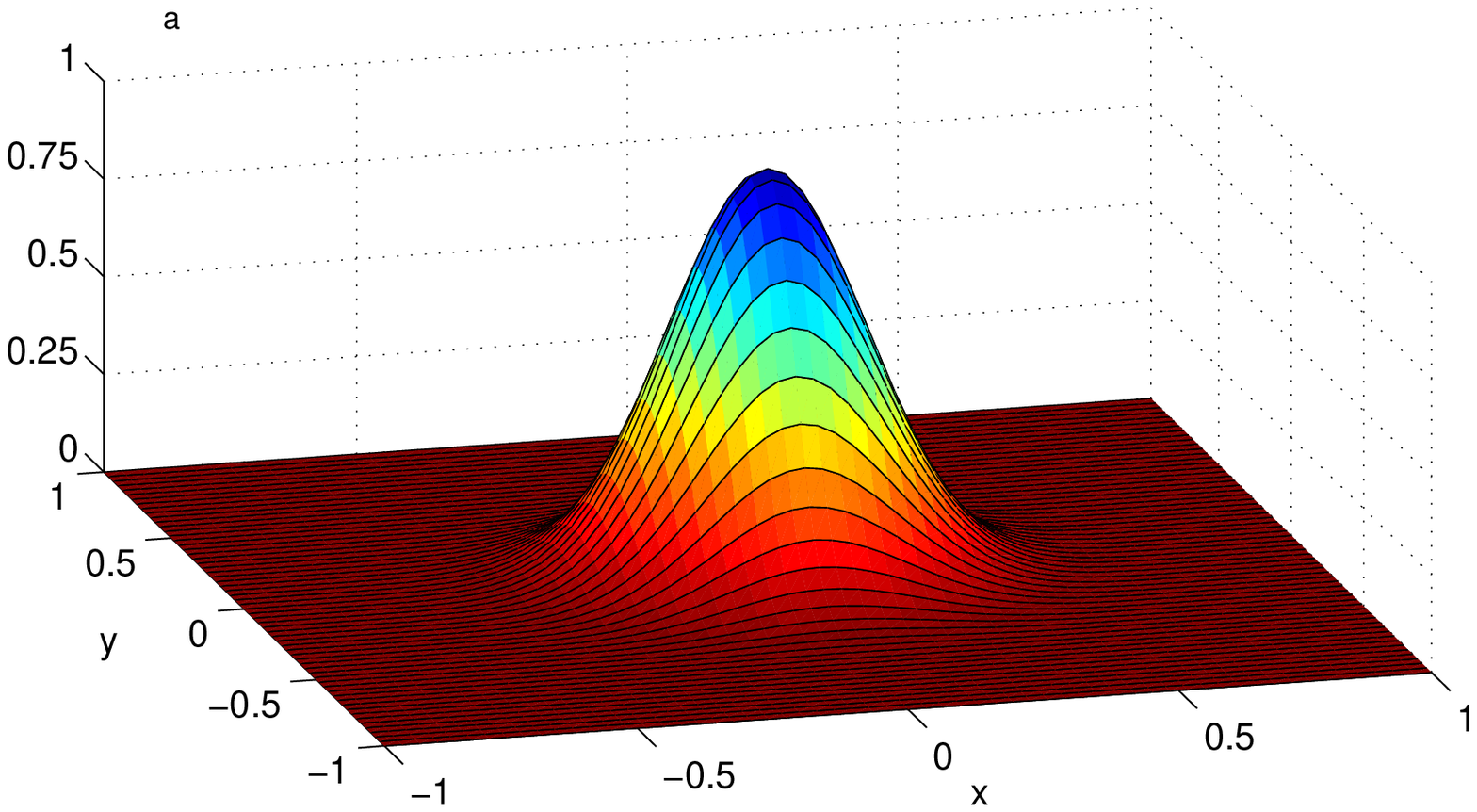}\hspace{0.1\textwidth}
\includegraphics[width=0.4\textwidth]{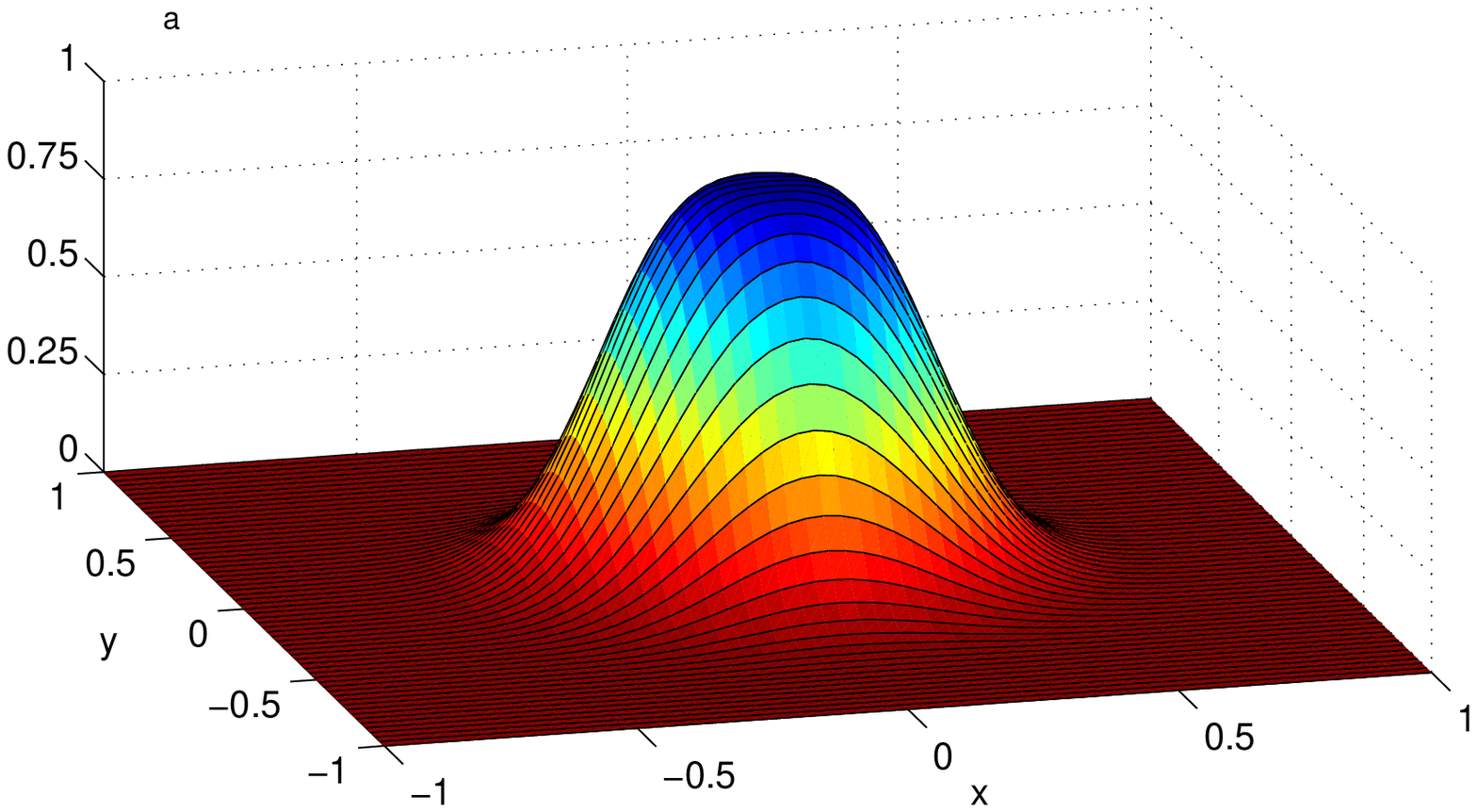}\\

\vspace{2mm}
\includegraphics[width=0.4\textwidth]{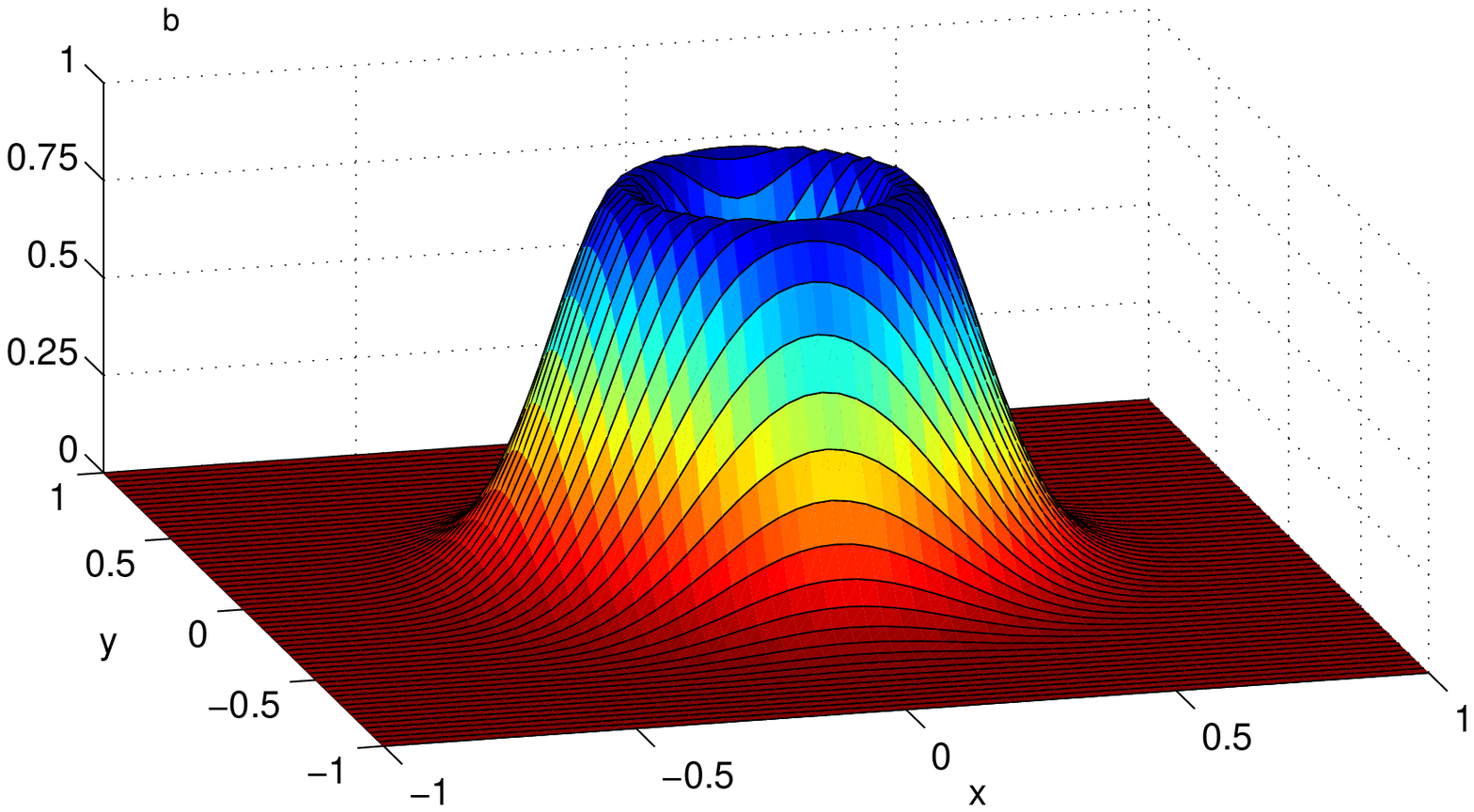}\hspace{0.1\textwidth}
\includegraphics[width=0.4\textwidth]{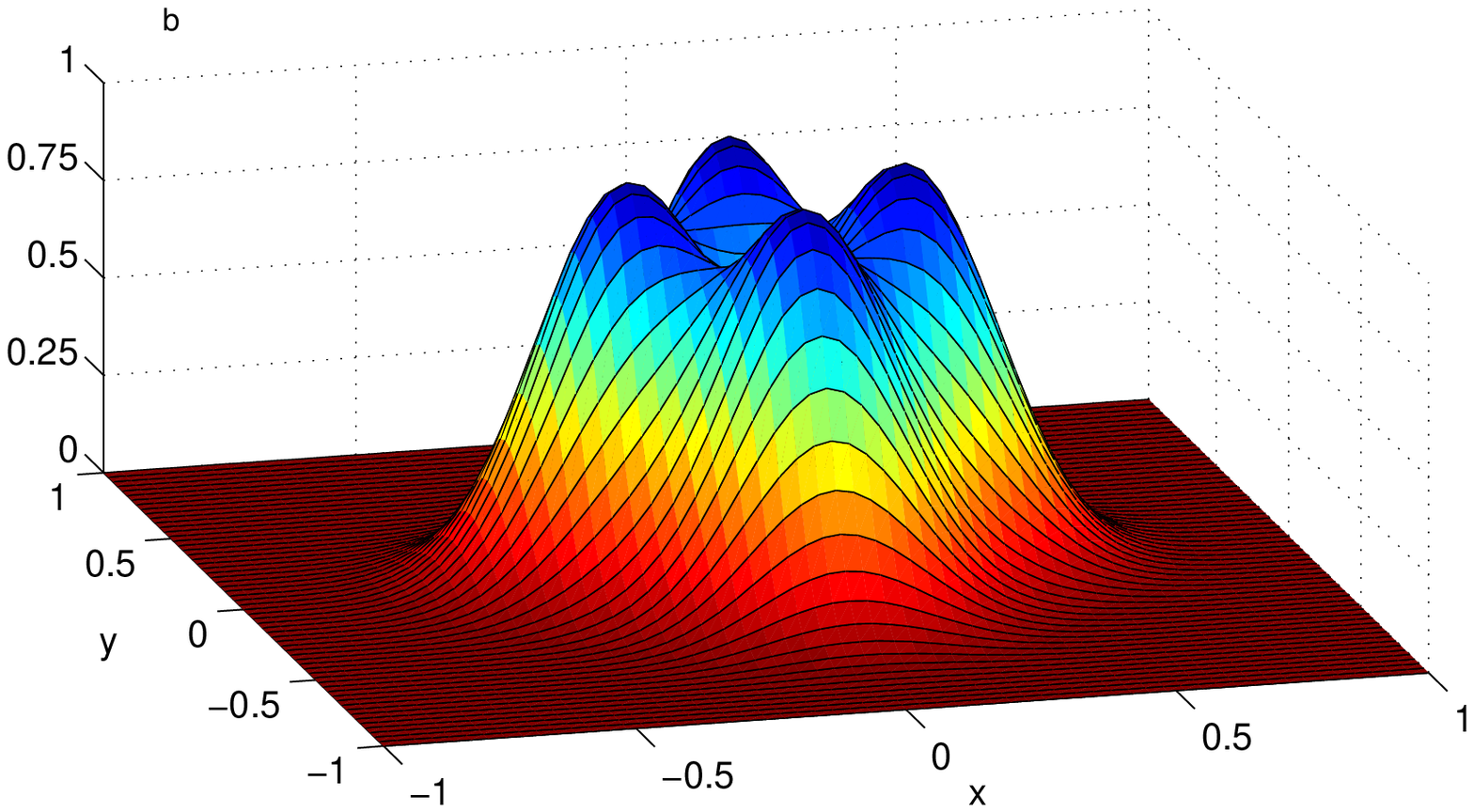}\\

\vspace{2mm}
\includegraphics[width=0.4\textwidth]{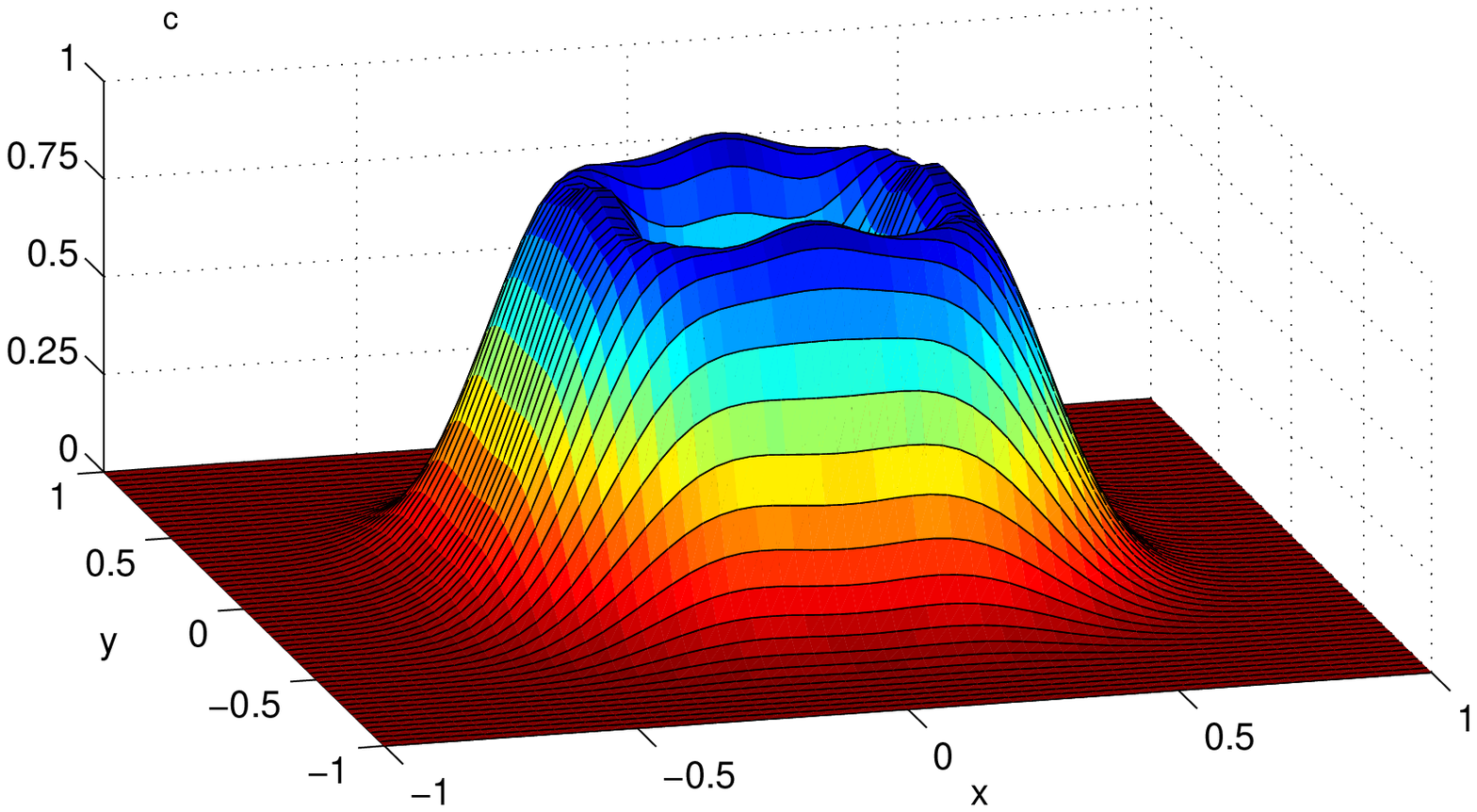}\hspace{0.1\textwidth}
\includegraphics[width=0.4\textwidth]{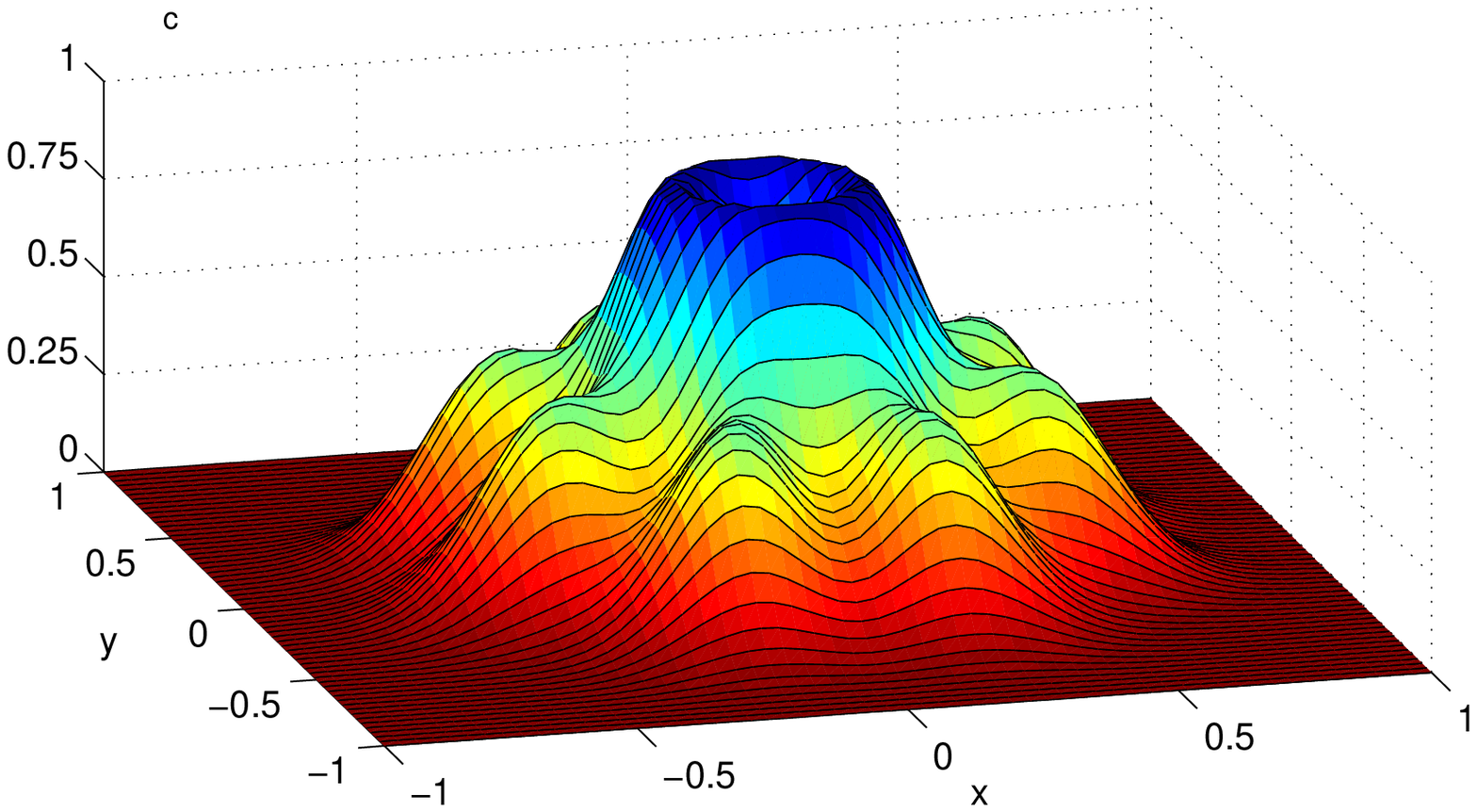}\\

\vspace{2mm}
\includegraphics[width=0.4\textwidth]{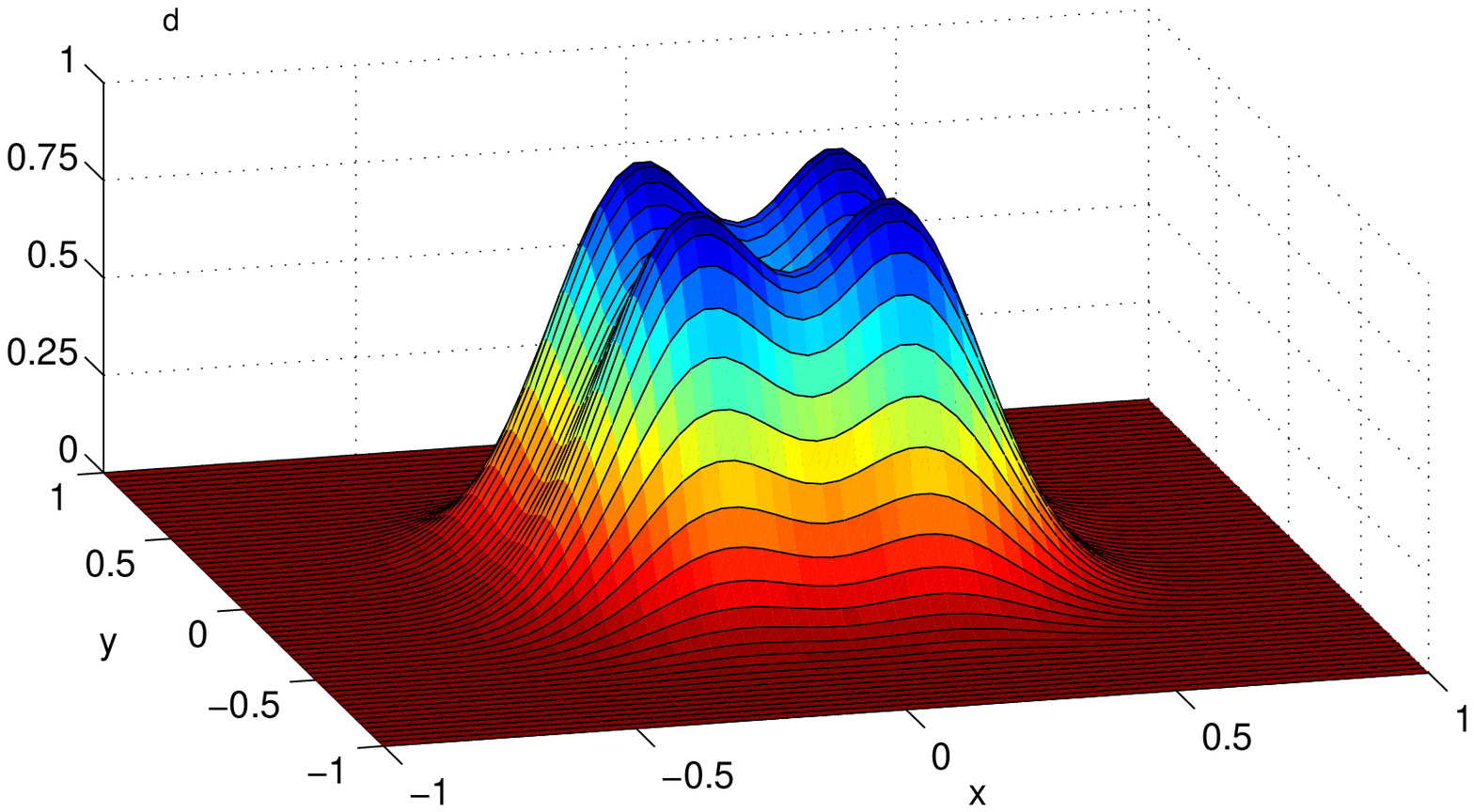}\hspace{0.1\textwidth}
\includegraphics[width=0.4\textwidth]{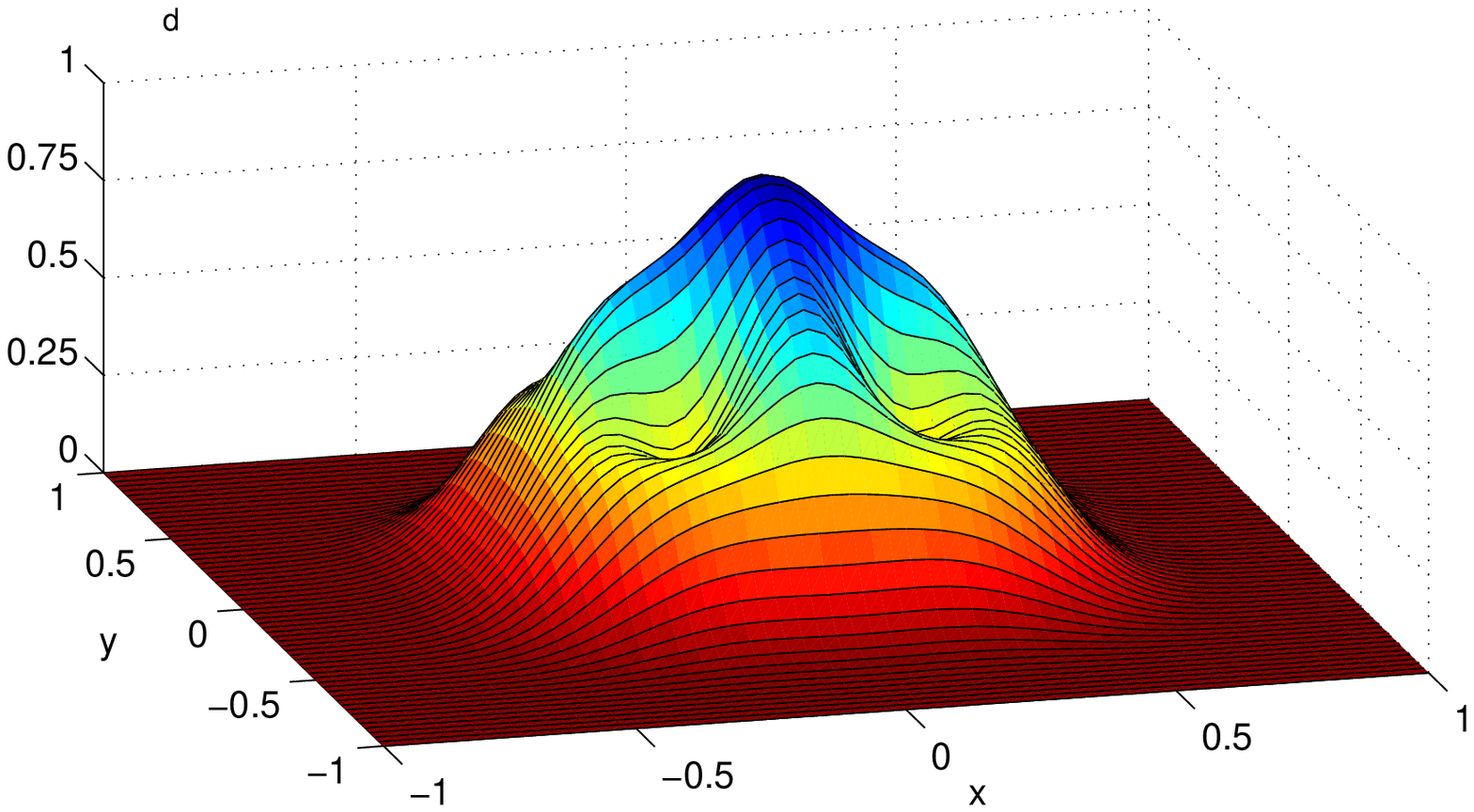}\\

\vspace{2mm}
\includegraphics[width=0.4\textwidth]{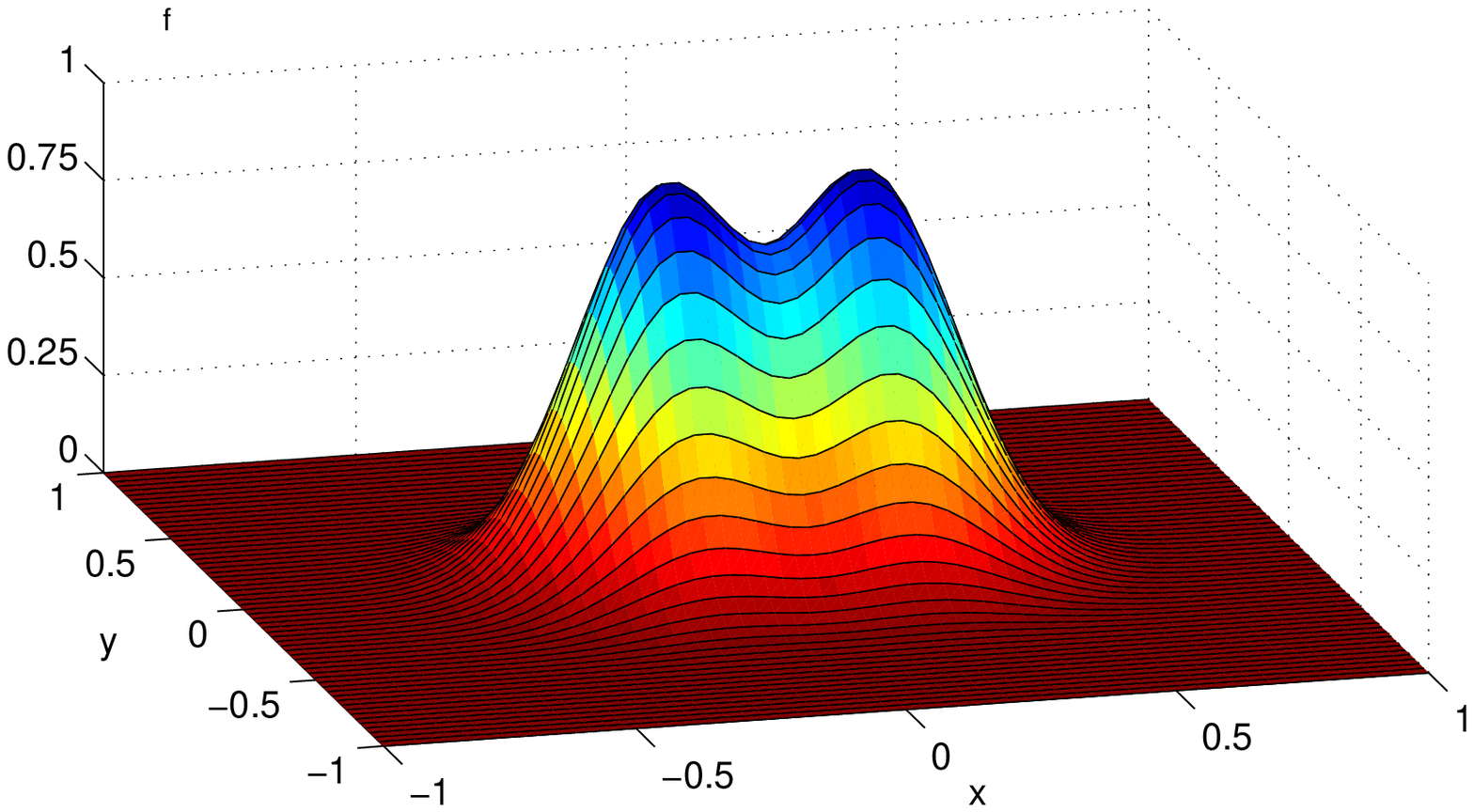}\hspace{0.1\textwidth}
\includegraphics[width=0.4\textwidth]{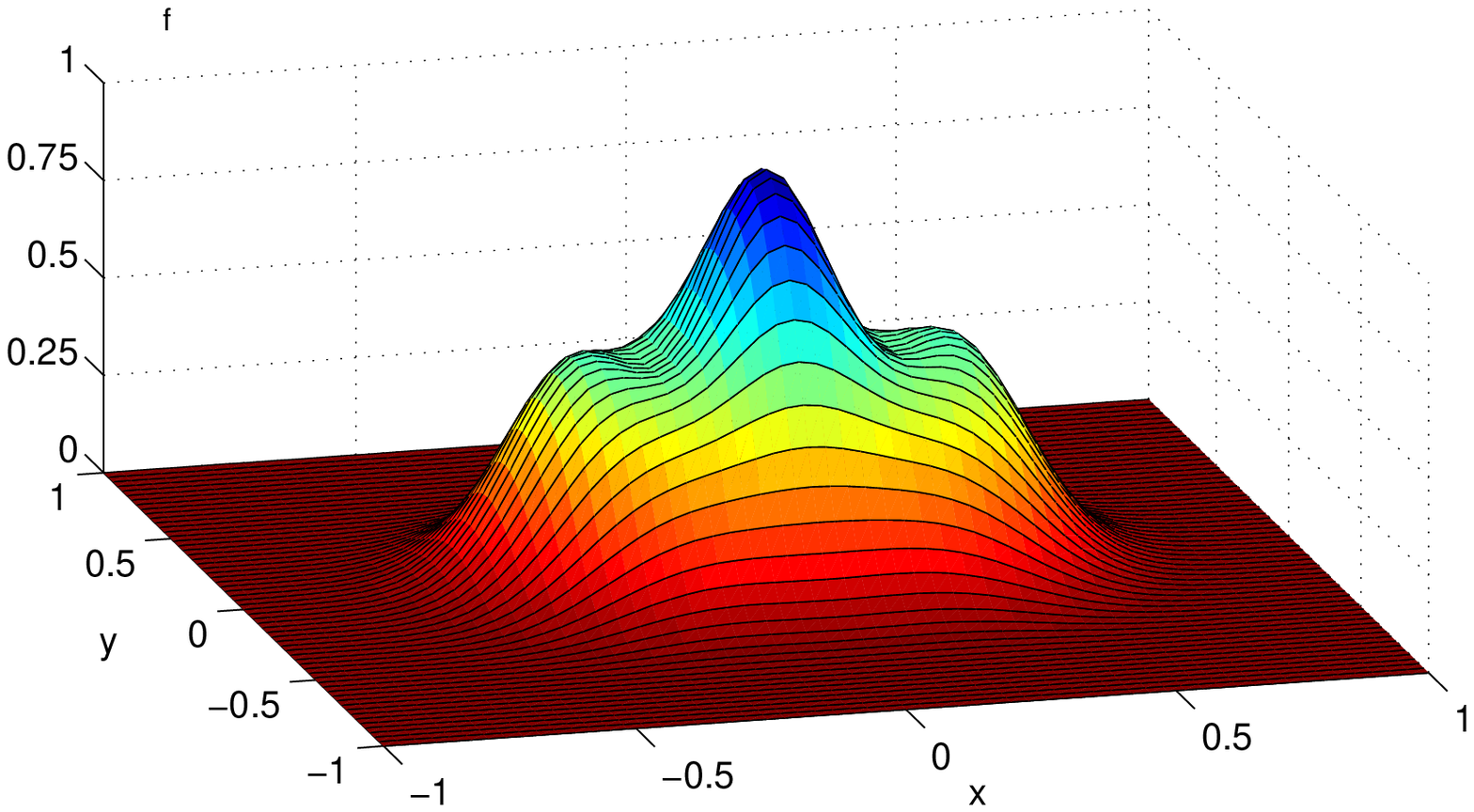}\\
\end{center}
\caption{\label{fig:hasonlit}(Color online) Normalized two-particle densities obtained from the lowest energy symmetric wave
  functions transforming as ${\bm\Gamma}^{1p}$, $p=1,\dots,5$. In case of $p=5$ the row index is $j=2$. The left (right)
  column shows the noninteracting (interacting) densities. Associated energy levels in the left (right) column from top to bottom
  are: 2.000000 (2.702), 4.000019 (4.059), 6.000214 (6.060), 4.000019 (4.745) and 3.000011 (3.724). These are taken from
  Tables~\ref{tab:noint} and~\ref{tab:int}.}
\end{figure}

\begin{figure}[p]
\psfrag{x}{$x$}
\psfrag{y}{$y$}
\psfrag{a}{$\Gamma^{11}$}
\psfrag{b}{$\Gamma^{12}$}
\psfrag{c}{$\Gamma^{13}$}
\psfrag{d}{$\Gamma^{14}$}
\psfrag{e}{${\bm\Gamma}^{15}$ $j=1$}
\psfrag{f}{${\bm\Gamma}^{15}$ $j=2$}
\psfrag{g}{${\bm\Gamma}^{21}$ $j=1$}
\psfrag{h}{${\bm\Gamma}^{21}$ $j=2$}
\psfrag{i}{${\bm\Gamma}^{22}$ $j=1$}
\psfrag{j}{${\bm\Gamma}^{22}$ $j=2$}
\psfrag{k}{${\bm\Gamma}^{23}$ $j=1$}
\psfrag{l}{${\bm\Gamma}^{23}$ $j=2$}
\psfrag{m}{${\bm\Gamma}^{24}$ $j=1$}
\psfrag{n}{${\bm\Gamma}^{24}$ $j=2$}
\psfrag{o}{$\Gamma^{41}$}
\psfrag{p}{$\Gamma^{42}$}
\psfrag{q}{$\Gamma^{43}$}
\psfrag{r}{$\Gamma^{44}$}
\psfrag{s}{${\bm\Gamma}^{45}$ $j=1$}
\psfrag{t}{${\bm\Gamma}^{45}$ $j=2$}
\begin{center}
\includegraphics[width=0.4\textwidth]{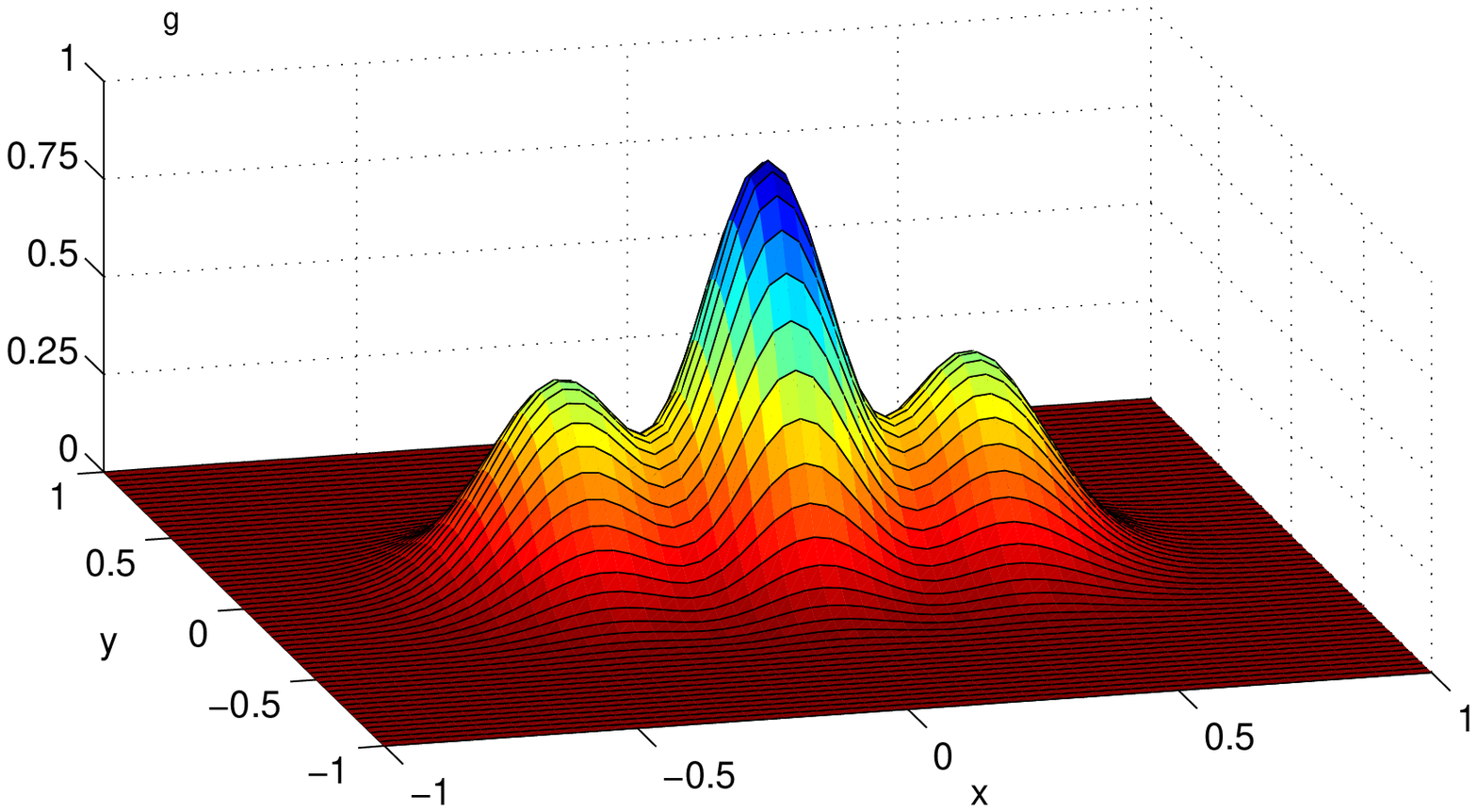}\hspace{0.1\textwidth}
\includegraphics[width=0.4\textwidth]{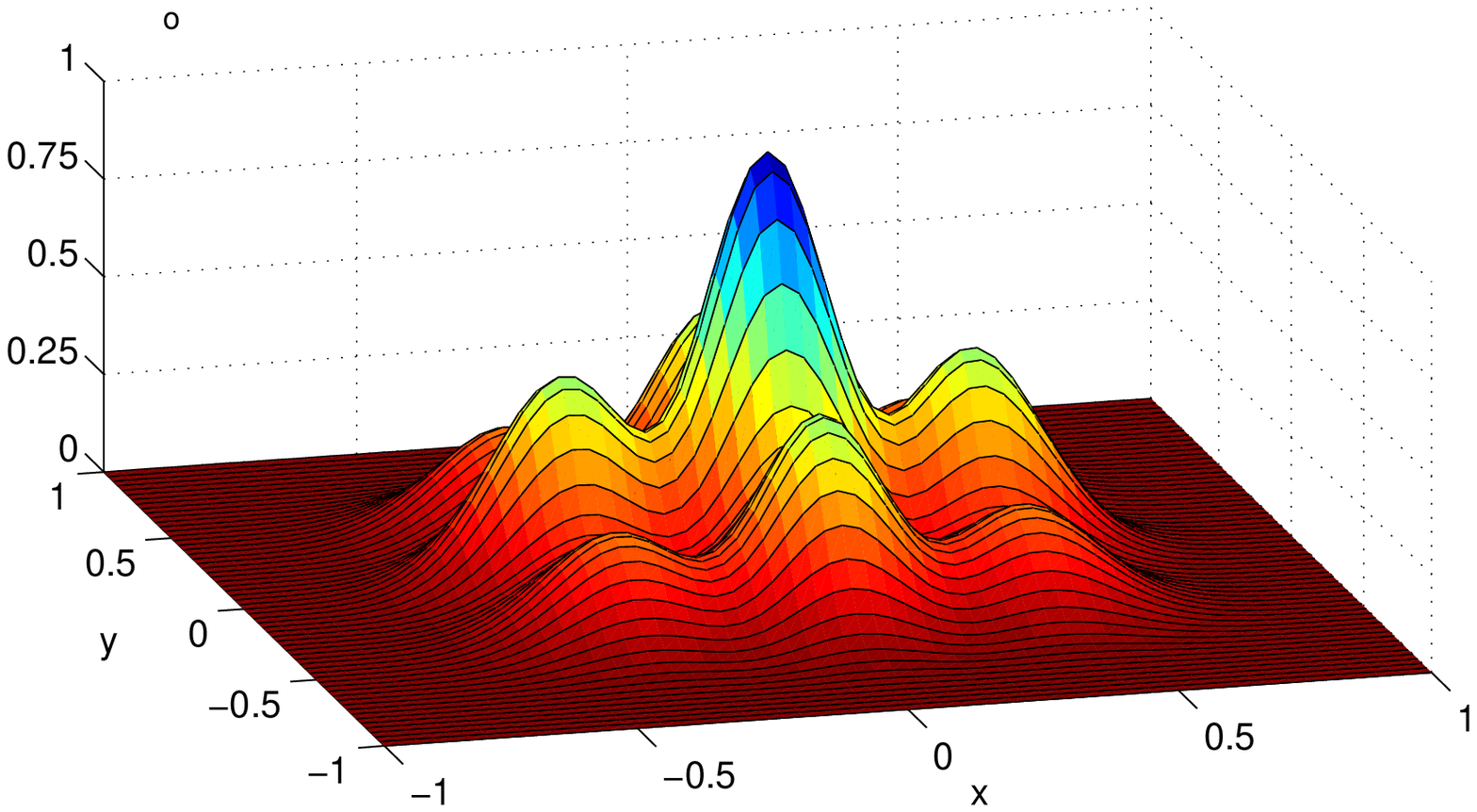}\\

\vspace{2mm}
\includegraphics[width=0.4\textwidth]{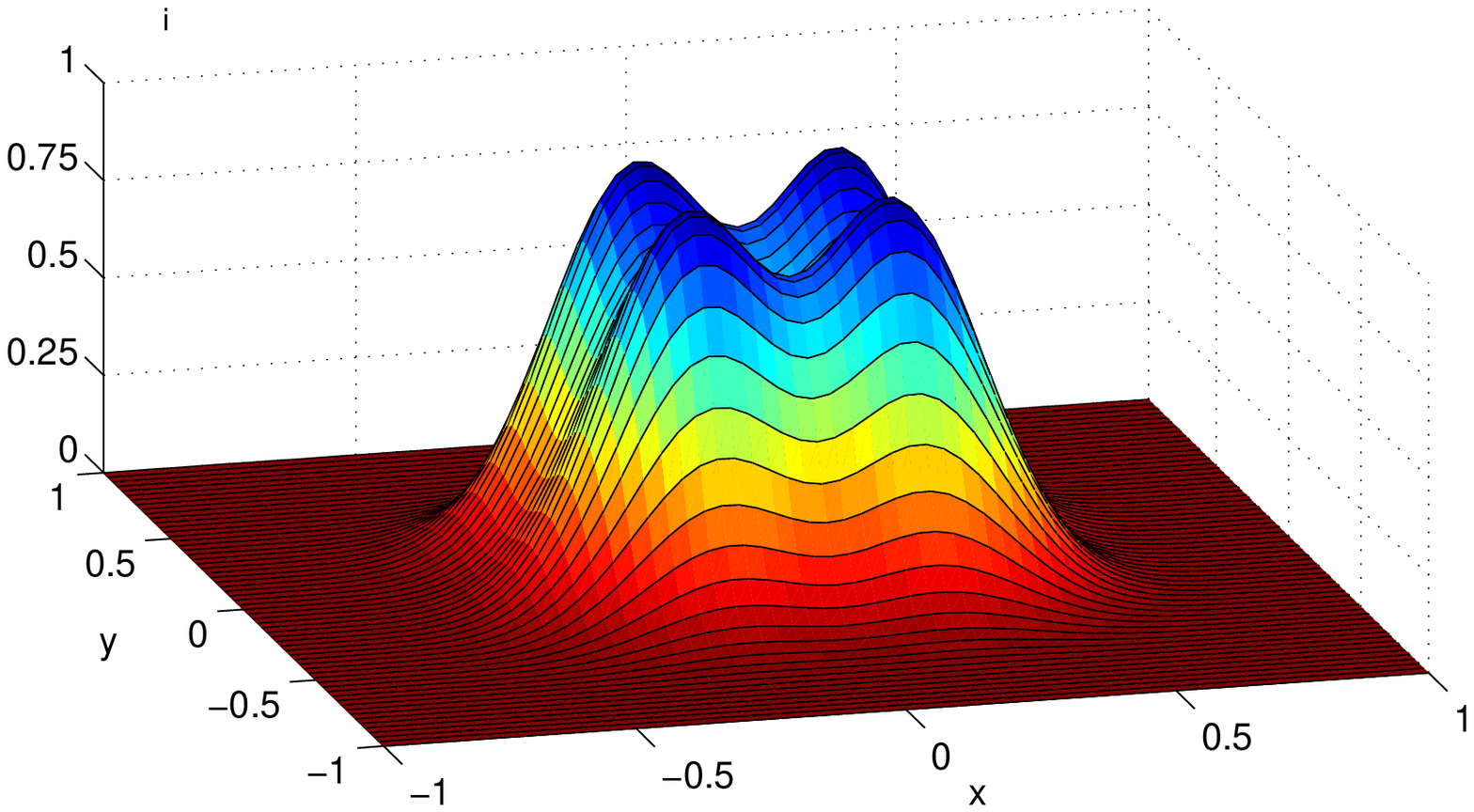}\hspace{0.1\textwidth}
\includegraphics[width=0.4\textwidth]{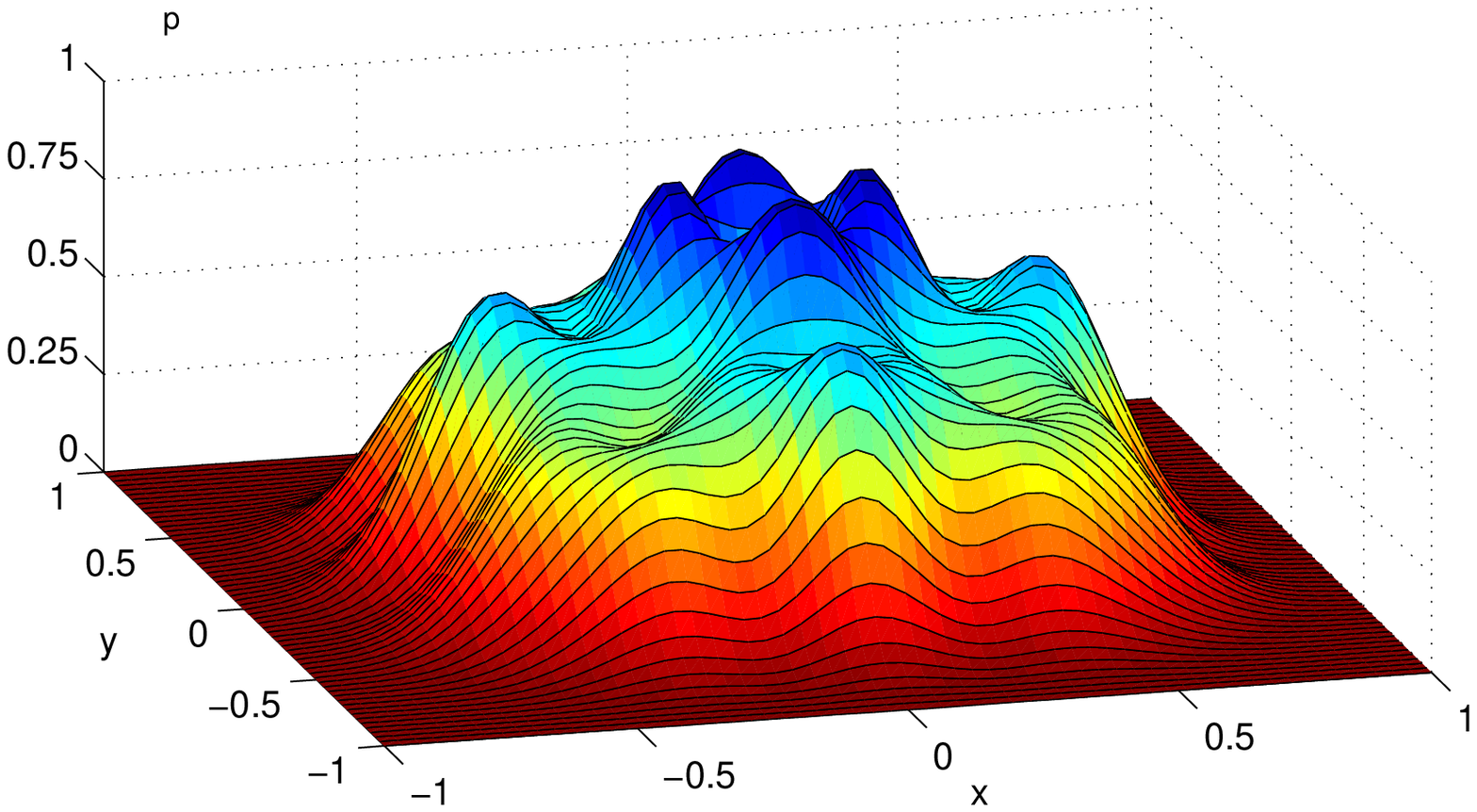}\\

\vspace{2mm}
\includegraphics[width=0.4\textwidth]{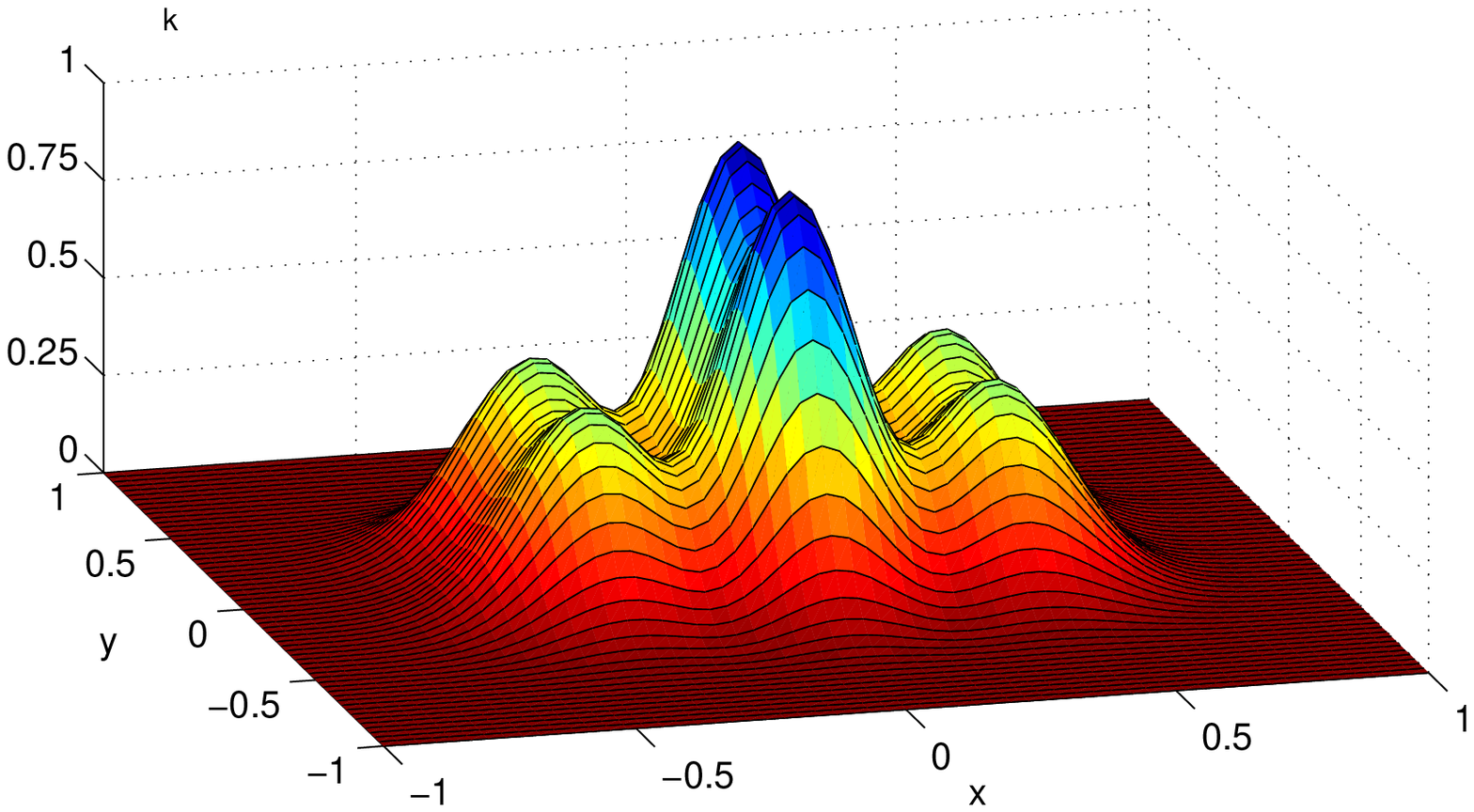}\hspace{0.1\textwidth}
\includegraphics[width=0.4\textwidth]{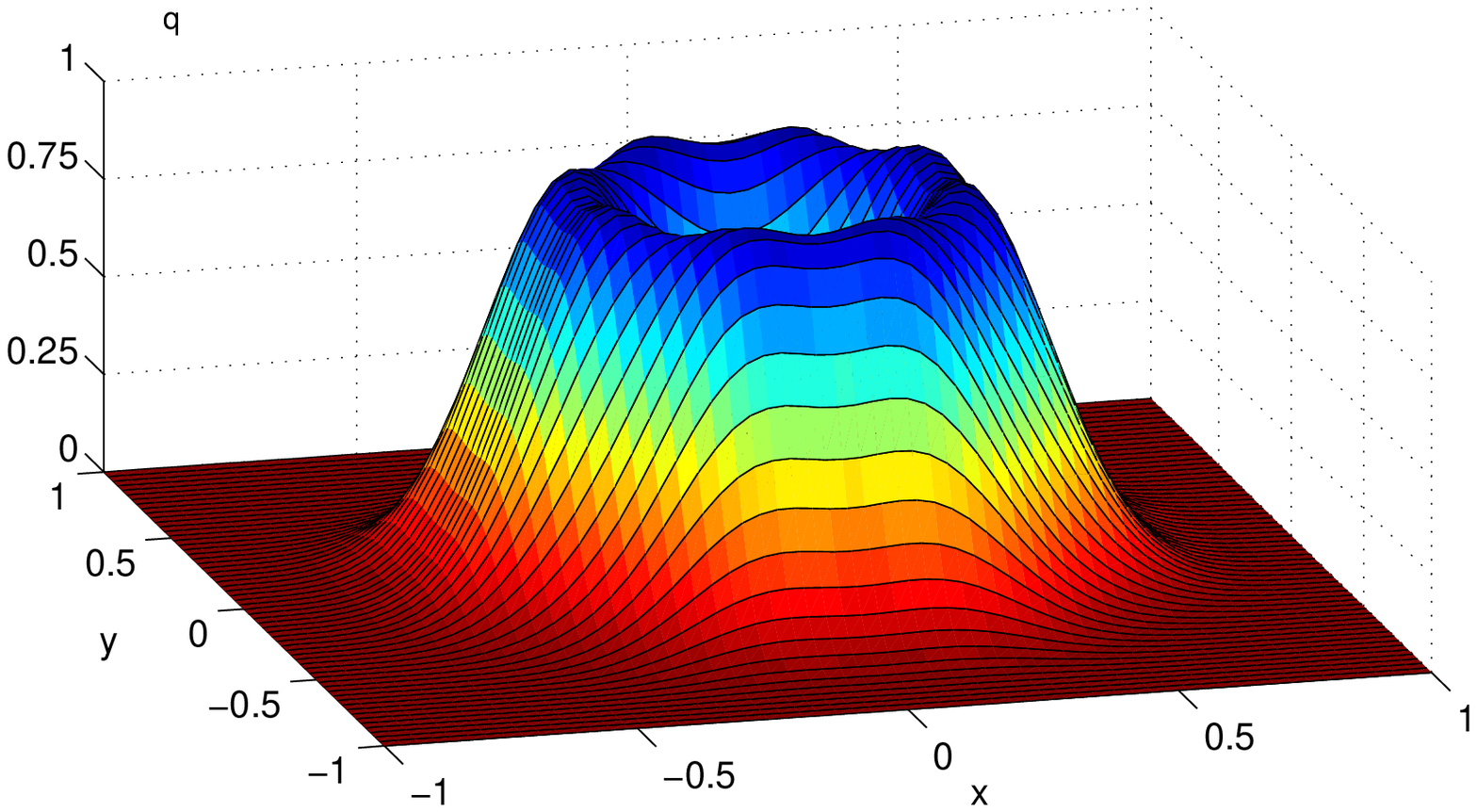}\\

\vspace{2mm}
\includegraphics[width=0.4\textwidth]{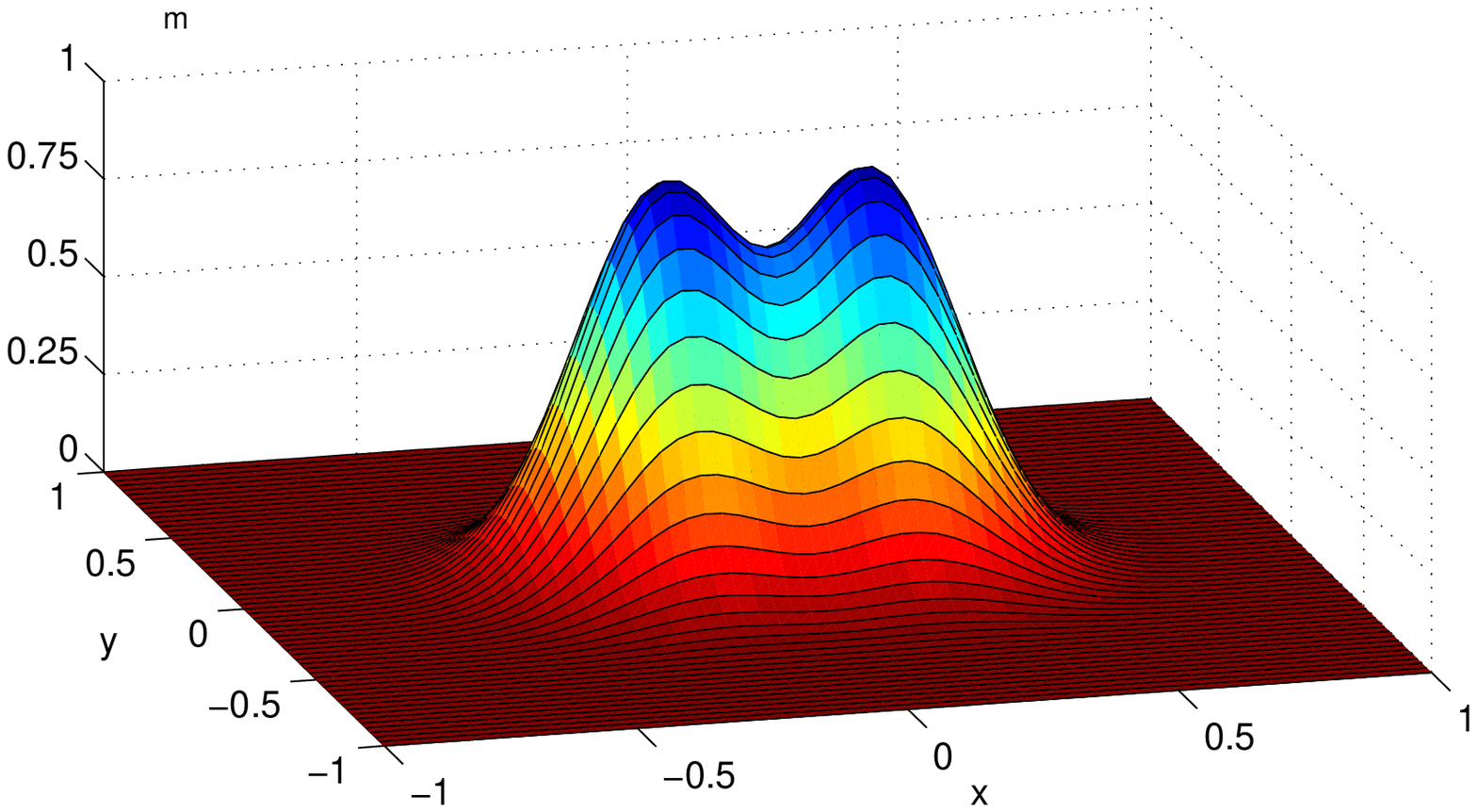}\hspace{0.1\textwidth}
\includegraphics[width=0.4\textwidth]{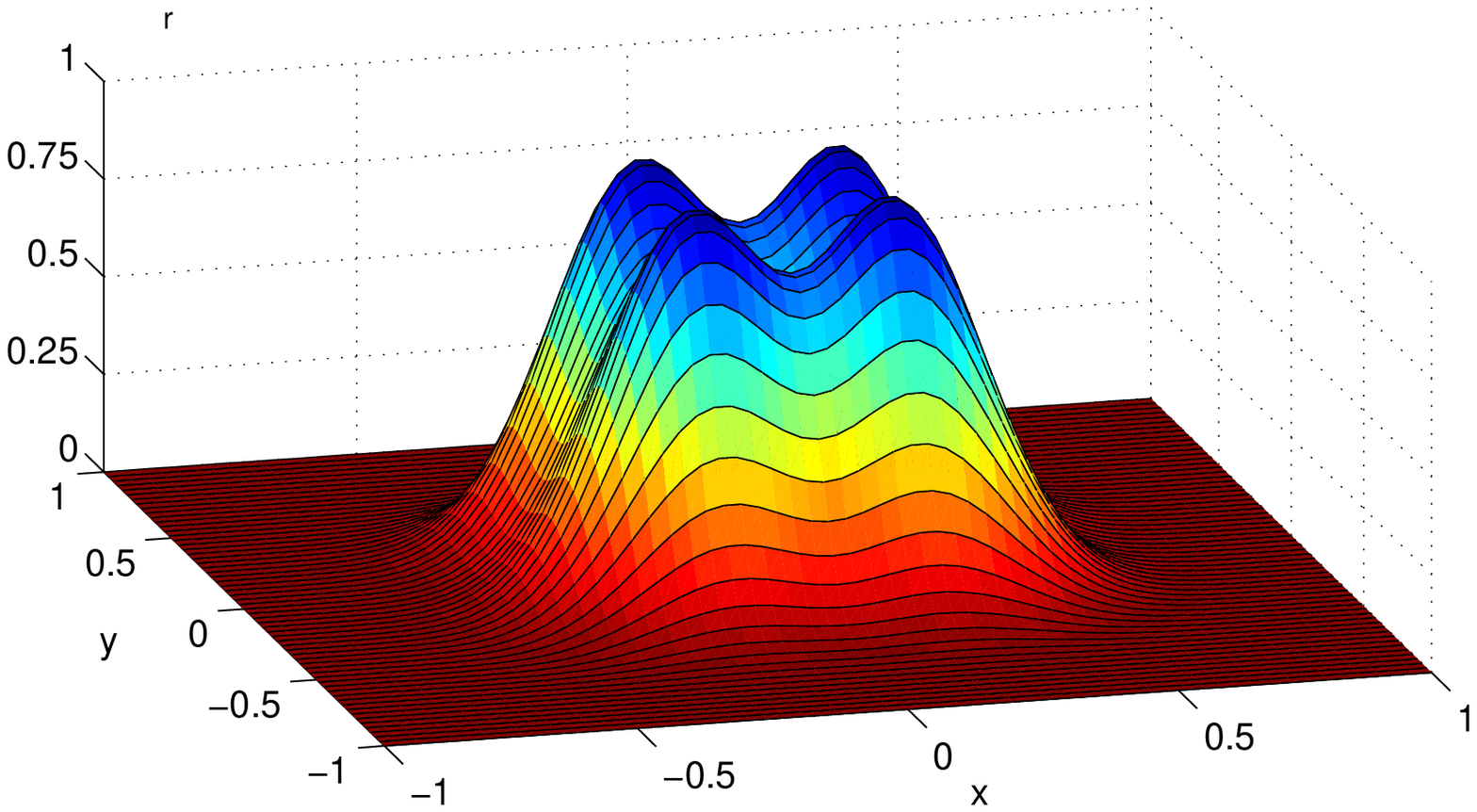}\\

\vspace{2mm}
\hspace{0.5\textwidth}
\includegraphics[width=0.4\textwidth]{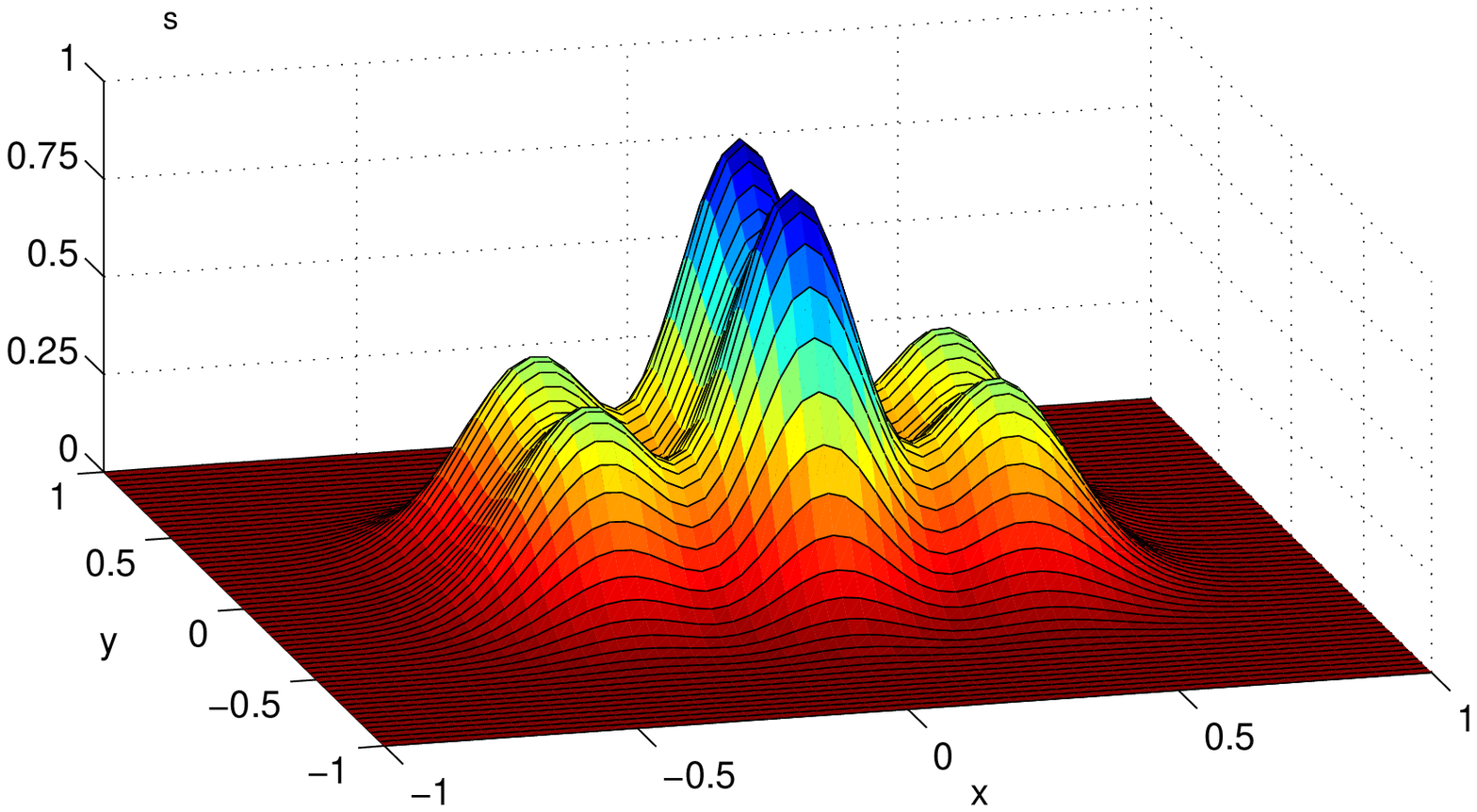}
\end{center}
\caption{\label{fig:csakint}(Color online) Normalized interacting two-particle densities obtained from the lowest energy wave
  functions. Left (right) column shows the antisymmetric (symmetric) densities belonging to ${\bm\Gamma}^{2p}$
  $({\bm\Gamma}^{4p})$. For two-dimensional representations the row index is $j=1$. Associated energies in the left column
  from top to bottom are: 4.078, 4.078, 5.079 and 3.078. In the right column from top to bottom: 6.059, 8.048, 6.044, 4.059 and
  5.059. These are taken from Tables~\ref{tab:noint} and~\ref{tab:int}.}
\end{figure}

Having obtained a particular pair of solutions $(E^{qp}_r,{\bm\psi}_{j,r}^{qp})$ from Eq.~\eqref{schqpj}, the corresponding
four-dimensional wave function $(\psi_{j,r}^{qp})_{pikl}$ is then obtained by rearranging the vector index by means of
Eqs.~\eqref{vectorindex} and~\eqref{indextransform}. From this the two-particle density in the 2D square domain is calculated as
\begin{equation}
n_{pi}=\sum_{k,l=1}^n(\psi_{j,r}^{qp})^2_{pikl}.\label{qpjdens}
\end{equation}
The lowest energy $(r=1)$ noninteracting ``symmetric'' densities belonging to ${\bm\Gamma}^{1p}$ can be seen in the left
column of Fig.~\ref{fig:hasonlit}. For $p=5$ the representation is two-dimensional and the figure shows the density associated
with the second row, $j=2$. For $j=1$ a very same structure is found that is elongated in direction $y$ instead of $x$. For the
other representations we do not show noninteracting densities separately, because they are only slightly different from those with
interaction depicted in Fig.~\ref{fig:csakint}.

\section{Numerical results for interacting particles}\label{sec:intres}
In this section we incorporate the effect of a repulsive Coulomb interaction and calculate some relevant physical
quantities. These are the interacting level structure, one- and two-particle density of states, wave functions and particle
densities categorized by representations and entanglement. The interaction itself reads
\begin{equation}
V(|\mathbf{r}_1-\mathbf{r}_2|)=\frac{c}{\sqrt{(x_1-x_3)^2+(x_2-x_4)^2}},\label{coulomb}
\end{equation}
where $c$ is a positive constant responsible for its strength. Appearance of this term precludes separation of variables in the
Schr\"odinger equation and all dynamical symmetries we encountered during the study of the noninteracting problem will be lost. In
addition to that many geometrical symmetries will be broken too as an arbitrarily chosen $\mathbf{R}\in\text{O}_4$ does not
necessarily commute with $\mathbf{Q}$ of Eq.~\eqref{q}. These considerations have led to the result that the invariance group
$\mathcal{G}$ of the interacting Hamiltonian consists of only geometrical symmetries, and as it turned out in
Subsection~\ref{sub:inv}, it is a subgroup of $\text{O}_4$ of order 32.

\begin{table}
\caption{\label{tab:int}Distribution of interacting two-particle energy eigenvalues $E^{qp}_r/\omega$ across representations of
  $\mathcal{G}$. The first twelve levels are shown for each representation. The strength of interaction is $c=1$ and the other
  parameters are the same as in Table~\ref{tab:noint}.}  \renewcommand*{\arraystretch}{1.1}
\begin{tabular*}{\textwidth}{@{\extracolsep{\fill}}c*{14}{r}}
\hline
&\multicolumn{14}{c}{$E^{qp}_r/\omega$}\\
$r$ & $\Gamma^{11}$ & $\Gamma^{12}$ & $\Gamma^{13}$ & $\Gamma^{14}$ & ${\bm\Gamma}^{15}$ &
${\bm\Gamma}^{21}$ & ${\bm\Gamma}^{22}$ & ${\bm\Gamma}^{23}$ & ${\bm\Gamma}^{24}$ &
$\Gamma^{41}$ & $\Gamma^{42}$ & $\Gamma^{43}$ & $\Gamma^{44}$ & ${\bm\Gamma}^{45}$\\
\hline
1& 2.702 &  4.059 &  6.060 &  4.745 &  3.724 &  4.078 &  4.078 &  5.079 &  3.078 &  6.059 &  8.048 &  6.044 &  4.059 &  5.059\\
2& 4.741 &  4.745 &  6.786 &  6.779 &  5.060 &  6.051 &  6.050 &  7.052 &  5.049 &  8.058 &  8.067 &  6.060 &  6.055 &  7.046\\
3& 4.843 &  6.056 &  8.063 &  6.917 &  5.760 &  6.073 &  6.069 &  7.073 &  5.069 &  8.068 & 10.057 &  8.044 &  6.061 &  7.057\\
4& 6.044 &  6.064 &  8.086 &  8.049 &  5.766 &  6.079 &  6.079 &  7.079 &  5.078 & 10.047 & 10.064 &  8.050 &  8.038 &  7.061\\
5& 6.063 &  6.802 &  8.842 &  8.076 &  5.881 &  6.087 &  6.080 &  7.087 &  5.080 & 10.062 & 10.095 &  8.065 &  8.049 &  7.067\\
6& 6.782 &  6.936 &  9.023 &  8.818 &  7.046 &  8.043 &  8.043 &  9.047 &  7.040 & 10.063 & 10.183 &  8.097 &  8.053 &  9.041\\
7& 6.803 &  8.038 & 10.048 &  8.842 &  7.060 &  8.052 &  8.053 &  9.056 &  7.048 & 10.075 & 12.052 & 10.036 &  8.062 &  9.049\\
8& 6.914 &  8.051 & 10.073 &  8.986 &  7.064 &  8.057 &  8.055 &  9.063 &  7.054 & 10.097 & 12.067 & 10.046 &  8.066 &  9.054\\
9& 6.981 &  8.058 & 10.079 &  9.130 &  7.080 &  8.062 &  8.063 &  9.069 &  7.055 & 10.184 & 12.089 & 10.053 &  8.097 &  9.058\\
10& 8.045 &  8.068 & 10.097 & 10.059 &  7.800 &  8.074 &  8.068 &  9.077 &  7.066 & 12.059 & 12.094 & 10.057 & 10.042 &  9.062\\
11& 8.052 &  8.079 & 10.140 & 10.080 &  7.819 &  8.078 &  8.080 &  9.081 &  7.072 & 12.069 & 12.103 & 10.064 & 10.047 &  9.065\\
12& 8.074 &  8.129 & 10.253 & 10.112 &  7.826 &  8.082 &  8.081 &  9.087 &  7.080 & 12.092 & 12.179 & 10.088 & 10.057 &  9.069\\
\hline
\end{tabular*}
\end{table}

\subsection{Interacting level structure: single- and two-particle density of states}\label{sub:intdos}
\begin{figure}
\psfrag{x}[t][b]{$\epsilon/\omega$}
\psfrag{y}[b][t]{$g_2(\epsilon)$}
\begin{center}
\includegraphics[width=0.45\textwidth]{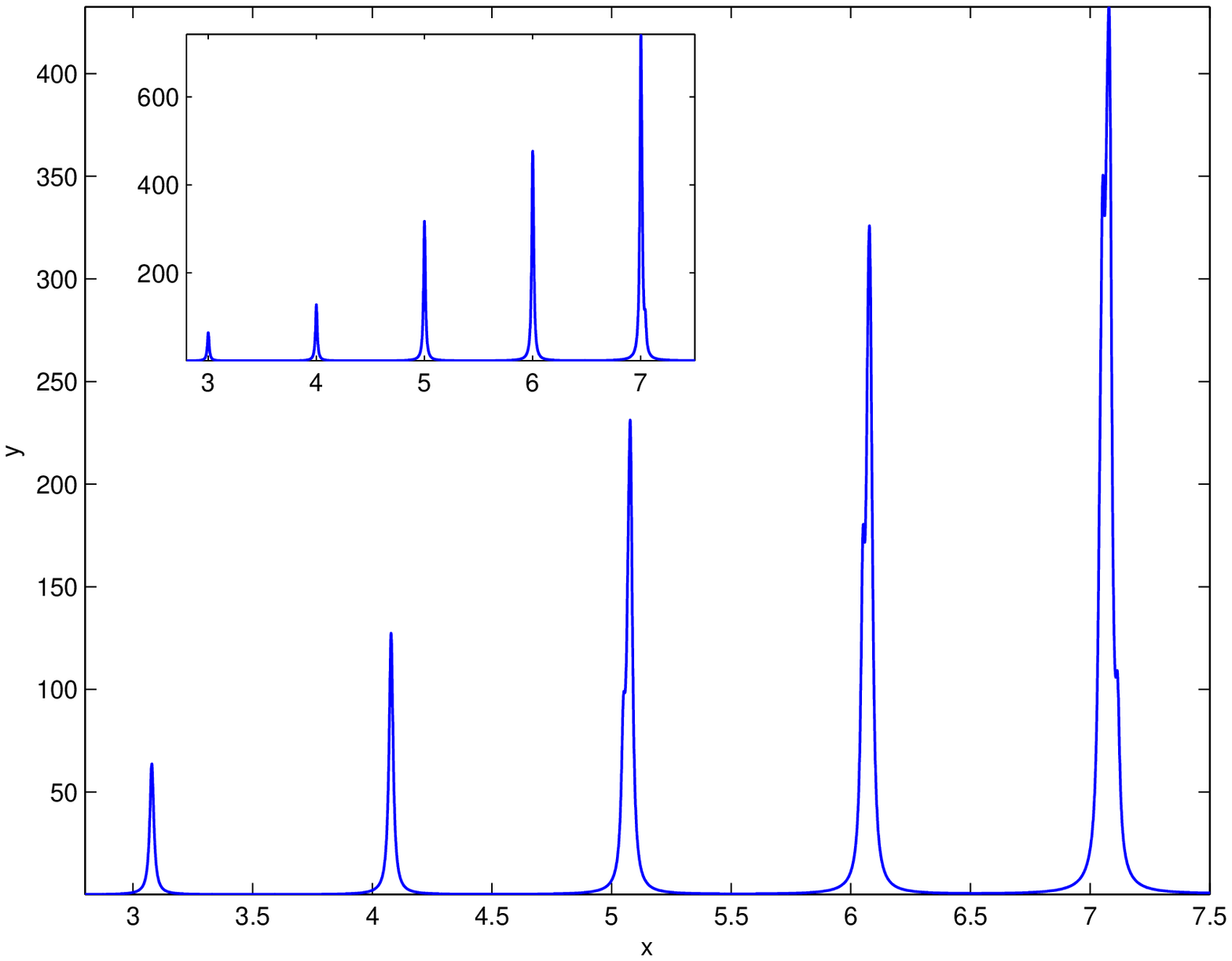}\hspace{0.05\textwidth}
\includegraphics[width=0.45\textwidth]{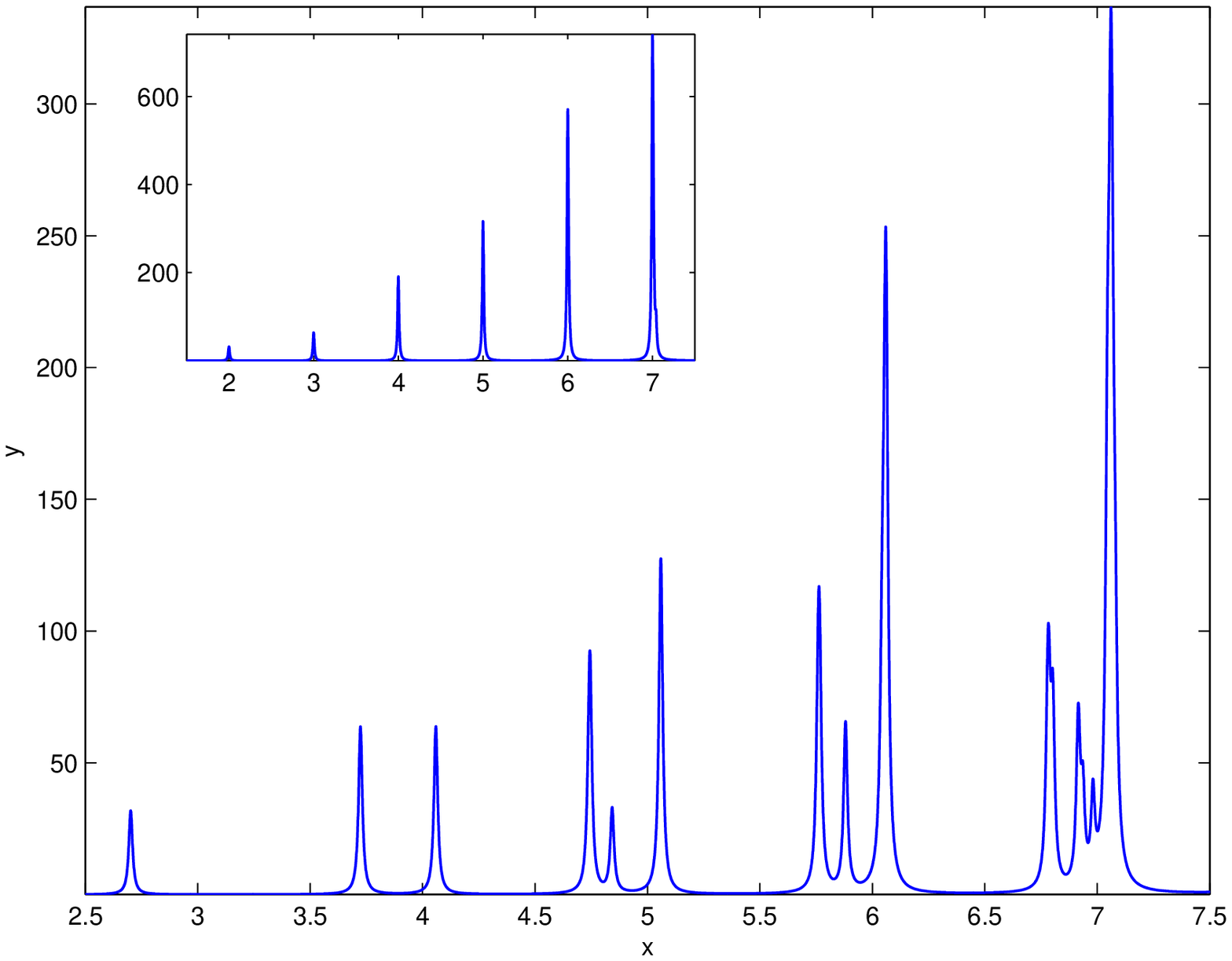}
\end{center}
\caption{\label{fig:dos2}(Color online) Two-particle DOS is shown for the antisymmetric (symmetric) subspace in the left (right)
  panel. Main figures show interacting data taken from Table~\ref{tab:int}. Insets are without interaction. For better visibility
  a small imaginary part $(\eta=0.01)$ was introduced by hand, i.e.\ the delta-function is approximated by
  $\delta(x)\approx(\eta/\pi)/(x^2+\eta^2)$.}
\end{figure}

Numerical solution of Eq.~\eqref{schqpj} for each irreducible representation and with interaction results in the interacting
spectrum. For a coupling strength of $c=1$ the first twelve levels in each symmetry channel are tabulated in
Table~\ref{tab:int}. In order to put it in a more expressive form, we have also plotted the true antisymmetric and symmetric
two-particle DOS in the left and right panels of Fig.~\ref{fig:dos2}, respectively. For evaluation Eq.~\eqref{dos2sym} was used
with eigenvalues taken from Table~\ref{tab:int}. The figures reflect clearly that the effect of Coulomb interaction in the
antisymmetric case is very weak, almost negligible compared to that observed in the symmetric channel. The deviation is easy to
understand: the antisymmetry criterion of a state $\psi_{pikl}$ leads to vanishing of the wave function whenever the two particles
are at the same site, $\psi_{pipi}=0$. This in turn means that the major contribution from the Coulomb potential is suppressed
heavily. In stark contrast to this, the diagonal part of the symmetric wave function can be quite large and the interaction,
either repulsive or attractive, can contribute easily.

Considering the two-particle problem, $g_2$ is actually nothing else than a nice representation of the level structure: positions
of the delta-peaks show available bound state energies if we were to introduce two interacting particles at the same time in an
empty system. The weight of the delta-function on the other hand, which is necessarily an integer, accounts for the total number
of independent (degenerate) states the particles can occupy. Bearing in mind all these, if we go on now with the exploration of
the one-particle DOS $g_1$, we immediately see that its definition requires a more elaborate analysis. It is because due to
interaction, introducing a single particle in an empty system or in a system that already has one particle, is not the
same. Moreover, depending on the symmetry of the two-particle state it can lead to quite different results. In the former problem
of a single particle living in the system, the interaction obviously does not matter as there is no partner to interact with, and
$g_1$ is given by Eq.~\eqref{dos1}. However, if we want to study the spectral properties when a second particle is added, we have
to apply techniques used in many-body physics. For this purpose we introduce the time-ordered zero temperature Green's
function \cite{fetter-book}
\begin{equation}
G(\mathbf{r},\mathbf{r'},t)=-i\langle\chi_m|T\Psi(\mathbf{r},t)\Psi^+(\mathbf{r'},0)|\chi_m\rangle,\label{green}
\end{equation}
where $|\chi_m\rangle$ is an occupied one-particle state. Its wave function in coordinate representation and energy are,
respectively, $\chi_m(\mathbf{r})=\phi_{m_1}(x)\phi_{m_2}(y)$ and $\epsilon_m=\omega(\nu_{m_1}+\nu_{m_2}+1)$. Indices $m_i$ can
take any nonnegative integer values, see Appendix~\ref{appsec:osc}. Also, the field operators are most conveniently expressed in
this basis: $\Psi(\mathbf{r})=\sum_m\chi_m(\mathbf{r})a_m$, where $a_m$ is the usual destruction operator. Now this is the point
where the discussion has to be split into two. If the two-particle state possesses an antisymmetric wave function, like a spin
triplet state, then $a_m$ is fermionic obeying the canonical anticommutation relations. On the other hand if the spatial state is
symmetric, like in the spin singlet state, then $a_m$ has bosonic nature satisfying the usual commutation relations. The
Lehmann-representation in each case reads
\begin{equation}
G(\mathbf{r},\mathbf{r'},\epsilon)=\sideset{}{'}\sum_{qp}\sum_{j=1}^{d_{qp}}\sum_{r=1}^{r^{qp}}
\frac{\langle\chi_m|\Psi(\mathbf{r})|\psi^{qp}_{j,r}\rangle\langle\psi^{qp}_{j,r}|\Psi^+(\mathbf{r'})|\chi_m\rangle}
{\epsilon-E^{qp}_r+\epsilon_m+i0}+\frac{\chi_m(\mathbf{r})\chi_m(\mathbf{r'})}{\epsilon-\epsilon_m-i0},
\end{equation}
where we have introduced a complete set of orthonormal two-particle states in the first term describing particle propagation
forward in time. For completeness we note that the second term describes the situation when instead of adding an extra particle we
remove the particle from its stationary state $\chi_m$. Since this is not what we want, from now on we will be interested in the
first (retarded) term only. Knowing the symmetries of the system it is not surprising that we chose the two-particle intermediate
states to be stationary states of the Schr\"odinger equation, so they are most conveniently labeled by representations to which
they belong. Here we would like to call the attention to an interesting property of $G$, which it does not exhibit in general. In
true many-body problems the Green's function is usually a very complicated object because neither the interacting (ground) state
of the $N$-particle system, nor the eigenstates of the $N\pm1$-particle subspaces are known. This is why one needs perturbative
treatments conventionally. The $N=1$ case is, however, exceptional. Exceptional because the interacting states are exactly the
same as the noninteracting ones. Furthermore, the 0-particle state is the trivial vacuum state and the 2-particle states are
obtained ``exactly'' from numerical computations. Thus everything is given for the construction of $G$.

\begin{figure}
\psfrag{x}[t][b]{$\epsilon/\omega$}
\psfrag{y}[b][t]{$g_1(\epsilon)$}
\begin{center}
\includegraphics[width=0.45\textwidth]{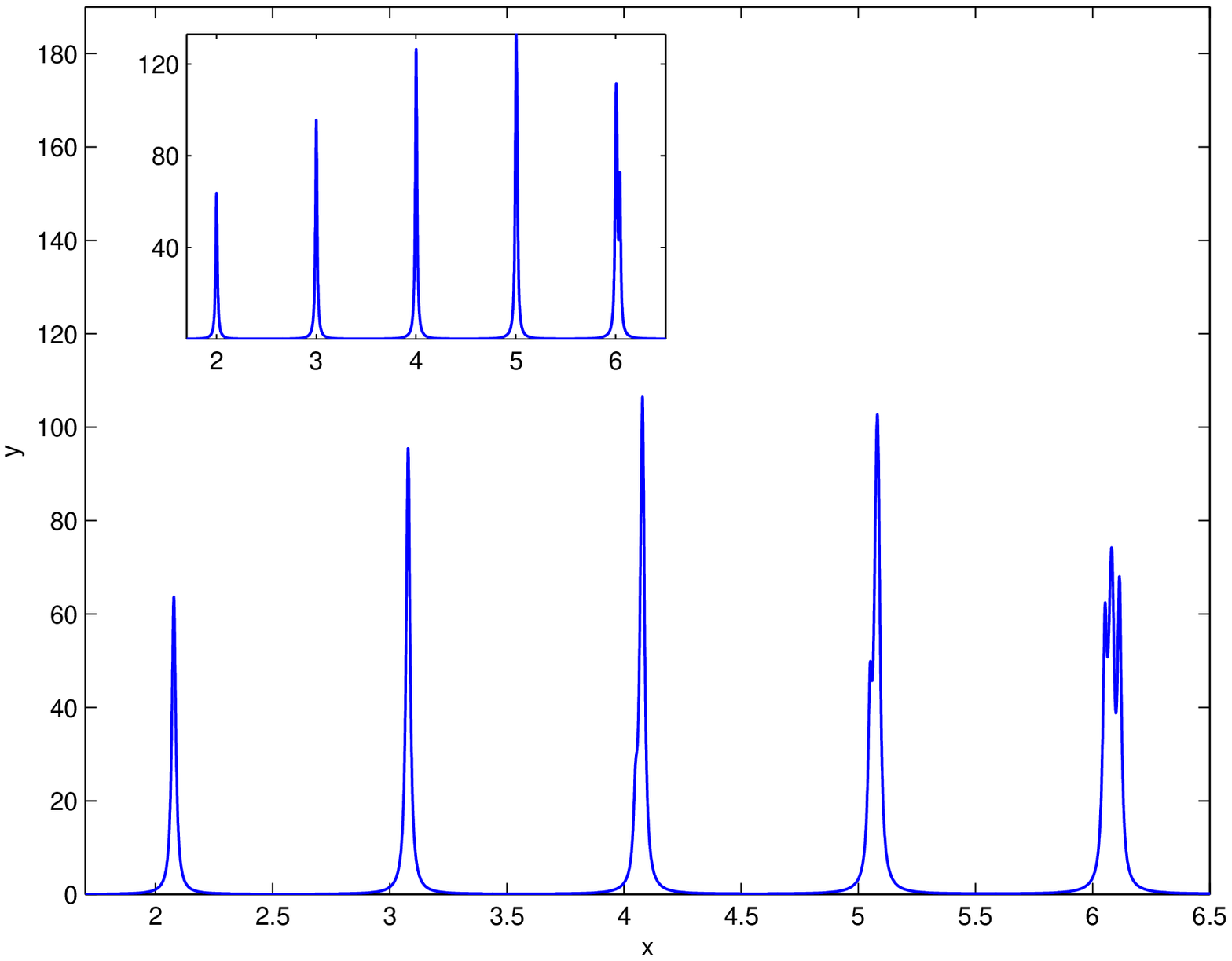}\hspace{0.05\textwidth}
\includegraphics[width=0.45\textwidth]{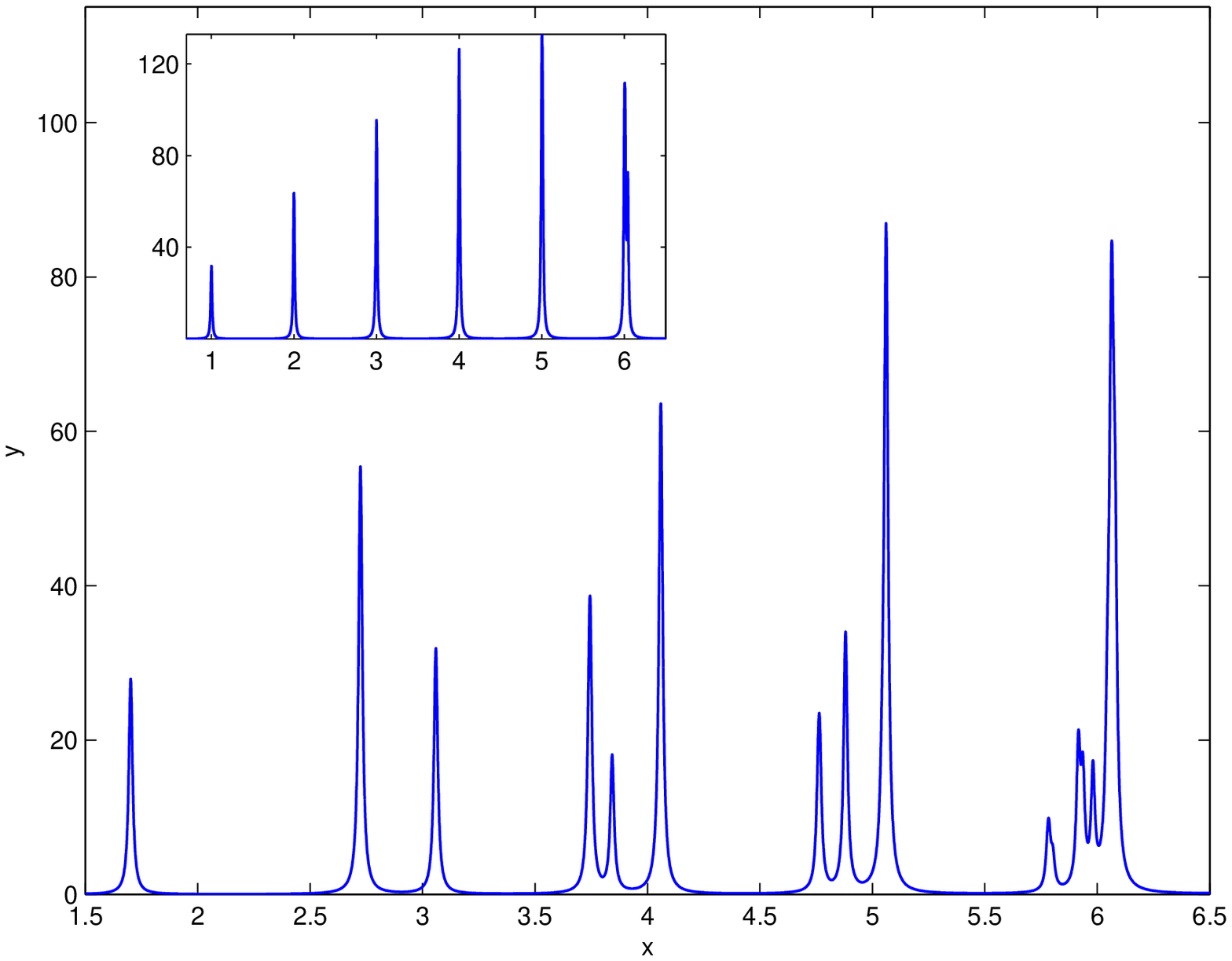}
\end{center}
\caption{\label{fig:dos1}(Color online) One-particle DOS is shown for the antisymmetric (symmetric) subspace in the left (right)
  panel.  Quantum numbers of the initially placed particle are $m_1=m_2=0$. Insets are without interaction. In order to better
  visualize the delta-functions, as in Fig.~\ref{fig:dos2}, a small imaginary part $(\eta=0.01)$ was included by hand.}
\end{figure}

From the knowledge of Green's function $g_1$ is expressed in the usual way \cite{fetter-book}. In the finite difference method,
where wave functions are computed only at discrete lattice points, it reads
\begin{equation}
g_1(\epsilon)=\sideset{}{'}\sum_{qp}\sum_{r=1}^{r^{qp}} w^{qp}_r\delta(\epsilon-E^{qp}_r+\epsilon_m),\label{dos1int}
\end{equation}
where the weight of the delta-peak for antisymmetric intermediate states is
\begin{equation}
w^{qp}_r=\sum_{j=1}^{d_{qp}}\sum_{p,i=1}^n
\left(\sum_{n_1,n_2=0}^{n-1}\sqrt{2}\phi_{n_1}(x_p)\phi_{n_2}(x_i)\left(S^{qp}_{j,r}\right)^{n_1n_2}_{m_1m_2}
-\sqrt{2}\phi_{m_1}(x_p)\phi_{m_2}(x_i)\left(S^{qp}_{j,r}\right)^{m_1m_2}_{m_1m_2}\right)^2.\label{fermiweight}
\end{equation}
For symmetric intermediate states it is slightly different
\begin{equation}
w^{qp}_r=\sum_{j=1}^{d_{qp}}\sum_{p,i=1}^n
\left(\sum_{n_1,n_2=0}^{n-1}\sqrt{2}\phi_{n_1}(x_p)\phi_{n_2}(x_i)\left(S^{qp}_{j,r}\right)^{n_1n_2}_{m_1m_2}
+\phi_{m_1}(x_p)\phi_{m_2}(x_i)\left(S^{qp}_{j,r}\right)^{m_1m_2}_{m_1m_2}\right)^2,\label{boseweight}
\end{equation}
and the matrix element is
\begin{equation}
\left(S^{qp}_{j,r}\right)^{n_1n_2}_{m_1m_2}=
\sum_{p,i,k,l=1}^n\phi_{n_1}(x_p)\phi_{n_2}(x_i)\phi_{m_1}(x_k)\phi_{m_2}(x_l)(\psi^{qp}_{j,r})_{pikl}.
\label{smatrix}
\end{equation}
The single-particle DOS associated with the antisymmetric and symmetric subspaces are illustrated in the left and right panels of
Fig.~\ref{fig:dos1}, respectively. The curves belong to that particular case when the first particle occupies the ground state
with quantum numbers $m_1=m_2=0$ and energy $\epsilon_0/\omega=2\nu_0+1\approx1$. Insets show the respective results without
interaction. From these figures the very same conclusion can be drawn as from Fig.~\ref{fig:dos2}. Namely, the Coulomb interaction
renormalizes the single particle energies in the symmetric subspace more markedly than in the antisymmetric.

\subsection{Interacting two-particle densities}\label{sub:densint}
The interacting four-dimensional wave functions $(\psi_{j,r}^{qp})_{pikl}$ can be used to compute two-particle densities. By means
of Eq.~\eqref{qpjdens} we determined these from the states with lowest energies $(r=1)$ and plotted them in the right columns of
Figs.~\ref{fig:hasonlit} and~\ref{fig:csakint}. From Fig.~\ref{fig:hasonlit}, showing densities calculated from symmetric wave
functions, we can observe again the fact that the Coulomb interaction not only affects the bound state energies but the wave
functions as well. This in turn leads to strongly modified densities. On the other hand, the interacting densities in the left
column of Fig.~\ref{fig:csakint}, calculated from antisymmetric states, look almost exactly the same as without interaction (these
are not shown), and this is in accord with our previous finding that the eigenvalues were not affected considerably either.

\subsection{Entanglement}\label{sub:ent}

In this subsection we calculate the effect of interaction on the entanglement of the two-particle state. From quantum mechanics it
is well known that whenever the total wave function $\psi_{AB}$ of a bipartite system is separable, that is it can be written as a
product $\psi_{AB}=\psi_A\psi_B$, then the total wave function not only provides complete description of the composite system but
does it also for both subsystems $A$ and $B$ separately. In this case $A$ and $B$ are independent and the terms $\psi_A$ and
$\psi_B$ are called pure states. In most cases, however, the knowledge of the total wave function does not necessarily involve
complete characterization of the subsystems as a factorization in general does not exist. If this happens, the subsystems are then
said to be superposed with one another, or in other words are entangled, and each of them is in a mixed state associated with
reduced density matrices $\rho_A$ and $\rho_B$, respectively \cite{landau-book}.

In our case the bipartite system is nothing else than a composite of two identical particles. Each of the two is confined in the
same 2D square domain so the one-particle Hilbert spaces are the same. Moreover, identity of particles imply that though their
reduced density matrices themselves may be different, their spectra and all other measures of entanglement are the same. Having
obtained a particular solution of Eq.~\eqref{schqpj} we can construct the density matrix for the ``first'' particle as
\begin{equation}
\rho_{pi,p'i'}=\sum_{k,l=1}^n(\psi^{qp}_{j,r})_{pikl}(\psi^{qp}_{j,r})_{p'i'kl},\label{rhodef}
\end{equation}
where the rows and columns are now labeled by a composite index. In a similar manner, for the ``second'' particle the square of
$\psi$ is integrated with respect to the coordinates of the first particle. It is well known, and can be seen from this expression
as well, that $\rho$ is a nonnegative Hermitian matrix with unit trace \cite{landau-book}. Also, its diagonal elements constitute
the usual static particle density we have already encountered in Eq.~\eqref{qpjdens} and in Figs.~\ref{fig:hasonlit}
and~\ref{fig:csakint}. One measure of entanglement is the quantity $\text{Tr}\rho^2$. This is unity for pure states but smaller
than one for mixed states, indicating entanglement of the two-particle state. Another characteristic measure of superposition
comes from the Schmidt decomposition (singular value decomposition) \cite{klyachko}
\begin{equation}
\psi_{pikl}=\sum_{j=1}^N\sqrt{\lambda_j}u^{(j)}_{pi}v^{(j)}_{kl}.\label{schmidt}
\end{equation}
Here $u^{(j)}$ and $v^{(j)}$ are, respectively, normalized one-particle eigenstates of the density matrices of the first and
second particles, each belonging to the same eigenvalue $\lambda_j$, the so-called Schmidt coefficient. Now, from this the von
Neumann entropy is found by
\begin{equation}
S=-\text{Tr}\rho\ln\rho=-\sum_j\lambda_j\ln\lambda_j.\label{neumann}
\end{equation}
Both $S$ and the number of nonzero Schmidt coefficients $N$ are widely used measures of entanglement.

In Table~\ref{tab:ent} we have tabulated some representative data reflecting the effect of interaction on entanglement. In the
third row we gave the shift in energy due to interaction for the lowest energy states of five particular representations. These
data are taken from Tables~\ref{tab:noint} and~\ref{tab:int}. Further down the table we show the different measures of
entanglement for noninteracting states as well as for their interacting counterparts. The first three and the last representation
describe systems with symmetric wave function under permutation of particles. The fourth describes an antisymmetric state. As the
data verify, the contribution of the repulsive Coulomb interaction in the antisymmetric subspace of the two-particle Hilbert space
is almost negligible, at least for a coupling strength of $c=1$. This is in accordance with our previous findings.

\begin{table}
\caption{\label{tab:ent}Measures of entanglement calculated from the lowest energy $(r=1)$ noninteracting and interacting
  eigenstates belonging to five representations of $\mathcal{G}$. Energy levels in the third row are taken from
  Tables~\ref{tab:noint} and~\ref{tab:int}. $N$ is the number of Schmidt coefficients $\lambda_j$ that are larger than the
  threshold $10^{-6}$. $S$ is the von Neumann entropy and $\Delta S$ is the difference between values with and without
  interaction. Representations $\Gamma^{11}$, $\Gamma^{13}$, ${\bm\Gamma}^{15}$ and $\Gamma^{42}$ characterize symmetric
  states, while ${\bm\Gamma}^{23}$ an antisymmetric state.} \renewcommand*{\arraystretch}{1.1}
\begin{tabular*}{\textwidth}{@{\extracolsep{\fill}}c*{10}{l}}
\hline
& \multicolumn{2}{c}{$\Gamma^{11}$} & \multicolumn{2}{c}{$\Gamma^{13}$} & \multicolumn{2}{c}{${\bm\Gamma}^{15}$}
& \multicolumn{2}{c}{${\bm\Gamma}^{23}$} & \multicolumn{2}{c}{$\Gamma^{42}$}\\
& no int.\ & int.\ & no int.\ & int.\ & no int.\ & int.\ & no int.\ & int.\ & no int.\ & int.\ \\
\hline
$E^{qp}_1/\omega$ &  2.000000 &  2.702 &  6.000214 &  6.060 &  3.000011 &  3.724 &  5.000220 &  5.079 &  8.002059 &  8.048\\
$\text{Tr}\rho^2$ &  1.000 &  0.756 &  0.187 &  0.140 &  0.500 &  0.417 &  0.250 &  0.250 &  0.125 &  0.082\\
$N$ &   1 &  30 &   6 &   8 &   2 &  34 &   4 &   8 &   8 &  14\\
$S$ &  0.000 &  0.587 &  1.733 &  2.015 &  0.693 &  1.079 &  1.386 &  1.387 &  2.079 &  2.557\\
$\Delta S$ & \multicolumn{2}{c}{0.587} & \multicolumn{2}{c}{0.282} & \multicolumn{2}{c}{0.386} & \multicolumn{2}{c}{0.001} &
\multicolumn{2}{c}{0.478}\\
\hline
\end{tabular*}
\end{table}

\section{Conclusions}\label{sec:conclusions}
We have studied numerically a Coulomb interacting two-particle system in two-dimensions. This quantum mechanical problem called
for the solution of the non-relativistic stationary Schr\"odinger equation, which is in fact an eigenvalue equation in a
four-dimensional configuration space. As to the domain of motion we have specified a square shaped potential well, a computational
box with infinitely high walls imposing closed boundary condition for the wave function. In addition to that, within the box we
have also specified a steeply increasing isotropic harmonic potential resembling that in the nucleus. The particles in question
are identical and can obey either Bose or Fermi statistics. Spin related effects were not considered in this work, we concentrated
only on the spatial part of the total two-particle state.

In Section~\ref{sec:sch} we have developed a fully discretized 89-point finite difference method of the Numerov-type that
approximates the four-dimensional Laplace operator, and thus the whole Schr\"odinger equation, with a local truncation error of at
most $\mathcal{O}(h^6)$. The errors of the energy eigenvalues were found to be the same order of magnitude. This result, which we
have proved analytically as well, can be considered as the cornerstone of the present work. It assures that the numerical
computations will converge very fast, or, in other words, calculations with rather low resolution might give as well reasonable
results. We found that the finite difference scheme can be put in a remarkably compact and concise form with the usage of matrices
with direct polynomial structure. Having obtained the correct matrix representation of the Laplacian we have constructed an
equivalent generalized matrix Schr\"odinger equation for the two-particle states and energies. In the course of its detailed
derivation we came to the conclusion that the required precision, with some effort, can be refined further along the same
lines. Also, the algorithm can be ported to other dimensions too.

In Section~\ref{sec:group} we proceeded with the analysis of internal symmetries of the problem. This we needed because the
eigenvalue equation, though involves only sparse matrices, is of dimension $n^4$ and therefore very memory consuming. More
importantly, we were interested not just in the ground state but also in many excited states, and a great deal of unnecessary
effort can be eliminated with the use of group representation theory and an appropriate similarity transformation that turns the
eigenvalue equation into block diagonal form. Hence, in this section we determined the invariance group of the interacting
Hamiltonian and all of its unitary irreducible representations. With the help of these and Wigner-Eckart theorem we performed the
transformation and obtained a block diagonal eigenvalue equation. Also, irreducible representations allowed for a convenient
distinction between the symmetric and antisymmetric solutions.

In Sections~\ref{sec:nointres} and~\ref{sec:intres} we presented the results of our numerical computations. At a resolution of
$n=30$, i.e.\ with a step size of $h\approx1/15$, the full two-particle Hilbert space is of dimension $n^4=810000$. At this point
we restricted our studies only to the low energy subspace. This we have chosen to be of dimension $8^4=4096$. In this space we
have computed the ground state and roughly the first 200 excited states together with their energies. It turned out that in case
of the noninteracting results, comparison with exact analytical formulas revealed that the numerical data are indeed very
accurate, to three or more digits of precision. We classified the level structure and the wave functions of the noninteracting as
well as the interacting system according to the irreducible representations to which they belong. We have computed the static
particle densities, the one- and two-particle density of states in both the symmetric and antisymmetric subspaces of the full
Hilbert space. Even investigated briefly the effect of Coulomb interaction on entanglement. To this end we have calculated the
reduced density matrix, its Schmidt decomposition and the von Neumann entropy. All these quantities show consistently that the
repulsive interaction affects the antisymmetric states and their energies only very mildly, because the vanishing of the wave
function when the particles are close to each other suppresses the contribution of the Coulomb potential. On the other hand, the
symmetric states were modified dramatically and the single-particle energies are renormalized considerably as well.

\section*{Acknowledgement}
We acknowledge stimulating discussions and suggestions to B.\ D\'ora, P.\ L\'evay and I.\ Nagy. This work was supported by the
Hungarian National Research Fund under Grants No.\ OTKA NI70594, T046269, K 72613 and T/F 038158.

\appendix
\section{Analysis of the matrices $\mathbf{M}_i$, $\mathbf{M}$ and $\mathbf{N}$}\label{appsec:matrix}
In this appendix we summarize some important linear algebraic results on the matrices $\mathbf{M}_i$ $(i=1,\dots,4)$, $\mathbf{M}$
and $\mathbf{N}$, that are encountered in Subsection~\ref{sub:free}. There, these properties were used repeatedly during the
discussion of the discrete Schr\"odinger equation, Eq.~\eqref{vegsosch}.

\subsection{Commutativity}\label{sub:comm}
Taking into account the defining equations~\eqref{m1}--\eqref{m4}, \eqref{avgood}, \eqref{m} and~\eqref{n} and the properties
of Kronecker-product, one finds that all these matrices can be put in the common form
\begin{equation}
\sum_{a,b,c,d}z_{abcd}\,\mathbf{A}^a\otimes\mathbf{A}^b\otimes\mathbf{A}^c\otimes\mathbf{A}^d,\label{kozos}
\end{equation}
where the indices can take nonnegative integer values. Clearly, in this decomposition it is the coefficients $z_{abcd}$ that are
characteristic to the matrices themselves. In case of $\mathbf{N}$ they even depend on $\gamma'$. The expression in
Eq.~\eqref{kozos} (written with four different matrices in general) is sometimes referred to as a direct polynomial, because it
has a multivariable polynomial structure where the variables are matrices and the product is the direct-product \cite{rozsa-book}.
Now, knowing the multiplication rule of matrices with direct-product structure, remembering that the commutator is a real bilinear
function and that every matrix commutes with itself, it follows naturally that every two matrices of the set mentioned above
commute. In particular
\begin{equation}
[\mathbf{M},\mathbf{N}]=\mathbf{0},\label{mn}
\end{equation}
and this is independent of $\gamma'$ and $n$. A similar result has been obtained in the usual one-dimensional fourth-order Numerov
method \cite{chawla}.

\subsection{Spectral representation of direct polynomials: the properties of $\mathbf{M}_i$}\label{sub:mi}
The direct polynomial structure makes it also rather easy to determine the spectrum and the spectral representation of such
matrices \cite{rozsa-book}. With this recognition, next we show that the matrices $\mathbf{M}_i$ are all negative definite. In our
case the basic ingredient is the symmetric $n$-by-$n$ matrix $\mathbf{A}$, given by Eq.~\eqref{a}, whose eigenvalues are known
analytically
\begin{equation}
\omega_{k}=2\cos\frac{k\pi}{n+1},\qquad k=1,\dots,n.\label{asajat}
\end{equation}
Since these are all distinct, with algebraic multiplicity of one, $\mathbf{A}$ is diagonalizable. Let $\mathbf{u}_k$ be a
normalized eigenvector belonging to $\omega_k$, that is $\mathbf{Au}_k=\omega_k\mathbf{u}_k$. Then
\begin{equation}
\mathbf{u}_k^\text{T}=\sqrt{\frac{2}{n+1}}\left(\sin\frac{k\pi}{n+1},\sin\frac{2k\pi}{n+1},\dots,\sin\frac{nk\pi}{n+1}\right),\label{u}
\end{equation}
and as is known, the set of these vectors form an orthonormal basis of $\mathbb{R}^n$. We note here that a vector $\mathbf{u}$ in
general will always be thought of as a $n$-by-$1$ column vector, the respective $1$-by-$n$ row form, whenever needed, will be
explicitly denoted by matrix transpose $\mathbf{u}^\text{T}$. With these we can write the spectral representation of $\mathbf{A}$
as
\begin{equation}
\mathbf{A}=\sum_{k=1}^n\omega_k\mathbf{u}_k\circ\mathbf{u}_k^\text{T},\label{spectrala}
\end{equation}
and the corresponding spectral representation of the direct polynomial in Eq.~\eqref{kozos} reads
\begin{equation}
\sum_\mathbf{k}\left(\sum_{a,b,c,d}z_{abcd}\,\omega_{k_1}^a\omega_{k_2}^b\omega_{k_3}^c\omega_{k_4}^d\right)
\left(\mathbf{u}_{k_1}\otimes\mathbf{u}_{k_2}\otimes\mathbf{u}_{k_3}\otimes\mathbf{u}_{k_4}\right)\circ
\left(\mathbf{u}_{k_1}^\text{T}\otimes\mathbf{u}_{k_2}^\text{T}\otimes\mathbf{u}_{k_3}^\text{T}
\otimes\mathbf{u}_{k_4}^\text{T}\right).\label{spektral}
\end{equation}
Here, $\mathbf{k}=(k_1,k_2,k_3,k_4)$ is a convenient composite index and in the sum each component runs in the range
$1,\dots,n$. This is a very useful result for it shows, that though our matrices are rather complex due to their multiple
direct-product structure, all eigenvalues can be obtained with practically no effort. For example the $n^4$ eigenvalues of
$\mathbf{M}_1$ are
\begin{equation}
m_1(\mathbf{k})=2\cos\frac{k_1\pi}{n+1}+2\cos\frac{k_2\pi}{n+1}+2\cos\frac{k_3\pi}{n+1}+2\cos\frac{k_4\pi}{n+1}-8,
\qquad k_i=1,\dots,n.\label{m1s}
\end{equation}
With a very same analysis the eigenvalues of $\mathbf{M}_2$, $\mathbf{M}_3$ and $\mathbf{M}_4$ can be given too. They are
\begin{align}
m_2(\mathbf{k})&=4\cos\frac{k_1\pi}{n+1}\cos\frac{k_2\pi}{n+1}+4\cos\frac{k_1\pi}{n+1}\cos\frac{k_3\pi}{n+1}
+4\cos\frac{k_1\pi}{n+1}\cos\frac{k_4\pi}{n+1}\notag\\
&\phantom{=}+4\cos\frac{k_2\pi}{n+1}\cos\frac{k_3\pi}{n+1}
+4\cos\frac{k_2\pi}{n+1}\cos\frac{k_4\pi}{n+1}+4\cos\frac{k_3\pi}{n+1}\cos\frac{k_4\pi}{n+1}-24,\label{m2s}
\end{align}
and
\begin{align}
m_3(\mathbf{k})&=8\cos\frac{k_1\pi}{n+1}\cos\frac{k_2\pi}{n+1}\cos\frac{k_3\pi}{n+1}
+8\cos\frac{k_1\pi}{n+1}\cos\frac{k_2\pi}{n+1}\cos\frac{k_4\pi}{n+1}\notag\\
&\phantom{=}+8\cos\frac{k_1\pi}{n+1}\cos\frac{k_3\pi}{n+1}\cos\frac{k_4\pi}{n+1}
+8\cos\frac{k_2\pi}{n+1}\cos\frac{k_3\pi}{n+1}\cos\frac{k_4\pi}{n+1}-32,\label{m3s}
\end{align}
and finally
\begin{align}
m_4(\mathbf{k})&=16\cos\frac{k_1\pi}{n+1}\cos\frac{k_2\pi}{n+1}\cos\frac{k_3\pi}{n+1}\cos\frac{k_4\pi}{n+1}-16\notag\\
&\phantom{=}+4\cos^2\frac{k_1\pi}{n+1}+4\cos^2\frac{k_2\pi}{n+1}+4\cos^2\frac{k_3\pi}{n+1}+4\cos^2\frac{k_4\pi}{n+1}-16.\label{m4s}
\end{align}
With all these, for any finite $n$ one readily finds
\begin{equation}
\max_\mathbf{k}m_i(\mathbf{k})<0,\qquad i=1,\dots,4.\label{neg}
\end{equation}
It shows that the matrices $\mathbf{M}_i$ are indeed negative definite. In addition to that
\begin{equation}
\lim_{n\to\infty}\max_\mathbf{k}m_i(\mathbf{k})=0,\label{limitsing}
\end{equation}
that is $\mathbf{M}_i$ become singular asymptotically.

\subsection{The properties of $\mathbf{M}$}\label{sub:m}
The positivity\footnote{A matrix, or more generally a linear operator is called, in brief, positive (negative) if it is positive
  (negative) definite.} of $\mathbf{M}$, that we have already adverted to in Subsection~\ref{sub:free}, follows immediately from
the negativity of $\mathbf{M}_i$ and Eq.~\eqref{m}, since all coefficients are negative. Let $\theta(\mathbf{k})$ be the
eigenvalue of $\mathbf{M}$, then
\begin{equation}
\theta(\mathbf{k})=-\frac{1}{30}(12m_1(\mathbf{k})+m_2(\mathbf{k})+m_3(\mathbf{k})),\label{ms}
\end{equation}
and its minimum and maximum are
\begin{equation}
\begin{pmatrix}
\theta_\text{min}\\
\theta_\text{max}
\end{pmatrix}
=
\begin{pmatrix}
\min_\mathbf{k}\theta(\mathbf{k})\\
\max_\mathbf{k}\theta(\mathbf{k})
\end{pmatrix}
=\mp\frac{16}{15}\cos^3\frac{\pi}{n+1}-\frac{4}{5}\cos^2\frac{\pi}{n+1}
\mp\frac{16}{5}\cos\frac{\pi}{n+1}+\frac{76}{15}.\label{minmax}
\end{equation}
Equation~\eqref{ms} comes from the general result of Eq.~\eqref{spektral} along the same line as $m_i(\mathbf{k})$ was obtained in
Subsection~\ref{sub:mi}. The latter result regarding extrema, however, needs some explanation. The allowed $\mathbf{k}$ vectors
are restricted to a cubic domain of $\mathbb{R}^4$, where their coordinates can only take integer values in the range
$1,\dots,n$. In order to find the global minimum and maximum we have to first look for possible local minima and maxima inside the
domain. The stationary points $\mathbf{k}^\star$ are defined by the solution of
$\nabla_\mathbf{k}\theta(\mathbf{k}^\star)=\mathbf{0}$. Now it is easy to show that at these points, wherever they are
\begin{equation}
\frac{\partial^2\theta(\mathbf{k}^\star)}{\partial k_i^2}=0,\qquad i=1,\dots,4,\label{masodikder}
\end{equation}
thus the Hessian matrix composed of the second derivatives is neither positive nor negative definite. This in turn leads to the
observation that there are no local extrema inside the domain. Consequently, the global extrema must be somewhere on the surface
of the hypercube. In a similar fashion, again by means of Eq.~\eqref{masodikder}, one can easily prove that there are no local
extrema on lower dimension parts of the surface either. Therefore, the global extrema must reside somewhere in the corners. Note
that the functions $m_i(\mathbf{k})$ are symmetric under all permutations of vector components. This feature calls for a
classification of the 16 corners and we find 5 classes with representatives: $\mathbf{k}_1=(1,1,1,1)$, $\mathbf{k}_2=(1,1,1,n)$,
$\mathbf{k}_3=(1,1,n,n)$, $\mathbf{k}_4=(1,n,n,n)$ and $\mathbf{k}_5=(n,n,n,n)$, respectively. With all these it is sufficient to
compare $\theta$ values at class representatives: let $\theta_i=\theta(\mathbf{k}_i)$, then
\begin{align}
\theta_1&=-\frac{16}{15}\cos^3\frac{\pi}{n+1}-\frac{4}{5}\cos^2\frac{\pi}{n+1}-\frac{16}{5}\cos\frac{\pi}{n+1}
+\frac{76}{15},\label{t1}\\
%
\theta_2&=\frac{8}{15}\cos^3\frac{\pi}{n+1}-\frac85\cos\frac{\pi}{n+1}+\frac{76}{15},\label{t2}\\
\theta_3&=\frac{4}{15}\cos^2\frac{\pi}{n+1}+\frac{76}{15},\label{t3}\\
\theta_4&=-\frac{8}{15}\cos^3\frac{\pi}{n+1}+\frac85\cos\frac{\pi}{n+1}+\frac{76}{15},\label{t4}\\
\theta_5&=\frac{16}{15}\cos^3\frac{\pi}{n+1}-\frac{4}{5}\cos^2\frac{\pi}{n+1}
+\frac{16}{5}\cos\frac{\pi}{n+1}+\frac{76}{15}.\label{t5}
\end{align}
Comparison of these formulas shows that $\theta_\text{min}=\theta_1$ and $\theta_\text{max}=\theta_5$ for all $n$, which proves
Eq.~\eqref{minmax} indeed.

We end the discussion of $\mathbf{M}$ with the presentation of some asymptotic formulas. From Eq.~\eqref{minmax} one can easily
extract the information that $\theta_\text{min}$ and $\theta_\text{max}$ are strictly decreasing and increasing functions of $n$,
respectively. In particular, asymptotic expansion yields
\begin{align}
\theta_\text{min}&=\frac{4\pi^2}{n^2}\left(1-\frac{2}{n}+\dots\right),\label{tmin}\\
\theta_\text{max}&=\frac{128}{15}\left(1-\frac{9\pi^2}{32n^2}+\dots\right),\label{tmax}
\end{align}
thus the spectrum obeys
\begin{equation}
\rho(\mathbf{M})=\{\theta_\text{min},\dots,\theta_\text{max}\}\subset[0,128/15].\label{sm}
\end{equation}
The most interesting asymptotic expression, however, is
\begin{equation}
(n+1)^2\theta(\mathbf{k})=\epsilon(\mathbf{k})\left(1-\frac{\epsilon(\mathbf{k})}{12n^2}
+\frac{\epsilon(\mathbf{k})}{6n^3}+\dots\right),\label{asympttheta}
\end{equation}
where
\begin{equation}
\epsilon(\mathbf{k})=\pi^2\sum_{i=1}^4k_i^2,\qquad k_i=1,2,\dots\label{boxenergy}
\end{equation}
is nothing else but the allowed energy levels of a particle in a four-dimensional infinite square well. This we have already
encountered apropos of Eq.~\eqref{spektrumlimit} in the main text. There we emphasized that these are the eigenvalues of the
four-dimensional negative Laplace operator supplied by the boundary condition of
Eq.~\eqref{4Dconstraint}. Equation~\eqref{asympttheta} thus indicates that $(n+1)^2\mathbf{M}$ evolves smoothly into $-\Delta$,
and the error is of order $\mathcal{O}(n^{-2})$. As we shall see next, this will be refined considerably, to
$\mathcal{O}(n^{-6})$, by taking into account the matrix $\mathbf{N}$ as well.

\subsection{The properties of $\mathbf{N}$ and the issue of $\gamma'$}\label{sub:n}
Wrapping up our discussion on matrix properties, we finally turn our attention to the properties of $\mathbf{N}$ and the
determination of $\gamma'$. Its eigenvalues $\lambda(\mathbf{k})$ are
\begin{equation}
\lambda(\mathbf{k})=1+\left(12\gamma'-\frac{1}{30}\right)m_1(\mathbf{k})+\left(\frac{1}{36}-4\gamma'\right)m_2(\mathbf{k})
+\gamma'm_3(\mathbf{k})-\frac{1}{240}m_4(\mathbf{k}).\label{ns}
\end{equation}
In Subsection~\ref{sub:free}, apropos of the derivation of the discrete Schr\"odinger equation and Eq.~\eqref{spektrumlimit}, we
saw that $\mathbf{N}$ plays a central role in the $\mathcal{O}(h^6)$ theory. In particular, it shows up in the discrete
realization of the Laplacian. According to arguments given there in 3.\ and 4.,\ $\gamma'$ must be chosen so that $\mathbf{N}$ is
positive definite \cite{chawla}. In other words, even its smallest eigenvalue must be positive. This criterion requires us to
calculate the global minimum
\begin{equation}
\lambda_\text{min}=\min_\mathbf{k}\lambda(\mathbf{k}).\label{lambdamin}
\end{equation}
The algorithm we applied in case of $\mathbf{M}$ in Subsection~\ref{sub:m} can again be utilized: stationary points are now
determined by $\nabla_\mathbf{k}\lambda(\mathbf{k}^\star)=\mathbf{0}$, and for the second derivatives we find
\begin{equation}
\frac{\partial^2\lambda(\mathbf{k}^\star)}{\partial k_i^2}=-\frac{\pi^2}{30(n+1)^2}
\sin^2\frac{k_i^\star\pi}{n+1},\qquad i=1,\dots,4.\label{lderiv}
\end{equation}
These are apparently strictly negative, wherever $\mathbf{k}^\star$ may be. The Hessian matrix is surely not positive definite,
$\lambda$ can have no local minima. The global minimum therefore must be somewhere at the corners of the hypercube. First of all,
$\lambda$ can be evaluated at the class representatives $\mathbf{k}_i$, defined in Subsection~\ref{sub:m}. Let
$\lambda_i=\lambda(\mathbf{k}_i)$, then
\begin{align}
\lambda_1&=-\frac{1}{15}\cos^4\frac{\pi}{n+1}+32\gamma'\cos^3\frac{\pi}{n+1}+\left(\frac35-96\gamma'\right)
\cos^2\frac{\pi}{n+1}+\left(96\gamma'-\frac{4}{15}\right)\cos\frac{\pi}{n+1}+\frac{11}{15}-32\gamma',\label{lambda1}\\
\lambda_2&=\frac{1}{15}\cos^4\frac{\pi}{n+1}-16\gamma'\cos^3\frac{\pi}{n+1}-\frac{1}{15}
\cos^2\frac{\pi}{n+1}+\left(48\gamma'-\frac{2}{15}\right)\cos\frac{\pi}{n+1}+\frac{11}{15}-32\gamma',\\
\lambda_3&=-\frac{1}{15}\cos^4\frac{\pi}{n+1}+\left(32\gamma'-\frac{13}{45}\right)\cos^2\frac{\pi}{n+1}+\frac{11}{15}-32\gamma',\\
\lambda_4&=\frac{1}{15}\cos^4\frac{\pi}{n+1}+16\gamma'\cos^3\frac{\pi}{n+1}-\frac{1}{15}
\cos^2\frac{\pi}{n+1}+\left(\frac{2}{15}-48\gamma'\right)\cos\frac{\pi}{n+1}+\frac{11}{15}-32\gamma',\\
\lambda_5&=-\frac{1}{15}\cos^4\frac{\pi}{n+1}-32\gamma'\cos^3\frac{\pi}{n+1}+\left(\frac35-96\gamma'\right)
\cos^2\frac{\pi}{n+1}+\left(\frac{4}{15}-96\gamma'\right)\cos\frac{\pi}{n+1}+\frac{11}{15}-32\gamma',\label{lambda5}
\end{align}
and with these we can write
\begin{equation}
\lambda_\text{min}=\min_i\lambda_i.\label{lmin}
\end{equation}
Analysis of $\lambda_i$ as functions of $n$ and $\gamma'$ reveals that as long as
\begin{equation}
\gamma'\le\frac{23}{3840}\approx0.0059,\label{gammaregion}
\end{equation}
$\lambda_i$ are positive for all $n$. Consequently, the smallest is positive too, resulting finally in the positivity of
$\mathbf{N}$. Now the ultimate question arises: Is there a specific value in this range whose choice would provide the best
possible description of the discrete Laplace operator, $(n+1)^2\mathbf{N}^{-1}\mathbf{M}$? The answer is simple: Yes, and there is
only one! To prove this statement and determine that distinguished value, we have to first study the structure of this matrix
product.

Due to commutativity, see Eq.~\eqref{mn}, it is obvious that they possess common eigenvectors and every eigenvalue of
$\mathbf{N}^{-1}\mathbf{M}$ is of the form $\lambda^{-1}\theta$, where $\lambda$ and $\theta$ are some eigenvalues of $\mathbf{N}$
and $\mathbf{M}$, respectively. Moreover, as they exhibit the same direct polynomial structure, we can immediately write
\begin{equation}
\mathbf{N}^{-1}\mathbf{M}=\sum_\mathbf{k}\frac{\theta(\mathbf{k})}{\lambda(\mathbf{k})}
\left(\mathbf{u}_{k_1}\otimes\mathbf{u}_{k_2}\otimes\mathbf{u}_{k_3}\otimes\mathbf{u}_{k_4}\right)\circ
\left(\mathbf{u}_{k_1}^\text{T}\otimes\mathbf{u}_{k_2}^\text{T}\otimes\mathbf{u}_{k_3}^\text{T}
\otimes\mathbf{u}_{k_4}^\text{T}\right).\label{laplacespektral}
\end{equation}
The spectral representation shows unambiguously that in the product $\lambda^{-1}\theta$ mentioned above, both terms have to be
taken at the very same $\mathbf{k}$. This in turn leads to the fact that every eigenvalue can be obtained in such a way, by simply
running $\mathbf{k}$ over all allowed values. Now, having obtained the whole spectrum, we can study its asymptotic behavior for
large $n$. We obtain
\begin{equation}
(n+1)^2\frac{\theta(\mathbf{k})}{\lambda(\mathbf{k})}=
\epsilon(\mathbf{k})\left(1-\frac{a(\mathbf{k})-b(\mathbf{k})\gamma'}{n^6}+\dots\right),\label{asymptlaplace}
\end{equation}
where $\epsilon(\mathbf{k})$ is given by Eq.~\eqref{boxenergy}. This is a fundamental result for it shows that the appearance of
$\mathbf{N}^{-1}$ in the discrete Laplace operator has led to a remarkable refinement of energy levels. Compared to
Eq.~\eqref{asympttheta} we see that the order of error dropped to $\mathcal{O}(n^{-6})$, which means that the numerical procedure
converges much faster to the exact analytical results. Equation~\eqref{asymptlaplace} can be regarded as the analytic proof for
that our $\mathcal{O}(h^6)$ theory is indeed that accurate. This proof can be considered as a generalization of the results of
Refs.~\cite{chawla} and~\cite{andrew}. These papers, based on a matrix norm approach, provide analytical proofs in
one-dimension for that the local truncation error in the standard Numerov method, which is $\mathcal{O}(h^4)$, determines the
accuracy of the eigenvalues as well. A more general but from practical point of view less constructive proof is given in
Ref.~\cite{keller} for the eigenvalue problem of linear second order self-adjoint and elliptic differential operators in
arbitrary dimensions.

Another interesting observation is that the order of error is not affected by $\gamma'$, there is still a large degree of
arbitrariness in its specific choice. Though we cannot make it smaller any further, we can still try to minimize its
prefactor. The right hand side of Eq.~\eqref{asymptlaplace} indicates there is, in principle, a perfect choice for $\gamma'$ for
any given $\mathbf{k}$, the one that would make the leading error vanish. However, detailed calculations show
\begin{equation}
\min_\mathbf{k}\left(\frac{a(\mathbf{k})}{b(\mathbf{k})}\right)
=\frac{a(\mathbf{k}_1)}{b(\mathbf{k}_1)}=\frac{269}{30240}\approx0.0088,\label{minab}
\end{equation}
so even the minimum of these values is slightly out of the range specified by Eq.~\eqref{gammaregion}. From this it follows that
the most precise approximation we can ever achieve in this theory is the one where $\gamma'$ is chosen to be closest to this
minimum, namely
\begin{equation}
\gamma'=\frac{23}{3840}.\label{gfinal}
\end{equation}
One can also prove the positivity of $b(\mathbf{k})$ for all $\mathbf{k}$, thus Eqs.~\eqref{asymptlaplace} and~\eqref{gfinal}
together imply that the exact energy levels are always approached from below.

Now that we have successfully obtained a very concrete value for $\gamma'$, we can return to the issue of
$\lambda_\text{min}$. Comparison of Eqs.~\eqref{lambda1}--\eqref{lambda5} reveals that for this choice, $\lambda_5$ is the
smallest for all $n$. Moreover, it can also be shown that in this case there are no local maxima of $\lambda$ in the range
$k_i=1,\dots,n$. Hence the global maximum must also be among $\lambda_i$, in fact it is $\lambda_\text{max}=\lambda_1$. Explicitly
\begin{equation}
\begin{pmatrix}
\lambda_\text{min}\\
\lambda_\text{max}
\end{pmatrix}
=-\frac{1}{15}\cos^4\frac{\pi}{n+1}\mp\frac{23}{120}\cos^3\frac{\pi}{n+1}+\frac{1}{40}
\cos^2\frac{\pi}{n+1}\mp\frac{37}{120}\cos\frac{\pi}{n+1}+\frac{13}{24}.\label{lminmax}
\end{equation}
From these it turns out that $\lambda_\text{min}$ and $\lambda_\text{max}$ are strictly decreasing and increasing functions of
$n$, respectively. For large arguments
\begin{align}
\lambda_\text{min}&=\frac{11}{20}\frac{\pi^2}{n^2}\left(1-\frac{2}{n}+\dots\right),\label{lminasympt}\\
\lambda_\text{max}&=1-\frac13\frac{\pi^2}{n^2}+\dots,\label{lmaxasympt}
\end{align}
thus the spectrum obeys
\begin{equation}
\rho\left(\mathbf{N}\left(\frac{23}{3840}\right)\right)=\{\lambda_\text{min},\dots,\lambda_\text{max}\}\subset[0,1].\label{sn}
\end{equation}
Although $\mathbf{N}$ is positive definite, asymptotically it becomes singular at the same rate as $\mathbf{M}$, see
Eq.~\eqref{tmin}. This is an important observation signalling the fact that the numerical solution of the discrete Schr\"odinger
equation in Eq.~\eqref{vegsosch} should not rely on $\mathbf{N}^{-1}$, because $\mathbf{N}$ becomes more and more ill-conditioned
as its size grows. In fact, its condition number is
\begin{equation}
\log\left(\frac{\lambda_\text{max}}{\lambda_\text{min}}\right)\sim\log n,
\end{equation}
which is practically $\mathcal{O}(1)$ in all relevant cases. It is by all means much much larger than the numeric (computational)
precision of its matrix entries.

\subsection{Ground state energy of the Laplace operator}\label{sub:gs}
Essentially all expressions and results we obtained in this appendix so far show up in the expression for the noninteracting
ground state energy. From Eq.~\eqref{boxenergy} we know the exact analytical result:
$\epsilon_0=\epsilon(\mathbf{k}_1)=4\pi^2$. Not surprisingly, it is the first energy level of the discrete Laplace operator that
is calculated with best precision
\begin{equation}
\epsilon_0(n)=(n+1)^2\min_\mathbf{k}\left(\frac{\theta(\mathbf{k})}{\lambda(\mathbf{k})}\right)
=(n+1)^2\frac{\theta(\mathbf{k}_1)}{\lambda(\mathbf{k}_1)}=(n+1)^2\frac{\theta_\text{min}}{\lambda_\text{max}},\label{gse}
\end{equation}
where $\theta_\text{min}$ and $\lambda_\text{max}$ are shown in Eqs.~\eqref{t1} and~\eqref{lminmax}, respectively. Asymptotic
expansion yields
\begin{equation}
\epsilon_0(n\to\infty)=4\pi^2\left(1-\frac{703}{60480}\frac{\pi^6}{n^6}+\dots\right).\label{gs2}
\end{equation}
The prefactor in the error is roughly 10, so even a modest resolution of $n=10$ can lead to a relative error of $10^{-5}$, and
this is very accurate indeed.

\section{Harmonic oscillator in a box}\label{appsec:osc}
This appendix is devoted to a short study of the one-dimensional quantum harmonic oscillator that is spatially confined in a
box. We collect some useful results and formulas and also tabulate the first few allowed discrete energy levels of the
system. Similar studies have been performed in Refs.~\cite{montgomery,consortini,zicovich,rau,mei}.

In Eq.~\eqref{harmonic} we defined the dimensionless quadratic potential we use for the numerics throughout the paper. In
one-dimension
\begin{equation}
U(x)=\frac14\omega^2x^2
\end{equation}
for $|x|<b$, otherwise $U(x)=\infty$. The parameter $\omega>0$ is a scaling constant responsible for the overall strength of the
potential. The condition that the oscillator is confined means there is an infinite repulsive wall at $|x|>b$, which the wave
function $\psi(x)$ cannot penetrate into. Within the box the wave equation reads
\begin{equation}
-\frac{\text{d}^2\psi}{\text{d}x^2}+\left(\frac14\omega^2x^2-\omega\left(\nu+\frac12\right)\right)\psi=0,\label{1dosc}
\end{equation}
where the energy eigenvalue is written in the form $E=\omega(\nu+1/2)$. Note that at this stage nothing is known about the range
of $\nu$, certainly except that $\nu\ge-1/2$, as the Hamiltonian is a nonnegative operator. If the barrier is removed by the
limiting procedure $b\to\infty$, the problem becomes that of the unconstrained oscillator. Then the boundary condition is related
to the asymptotic behavior of $\psi$, namely, bound state wave functions must be square integrable and this leads finally to the
familiar result $\nu=n$ with $n=0,1,2,\dots$.

Coming back to Eq.~\eqref{1dosc} we see that it is the differential equation of parabolic cylinder functions $D_\nu$, sometimes
referred to as Weber equation \cite{abramowitz}. Two linearly independent solutions would be naturally $D_\nu(\sqrt{\omega}x)$ and
$D_\nu(-\sqrt{\omega}x)$. However, these are neither odd nor even functions of $x$, hence they are not suited well for the
particular problem. Nevertheless, we can construct appropriate linear combinations that are eigenfunctions of parity
\begin{align}
\psi_{1,\nu}(x)&=\sqrt{\omega}xe^{-\omega x^2/4}
M\left(\frac12-\frac{\nu}{2},\frac32,\frac{\omega x^2}{2}\right),\label{ptlan}\\
\psi_{2,\nu}(x)&=e^{-\omega x^2/4}M\left(-\frac{\nu}{2},\frac12,\frac{\omega x^2}{2}\right),\label{paros}
\end{align}
so $\psi_{1,\nu}$ is odd and $\psi_{2,\nu}$ is even. Here $M(a,b,z)$ is the confluent hypergeometric function of the first
kind \cite{abramowitz} defined by the hypergeometric series
\begin{equation}
M(a,b,z)=1+\frac{a}{b}z+\frac{a(a+1)}{b(b+1)}\frac{z^2}{2!}+\dots.\label{mdef}
\end{equation}
It is an entire function of $z$ provided that $b\ne0,-1,-2,\dots$. Now the general solution of Eq.~\eqref{1dosc} is
\begin{equation}
\psi_\nu(x)=A\psi_{1,\nu}(x)+B\psi_{2,\nu}(x),\label{general}
\end{equation}
with $A$ and $B$ being arbitrary constants.

The existence of discrete energy levels is related to the boundary condition imposed by the barrier
\begin{align}
0&=\psi_\nu(b)=A\psi_{1,\nu}(b)+B\psi_{2,\nu}(b),\label{e1}\\
0&=\psi_\nu(-b)=-A\psi_{1,\nu}(b)+B\psi_{2,\nu}(b).\label{e2}
\end{align}
Since this is a homogeneous system of linear equations, in order for a nontrivial solution for the coefficients $A$ and $B$ to
exist, the determinant must vanish. This in turn leads to
\begin{equation}
M\left(\frac12-\frac{\nu}{2},\frac32,\frac{\omega b^2}{2}\right)
M\left(-\frac{\nu}{2},\frac12,\frac{\omega b^2}{2}\right)=0.\label{determinant}
\end{equation}
This is the final result that implicitly determines $\nu$ and such the allowed energy levels. Clearly, this product vanishes if
either of the two terms becomes zero, thus we have to distinguish two cases:

\begin{table}
\caption{\label{tab:osc}Allowed energy levels of the one-dimensional confined harmonic oscillator, $E_n=\omega(\nu_n+1/2)$, for
  the case of $b=1$ and $\omega^2/2=500$.}
\begin{center}
\renewcommand*{\arraystretch}{1.1}
\begin{tabular}{cp{1cm}r@{}lp{1cm}cp{1cm}r@{}l}
\hline
$n$ & & & $\nu_n$ & & $n$ & & & $\nu_n$\\
\hline
0 & & 0 & .000001 & & 1 & & 1 & .000017\\
2 & & 2 & .000235 & & 3 & & 3 & .001945\\
4 & & 4 & .010898 & & 5 & & 5 & .043776\\
6 & & 6 & .132232 & & 7 & & 7 & .315886\\
8 & & 8 & .628132 & & 9 & & 10 & .088573\\
10 & & 11 & .705530 & & 11 & & 13 & .481490\\
12 & & 15 & .416694 & & 13 & & 17 & .510727\\
14 & & 19 & .763071 & & 15 & & 22 & .173266\\
\hline
\end{tabular}
\end{center}
\end{table}

\begin{enumerate}
\item Let's consider first
\begin{equation}
M\left(\frac12-\frac{\nu}{2},\frac32,\frac{\omega b^2}{2}\right)=0.\label{feltetel1}
\end{equation}
It has infinitely many positive roots which we index with nonnegative odd integers: $\nu_1<\nu_3<\nu_5\dots$. For the particular
case of $b=1$ and $\omega^2/2=500$ the first eight of these are tabulated in Table~\ref{tab:osc}. The choice for these parameters
is the same we used for numerical computations in Sections~\ref{sec:nointres} and~\ref{sec:intres}. Going back to Eq.~\eqref{e1}
or~\eqref{e2} and setting $\nu=\nu_n$ it is easy to see that $B=0$ must be. The corresponding normalized eigenstate is
\begin{equation}
\phi_n(x)=\frac{\psi_{1,\nu_n}(x)}{\|\psi_{1,\nu_n}\|},\qquad n=1,3,5,\dots,\label{ptlanosc}
\end{equation}
where the norm is as usual
\begin{equation}
\|\psi_{1,\nu_n}\|^2=\int_{-b}^b\psi^2_{1,\nu_n}(x)\text{d}x.\label{norm}
\end{equation}
It is interesting to examine Eq.~\eqref{feltetel1} for large $\nu$. Asymptotic expansion of $M$ reveals \cite{abramowitz}
\begin{equation}
\omega\left(\nu+\frac12\right)=\frac{\pi^2}{(2b)^2}(2m+2)^2,\label{dobozparos}
\end{equation}
where $m$ is a large but otherwise arbitrary integer. This shows that at high energy the spectrum reproduces half of the familiar
energy levels of a particle in a one-dimensional box of size $2b$, because $2m+2$ is always even.

\item Consider now the other equation
\begin{equation}
M\left(-\frac{\nu}{2},\frac12,\frac{\omega b^2}{2}\right)=0.\label{feltetel2}
\end{equation}
In complete analogy with 1.\ this has infinitely many positive roots which we now index with nonnegative even integers:
$\nu_0<\nu_2<\nu_4\dots$. For the same parameter values as in (i) the first eight of these are shown in Table~\ref{tab:osc}. Also,
from Eq.~\eqref{e1} we find $A=0$ and the normalized eigenstate is
\begin{equation}
\phi_n(x)=\frac{\psi_{2,\nu_n}(x)}{\|\psi_{2,\nu_n}\|},\qquad n=0,2,4,\dots,\label{psosc}
\end{equation}
where the norm is analog to that of Eq.~\eqref{norm}. For completeness we examine Eq.~\eqref{feltetel2} for large $\nu$
too. Asymptotic expansion again provides
\begin{equation}
\omega\left(\nu+\frac12\right)=\frac{\pi^2}{(2b)^2}(2m+1)^2.\label{dozozptlan}
\end{equation}
This result complements Eq.~\eqref{dobozparos} as it accounts for that part of the high energy spectrum that is related to the
square of odd integers.
\end{enumerate}

Having obtained the allowed energy levels and the corresponding normalized wave functions we can now construct the Hilbert space
of this one-dimensional problem. As eigenfunctions of the Schr\"odinger equation belonging to different eigenvalues are
necessarily orthogonal, the functions $\phi_n(x)$ defined by Eqs.~\eqref{ptlanosc} and \eqref{psosc} form an orthonormal set. If
the Hilbert space is defined as all possible linear combinations of this set, then completeness follows naturally.

Summing up the results, we found that the infinite potential barrier superimposed on the quadratic potential shifts the allowed
energy levels upwards, but otherwise, as expected, leaves the discrete feature of the spectrum unaffected. At high energy the
spectrum and wave functions turn into those of the ``particle in a box'' problem. For the particular choice of parameters we found
that close to the ground state the wave functions and their energies are practically those of the unconstrained
oscillator. Nevertheless, in higher dimensions these shifts, whatever small they are, will lift certain degeneracies of the
excited energy levels and so lead to qualitative changes.

\section{Irreducible representations of $\mathcal{G}$}\label{appsec:irr}
In this appendix we explicitly construct all inequivalent unitary irreducible representations of the group of the Schr\"odinger
equation. In Eq.~\eqref{semi} we have already pointed out that $\mathcal{G}$ has a semi-direct product structure. This is a rather
satisfactory situation, because in this special case the knowledge of all irreducible representations of $\mathcal{A}$ and
$\mathcal{B}$ is sufficient to induce those of $\mathcal{G}$ by means of general theorems of group theory. The method itself is
called induction. Before that, however, we shall give a very brief overview of the representations of $\mathcal{A}$ and
$\mathcal{B}$. Representations and characters will be denoted by ${\bm\Gamma}$ and $\chi$, respectively.

\subsection{Representations of $\mathcal{A}$ and $\mathcal{B}$}\label{sub:abrep}
Using the matrices of Eq.~\eqref{amembers} direct calculations may verify that $\mathcal{A}$ is indeed Abelian. Irreducible
representations are necessarily one-dimensional. Furthermore, as $\mathbf{R}_2=\mathbf{R}_3\mathbf{R}_4$ and the square of any
member equals identity, it has a direct-product structure
\begin{equation}
\mathcal{A}\sim\{\mathbf{R}_1,\mathbf{R}_3\}\otimes\{\mathbf{R}_1,\mathbf{R}_4\}\sim C_2^2.\label{ac2}
\end{equation}
It shows that $\mathcal{A}$ is isomorphic to $C_2^2$, with $C_2$ being the cyclic group of order 2. From this the character table
follows naturally, see Table~\ref{tab:char}.
\begin{table}
\caption{\label{tab:char}Character tables of $\mathcal{A}$ (left) and $\mathcal{B}$ (right). As $\mathcal{A}$ is Abelian all
  irreducible representations are one dimensional and the characters themselves are the matrix elements. Conjugacy classes are
  denoted by $\mathcal{C}_i$.}
\renewcommand*{\arraystretch}{1.1}
\begin{tabular*}{5.2cm}{@{\extracolsep{\fill}}lrrrr}
\hline
& $\mathcal{C}_1=\mathbf{R}_1$ & $\mathcal{C}_2=\mathbf{R}_2$ & $\mathcal{C}_3=\mathbf{R}_3$ & $\mathcal{C}_4=\mathbf{R}_4$\\ 
\hline
$\chi_\mathcal{A}^1$ & 1 &  1 &  1 &  1\\
$\chi_\mathcal{A}^2$ & 1 &  $-1$ & 1 & $-1$\\
$\chi_\mathcal{A}^3$ & 1 & $-1$ &  $-1$ & 1\\
$\chi_\mathcal{A}^4$ & 1 & 1 & $-1$ & $-1$\\
& & & &\\
\hline
\end{tabular*}
\hspace{\fill}
\begin{tabular*}{9cm}{@{\extracolsep{\fill}}lrrrrr}
\hline
& $\mathcal{C}_1=\mathbf{R}_1$ & $\mathcal{C}_2=\{\mathbf{R}_4,\mathbf{R}_8\}$
& $\mathcal{C}_3=\mathbf{R}_5$ & $\mathcal{C}_4=\{\mathbf{R}_3,\mathbf{R}_7\}$
& $\mathcal{C}_5=\{\mathbf{R}_2,\mathbf{R}_6\}$\\
\hline
$\chi_\mathcal{B}^1$ & 1 &  1 &  1 &  1 & 1\\
$\chi_\mathcal{B}^2$ & 1 &  1 & 1 & $-1$ & $-1$\\
$\chi_\mathcal{B}^3$ & 1 & $-1$ & 1 &  1 & $-1$\\
$\chi_\mathcal{B}^4$ & 1 & $-1$ & 1& $-1$ &  1\\
$\chi_\mathcal{B}^5$ & 2 & 0 & $-2$ & 0 & 0\\
\hline
\end{tabular*}
\end{table}

The matrix group $\mathcal{B}$ was defined by its members in Eq.~\eqref{bmembers}. As noted there, it is essentially nothing else
than a faithful four-dimensional representation of the point symmetry group of a square, $C_{4v}$. The fact it is a group of order
8 and has 5 classes involves it has 5 irreducible representations that are unique up to relabelling. Its character system is shown
in Table~\ref{tab:char}. The first four are one-dimensional, completely described by their characters. The fifth is
two-dimensional and a concrete realization with unitary matrices is
\begin{alignat}{8}
{\bm\Gamma}_\mathcal{B}^5(\mathbf{R}_1)&=
\begin{pmatrix}
1 &\\
& 1
\end{pmatrix},
&\qquad
{\bm\Gamma}_\mathcal{B}^5(\mathbf{R}_2)&=
\begin{pmatrix}
& 1\\
1 & 
\end{pmatrix},
&\qquad
{\bm\Gamma}_\mathcal{B}^5(\mathbf{R}_3)&=
\begin{pmatrix}
& -1\\
1  &
\end{pmatrix},
&\qquad
{\bm\Gamma}_\mathcal{B}^5(\mathbf{R}_4)&=
\begin{pmatrix}
1  &\\
& -1
\end{pmatrix},\notag\\
{\bm\Gamma}_\mathcal{B}^5(\mathbf{R}_5)&=-
\begin{pmatrix}
1 &\\
& 1
\end{pmatrix},
&\qquad
{\bm\Gamma}_\mathcal{B}^5(\mathbf{R}_6)&=-
\begin{pmatrix}
& 1\\
1 &
\end{pmatrix},
&\qquad
{\bm\Gamma}_\mathcal{B}^5(\mathbf{R}_7)&=-
\begin{pmatrix}
& -1\\
1 &
\end{pmatrix},
&\qquad
{\bm\Gamma}_\mathcal{B}^5(\mathbf{R}_8)&=-
\begin{pmatrix}
1 &\\
& -1
\end{pmatrix}.\label{b5}
\end{alignat}
Comparison with Eq.~\eqref{bmembers} shows that these are nothing else but the upper left (or bottom right) 2-by-2 blocks of
$\mathbf{R}_i$, so ${\bm\Gamma}_\mathcal{B}^5$ is a faithful representation.

\subsection{Induced representations of $\mathcal{G}=\mathcal{A}\circledS\mathcal{B}$}\label{sub:ind}
In order to obtain the representations ${\bm\Gamma}$ of $\mathcal{G}$ we have to first explore the little groups of
$\mathcal{B}$. In what follows we apply the same notation as that of Ref.~\cite{cornwell-book}. Let $\mathcal{B}(q)$ be the
subset of elements $\mathbf{R}_b$ of $\mathcal{B}$ such that
\begin{equation}
\chi_\mathcal{A}^q(\mathbf{R}_b\mathbf{R}_a\mathbf{R}_b^{-1})=\chi_\mathcal{A}^q(\mathbf{R}_a)\label{little}
\end{equation}
for all $\mathbf{R}_a\in\mathcal{A}$. Then $\mathcal{B}(q)$ is a subgroup of $\mathcal{B}$ and it is called the $q$th little
group. Straightforward calculation yields
\begin{align}
\mathcal{B}(1)&=\mathcal{B}(4)=\mathcal{B},\label{14}\\
\mathcal{B}(2)&=\mathcal{B}(3)=\left\{\mathbf{R}_1,\mathbf{R}_4,\mathbf{R}_5,\mathbf{R}_8\right\}.\label{23}
\end{align}
For explicit forms of matrices see Eq.~\eqref{bmembers}. According to Eq.~\eqref{14} the little groups of $q=1$ and 4 are
$\mathcal{B}$ itself thus the orbit of $q=1$ is the set $\{1\}$, whereas that of $q=4$ is $\{4\}$. Irreducible representations of
$\mathcal{G}$ belonging to these values are found easily \cite{cornwell-book}
\begin{equation}
{\bm\Gamma}^{qp}(\mathbf{R}_a\mathbf{R}_b)=\chi_\mathcal{A}^q(\mathbf{R}_a)
{\bm\Gamma}_\mathcal{B}^p(\mathbf{R}_b),\qquad q=1,4,\quad p=1,\dots,5.\label{rep14}
\end{equation}
Essentially only $p=5$ is a real matrix representation because it is two-dimensional, see Eq.~\eqref{b5}. The others are
one-dimensional and their matrix elements are tabulated in Table~\ref{tab:char}.

On the other hand it turns out that the orbit of $q=2$ as well as $q=3$ is $\{2,3\}$. This is the case where the method of
induction is actually made use of. As it is sufficient to consider only one element in each orbit, we choose $q=2$. We will need
shortly the coset representatives for the decomposition of $\mathcal{B}$ into right cosets with respect to $\mathcal{B}(2)$: they
are $\mathbf{R}_1$ and $\mathbf{R}_7$ of $\mathcal{B}$, respectively.  Also, inspection of $\mathcal{B}(2)$ reveals that it is
Abelian and isomorphic to $\mathcal{A}$
\begin{gather}
\mathbf{R}_1\in\mathcal{B}\longleftrightarrow\mathbf{R}_1\in\mathcal{A},\\
\mathbf{R}_4\in\mathcal{B}\longleftrightarrow\mathbf{R}_4\in\mathcal{A},\\
\mathbf{R}_5\in\mathcal{B}\longleftrightarrow\mathbf{R}_3\in\mathcal{A},\\
\mathbf{R}_8\in\mathcal{B}\longleftrightarrow\mathbf{R}_2\in\mathcal{A}.
\end{gather}
Because of this the representations of $\mathcal{B}(2)$ are the same as those of $\mathcal{A}$ shown in Table~\ref{tab:char}. With
all these information at hand we are now in a position to explicitly give the matrix elements of the remaining four
two-dimensional irreducible representations of $\mathcal{G}$. The upper left elements read
\begin{align}
\Gamma^{2p}(\mathbf{R}_a\mathbf{R}_b)_{11}&=
\begin{cases}
\chi_\mathcal{A}^2(\mathbf{R}_a)\chi_{\mathcal{B}(2)}^p(\mathbf{R}_b),
&\text{if $\mathbf{R}_b\in\mathcal{B}(2)$},\\
0,&\text{if $\mathbf{R}_b\notin\mathcal{B}(2)$},
\end{cases}
\qquad p=1,\dots,4,
\end{align}
whereas the upper right are
\begin{align}
\Gamma^{2p}(\mathbf{R}_a\mathbf{R}_b)_{12}&=
\begin{cases}
\chi_\mathcal{A}^2(\mathbf{R}_a)\chi_{\mathcal{B}(2)}^p(\mathbf{R}_b\mathbf{R}_7^{-1}),
&\text{if $\mathbf{R}_b\mathbf{R}_7^{-1}\in\mathcal{B}(2)$},\\
0,&\text{if $\mathbf{R}_b\mathbf{R}_7^{-1}\notin\mathcal{B}(2)$},
\end{cases}
\qquad p=1,\dots,4.
\end{align}
The lower left are given by
\begin{align}
\Gamma^{2p}(\mathbf{R}_a\mathbf{R}_b)_{21}&=
\begin{cases}
\chi_\mathcal{A}^3(\mathbf{R}_a)\chi_{\mathcal{B}(2)}^p(\mathbf{R}_7\mathbf{R}_b),
&\text{if $\mathbf{R}_7\mathbf{R}_b\in\mathcal{B}(2)$},\\
0,&\text{if $\mathbf{R}_7\mathbf{R}_b\notin\mathcal{B}(2)$},
\end{cases}
\qquad p=1,\dots,4,
\end{align}
and finally the lower right are obtained as
\begin{align}
\Gamma^{2p}(\mathbf{R}_a\mathbf{R}_b)_{22}&=
\begin{cases}
\chi_\mathcal{A}^3(\mathbf{R}_a)\chi_{\mathcal{B}(2)}^p(\mathbf{R}_7\mathbf{R}_b\mathbf{R}_7^{-1}),
&\text{if $\mathbf{R}_7\mathbf{R}_b\mathbf{R}_7^{-1}\in\mathcal{B}(2)$},\\
0,&\text{if $\mathbf{R}_7\mathbf{R}_b\mathbf{R}_7^{-1}\notin\mathcal{B}(2)$},
\end{cases}
\qquad p=1,\dots,4.
\end{align}

Perhaps the most important irreducible representation of all, though mathematically trivial, is the so-called completely symmetric
representation given by
\begin{equation}
{\bm\Gamma}^{11}(\mathbf{R})=1,\qquad\text{for all $\mathbf{R}\in\mathcal{G}$}.\label{compsymm}
\end{equation}

Summarizing the results achieved in this appendix we can say there are altogether $5+5+4=14$ inequivalent unitary irreducible
representations of the group of the Schr\"odinger equation. Because of their intrinsic product structure they are conveniently
labeled by a composite index $qp$ (no multiplication). If $q=1$ or 4 then $p=1,\dots,5$, while if $q=2$ then $p=1,\dots,4$. Among
the fourteen representations there are six two-dimensional: $qp=15$, 45, 21, 22, 23 and 24, respectively. All others are
one-dimensional. Since all representations involve only real matrices the unitary property is equivalent to
orthogonality. Further, from the character tables and the formulas above it is apparent that any matrix element can only be 0, 1
or $-1$. This, in conjunction with orthogonality results in the fact that each ${\bm\Gamma}^{qp}(\mathbf{R})$ is a so-called
signed permutation matrices: there is exactly one nonzero entry in each row and column and these are either 1 or $-1$.

\section{Symmetry adapted basis of the Hilbert space}\label{appsec:basis}
In Subsection~\ref{sub:pro} we introduced a low-energy subspace $\mathcal{H}_m$ in the full Hilbert space. This is spanned by all
real linear combinations of the vectors $\mathbf{v}(\mathbf{k})$ given by Eq.~\eqref{linalgbasis}. In this appendix we show that
the method of projections results in a new orthogonal basis in it, every member of which transforms as some row of some
irreducible representation of $\mathcal{G}$. This newly formed set is called the symmetry adapted basis and will be used in the
solution of the eigenvalue problem.

The four components of $\mathbf{k}$ take integer values in the range $0,\dots,m-1\le n$, there are thus altogether $m^4$ such
vectors. Let us denote the set of all $\mathbf{k}$ by $K$. As to the scalar product we find
\begin{equation}
\left(\mathbf{v}(\mathbf{k}),\mathbf{v}(\mathbf{k'})\right)=\sum_{\mu=1}^{n^4}v_\mu(\mathbf{k})v_\mu(\mathbf{k'})
=\delta_{\mathbf{k}\mathbf{k'}},\label{ortonormv}
\end{equation}
indicating these vectors are indeed orthonormal.

Consider next the group of the Schr\"odinger equation. From either Eqs.~\eqref{amembers}, \eqref{bmembers} and~\eqref{invgroup},
or from the fact that $\mathcal{G}$ is a subgroup of $\text{O}_4$ it follows
\begin{equation}
\mathbf{R}=\mathbf{d}(\mathbf{R}){\bm\sigma}(\mathbf{R})\label{rdecomp}
\end{equation}
for all $\mathbf{R}\in\mathcal{G}$. Here $\mathbf{d}$ is a diagonal matrix with $\pm1$ in the diagonal and ${\bm\sigma}$ is a
permutation matrix with only one nonzero element in each row and column, which is 1. Actually, ${\bm\sigma}$ is nothing else than
$|\mathbf{R}|$ where the absolute value should be taken element-wise. This decomposition is unique and the order of terms is
important. Writing the terms in opposite order yields
\begin{equation}
{\bm\sigma}\mathbf{d}=\text{diag}\left({\bm\sigma}\text{diag}(\mathbf{d})\right){\bm\sigma}\ne\mathbf{d}{\bm\sigma},
\label{opp}
\end{equation}
where $\text{diag}(\mathbf{a})$ is either a diagonal matrix composed of the column vector $\mathbf{a}$ or a column vector
extracted from the diagonal matrix $\mathbf{a}$. With the aid of this equation it is now not too difficult to verify that the
mapping ${\bm\sigma}(\mathbf{R})$ from $\mathcal{G}$ is a four-to-one homomorphism onto a group of permutations
$\mathcal{G}^\star$ of order 8. Applying the scalar transformation operator of Eq.~\eqref{opkron} to a basis vector we get
\begin{equation}
\mathbf{P}(\mathbf{R})\mathbf{v}(\mathbf{k})=f(\mathbf{R},\mathbf{k})\mathbf{v}({\bm\sigma}(\mathbf{R})\mathbf{k}),\label{basisop}
\end{equation}
where the coefficient is
\begin{equation}
f(\mathbf{R},\mathbf{k})=\prod_{i=1}^4\left(\sum_{p=1}^4R_{pi}\right)^{k_i}.
\end{equation}
Careful inspection justifies the conjecture that $f$ is also a homomorphic mapping of $\mathcal{G}$
\begin{equation}
f(\mathbf{R'R},\mathbf{k})=f(\mathbf{R'},{\bm\sigma}(\mathbf{R})\mathbf{k})f(\mathbf{R},\mathbf{k}).\label{fhom}
\end{equation}
This time it is sixteen-to-one and the mapping is onto the group $\{1,-1\}$. Going back to Eq.~\eqref{basisop} we see that the
basis vectors have a characteristic feature: under coordinate transformations, apart from potential sign changes, they transform
among themselves. This indicates that $\mathcal{H}_m$ is an invariant subspace for all $\mathbf{P}(\mathbf{R})$.

Next we define certain subsets of $K$. Let $K(\mathbf{k})$ be the subset including the following permutations
\begin{equation}
K(\mathbf{k})=\left\{{\bm\sigma}\mathbf{k}\mid{\bm\sigma}\in\mathcal{G}^\star\right\},\label{kk}
\end{equation}
and an element is contained only once. It is obvious that every element of $K$ is in exactly one such class and the set of all
different classes forms a complete disjoint decomposition of $K$. Combinatorial calculation gives for the total number of
different classes
\begin{equation}
N=
\begin{pmatrix}
m\\1
\end{pmatrix}+4
\begin{pmatrix}
m\\2
\end{pmatrix}+6
\begin{pmatrix}
m\\3
\end{pmatrix}+3
\begin{pmatrix}
m\\4
\end{pmatrix}.\label{totaln}
\end{equation}
Let us denote the $i$th class by $K_i$ and its order by $z_i$. Thus we obviously have
\begin{equation}
\bigcup_{i=1}^NK_i=K,\label{union}
\end{equation}
and
\begin{equation}
\sum_{i=1}^Nz_i=m^4.\label{dimensions}
\end{equation}
Now apply the projection operators as prescribed in Eq.~\eqref{newbasis}. Using Eq.~\eqref{basisop} we get
\begin{equation}
\mathbf{w}_j^{qp}(\mathbf{k})=\mathbf{P}_{jj}^{qp}\mathbf{v}(\mathbf{k})=
(d_{qp}/g)\sum_{\mathbf{R}\in\mathcal{G}}\Gamma^{qp}(\mathbf{R})_{jj}f(\mathbf{R},\mathbf{k})
\mathbf{v}({\bm\sigma}(\mathbf{R})\mathbf{k}),\label{newbasis2}
\end{equation}
showing that for any given $\mathbf{k}$ the new function is either zero or a nonzero linear combination of the original basis
elements belonging to the class $K(\mathbf{k})$. This means that not just the whole space associated with $K$ but also each
smaller dimensional subspace associated with $K_i$ are invariant subspaces of the scalar transformation operators as well as the
projections. For the moment let us assume that the vector in Eq.~\eqref{newbasis2} does not vanish. Therefore, for any other
$\mathbf{k'}$ that is not contained in $K(\mathbf{k})$ the resulting $\mathbf{w}_j^{qp}(\mathbf{k'})$, if not zero, must be
necessarily orthogonal to $\mathbf{w}_j^{qp}(\mathbf{k})$ and so linearly independent. On the other hand, should it be any other
member of the same class we would obtain
\begin{equation}
\mathbf{w}_j^{qp}(\mathbf{k'})=\mathbf{P}_{jj}^{qp}\mathbf{v}(\mathbf{k'})=f(\mathbf{R}^\star,\mathbf{k})
\sum_{i=1}^{d_{qp}}\Gamma^{qp}(\mathbf{R}^\star)_{ji}\mathbf{P}^{qp}_{ji}\mathbf{v}(\mathbf{k}),\label{masik}
\end{equation}
where we used that $f$, ${\bm\sigma}$ and the representation ${\bm\Gamma}^{qp}$ are all homomorphisms and $\mathbf{R}^\star$
is any element of $\mathcal{G}$ such that $\mathbf{k'}={\bm\sigma}(\mathbf{R}^\star)\mathbf{k}$. Now it is easy to see that if the
representation is one-dimensional, that is $d_{qp}=i=j=1$, this vector is proportional to $\mathbf{w}_j^{qp}(\mathbf{k})$, which
in turn means linear dependence.

If $d_{qp}$ is greater than one the analysis of independence is not so trivial and we can only quote here the result: there might
be more than one linearly independent vectors, but in any case they are mutually orthogonal. Though the proof of orthogonality
needs some extra effort, the number of these functions can be obtained quite easily as follows. Introduce the quantity $r_i^{qp}$,
which is by definition an integer and measures the number of linearly independent functions transforming as, say the $j$th row of
${\bm\Gamma}^{qp}$ and result from the projections applied on the subspace associated with $K_i$. They are
$\mathbf{w}_j^{qp}(\mathbf{k}_s)$, $s=1,\dots,r_i^{qp}$. Let us now rewrite Eq.~\eqref{basisop} as
\begin{equation}
\mathbf{P}(\mathbf{R})\mathbf{v}(\mathbf{k})=\sum_{\mathbf{k'}\in K_i}\Gamma(\mathbf{R})_{\mathbf{k'}\mathbf{k}}
\mathbf{v}(\mathbf{k'}),\label{basisop2}
\end{equation}
where $\mathbf{k}\in K_i$ as well and
\begin{equation}
\Gamma(\mathbf{R})_{\mathbf{k'}\mathbf{k}}=f(\mathbf{R},\mathbf{k})
\delta_{\mathbf{k'},{\bm\sigma}(\mathbf{R})\mathbf{k}}.\label{redgamma}
\end{equation}
As the notation suggests, this is a representation of $\mathcal{G}$ of dimension $z_i$. Also, because of Eq.~\eqref{ortonormv} the
representation consists of orthogonal matrices only. This, in conjunction with Eq.~\eqref{redgamma} shows that the matrices are
actually signed permutation matrices. The number of times $r_i^{qp}$ that an irreducible representation ${\bm\Gamma}^{qp}$
appears in it is given explicitly by the respective characters
\begin{equation}
r_i^{qp}=\frac{1}{g}\sum_{\mathbf{R}\in\mathcal{G}}\chi_i(\mathbf{R})\chi^{qp}(\mathbf{R}),\label{rqp}
\end{equation}
where $\chi^{qp}$ is the character of ${\bm\Gamma}^{qp}$ (see Appendix~\ref{appsec:irr}) and
$\chi_i(\mathbf{R})=\sum_{\mathbf{k}\in K_i}\Gamma(\mathbf{R})_{\mathbf{kk}}$. These also obey
\begin{equation}
z_i=\sum_{qp}d_{qp}r_i^{qp}.\label{dimi}
\end{equation}
A very interesting result is found finally: in any linearly independent set $\left\{\mathbf{w}_j^{qp}(\mathbf{k})\right\}$ (here
the indices are fixed and $\mathbf{k}$ varies) the members are mutually orthogonal. Remember that independence of vectors in inner
product spaces is an obvious consequence of orthogonality, but the reverse statement does not hold usually. Now that we have
explored these characteristics we can give an efficient algorithm for the explicit construction of the symmetry adapted
basis. This method is also easy to implement numerically.
\begin{enumerate}
\item Consider a class $K_i$ of $K$ and a projection $\mathbf{P}^{qp}_{jj}$ with indices fixed. Apply it to the elements of the
  class one by one. According to the results above, in the newly formed set $\left\{\mathbf{w}_j^{qp}(\mathbf{k})\right\}$, which
  is at most of order $z_i$, there is at most one linearly independent vector if the representation is one-dimensional. In this
  case one should keep the first and forget about the rest. Note, that it might happen that the projection results in zero vectors
  only. In this case go to (ii) because we don't need them: members of a basis can only be nonzero vectors. On the other hand, if
  a representation of dimension greater than one is considered the results above tell that there are exactly $r_i^{qp}$
  independent functions in the set and these are orthogonal. As a result the selection mechanism of independence can be equally
  based on orthogonality which is much faster numerically, at least when there are many (and large) vectors to deal with.
\item Repeat (i) for every $N$ classes of $K$, all irreducible representations $qp$ and all rows $j$.
\end{enumerate}
The vectors kept in this algorithm necessarily form a maximal linearly independent set, all of which are members of
$\mathcal{H}_m$. The crucial observation is now the fact that their number is exactly $m^4$, which follows from the completeness
of projections \cite{cornwell-book}
\begin{equation}
\sum_{qp}\sum_{j=1}^{d_{qp}}\mathbf{P}^{qp}_{jj}=\mathbf{E}_{n^4}.\label{teljesseg}
\end{equation}
This means that the symmetry adapted vectors form orthogonal but not necessarily normalized basis for $\mathcal{H}_m$ and this is
what we wanted to prove. With the aid of Eq.~\eqref{rqp} the total number of independent elements in the new basis transforming as
say the $j$th row of ${\bm\Gamma}^{qp}$ reads
\begin{equation}
r^{qp}=\sum_{i=1}^Nr_i^{qp}.\label{dim}
\end{equation}
The vectors themselves will be denoted by $\mathbf{w}_j^{qp}(\mathbf{k}_s)$, $s=1,\dots,r^{qp}$. Similarly to Eq.~\eqref{dimi} the
dimensions satisfy an analogue constraint
\begin{equation}
m^4=\sum_{qp}d_{qp}r^{qp},\label{vegsodim}
\end{equation}
which follows immediately from Eq.~\eqref{dimensions} and expresses the fact that the whole subspace associated with $K$
decomposes into a direct sum of smaller dimensional subspaces, each belonging to a specific irreducible representation.

\bibliography{schrodinger}
\bibliographystyle{elsart-num.bst}

\end{document}